%% file: iCIPT2.tex
\documentclass[journal=jctcce,manuscript=article]{achemso}

\usepackage[T1]{fontenc} 
\usepackage[version=3]{mhchem} 

\usepackage{amsfonts}
\usepackage{amsmath}
\usepackage{booktabs}
\usepackage{tikz}
\usepackage{threeparttable}
\usepackage{multirow}
\usepackage{appendix}
\usepackage{amssymb,amsmath}
\usepackage{graphicx}
\usepackage{subfigure}
\usepackage{enumerate}
\usepackage{colortbl}
\usepackage{framed}
\usepackage{verbatim}
\usepackage{amsbsy}
\usepackage{bm}
\usepackage{bbm}
\usepackage{ulem}
\usepackage{float}
\usepackage[linesnumbered,boxed]{algorithm2e}
\usepackage{algorithm2e}
\usepackage{lscape}

\newfloat{Algorithm}{htbp}{alg}
\floatname{Algorithm}{Algorithm}

\newcounter{exe}[figure]
\newcommand{\iexe}{\refstepcounter{exe}\the\value{exe}:}

\setkeys{acs}{maxauthors = 0} 

\author{Ning Zhang}
\affiliation{Beijing National Laboratory for Molecular Sciences, Institute of Theoretical and Computational Chemistry,
College of Chemistry and Molecular Engineering, Beijing 100871, China}
\author{Wenjian Liu}\email{liuwj@sdu.edu.cn}
\affiliation{Qingdao Institute for Theoretical and Computational Sciences,
Shandong University, Qingdao, Shandong 266237, China}
\author{Mark R. Hoffmann}
\affiliation{Chemistry Department, University of North Dakota, Grand Forks, ND 58202-9024, U.S.A.}

\title[\texttt{achemso} demonstration]
{Iterative Configuration Interaction with Selection}

\setkeys{acs}{articletitle=true}

\begin{document}

\begin{abstract}
Even when starting with a very poor initial guess,
the iterative configuration interaction (iCI) approach [J. Chem. Theory Comput. 12, 1169 (2016)]
for strongly correlated electrons can converge from above to full CI (FCI) very quickly by constructing and diagonalizing a very small Hamiltonian matrix at each macro/micro-iteration.
However, as a direct solver of the FCI problem, iCI scales exponentially with respect to the numbers of electrons and orbitals.
The problem can be mitigated by observing that a vast number of configurations have little weights in the wave function and hence
do not contribute discernibly to the correlation energy. The real questions are then (a) how to identify those important configurations as early as possible in the calculation
and (b) how to account for the residual contributions of those unimportant configurations. It is generally true that
if a high-quality yet compact variational space can be determined for describing the static correlation, a low-order treatment of
the residual dynamic correlation would then be sufficient. While this is common to all selected CI schemes, the `iCI with selection' scheme
presented here has the following distinctive features:
(1) Full spin symmetry is always maintained by taking configuration state functions (CSF) as the many-electron basis.
(2) Although the selection is performed on individual CSFs, it is orbital configurations (oCFG) that are used as the organizing units.
(3) Given a coefficient pruning-threshold $C_{\mathrm{min}}$ (which determines the size of the variational space for static correlation),
the selection of important oCFGs/CSFs is performed iteratively until convergence.
(4) At each iteration for the growth of the wave function, the first-order interacting space is decomposed into disjoint subspaces, so as to reduce memory requirement on one hand and facilitate parallelization on the other.
(5) Upper bounds (which involve only two-electron integrals) for the interactions between doubly connected oCFG pairs are
used to screen each first-order interacting subspace before the first-order coefficients of individual CSFs are evaluated.
(6) Upon convergence of the static correlation for a given $C_{\mathrm{min}}$, dynamic correlation is estimated by using the state-specific Epstein-Nesbet second-order perturbation theory (PT2). The efficacy of the iCIPT2 scheme is demonstrated numerically by benchmark examples, including \ce{C2}, \ce{O2}, \ce{Cr2} and \ce{C6H6}.
\end{abstract}
\newpage
\section{Introduction}
The common paradigm for treating strongly correlated systems of electrons is to decompose
the overall electron correlation into static/nondynamic and dynamic components and treat them differently.
While such decomposition is not unique and is sometimes even impossible, it provides insights and hence guidance for designing various highly accurate correlation methods.
As a matter of fact, all the available wave function-based correlation methods
can, in this context, be classified into three families according to when the static and dynamic components of electron correlation
are handled, viz. ``static-then-dynamic'', ``dynamic-then-static'' and ``static-dynamic-static'' (SDS)\cite{iCI}. Conceptually, the only exception is full configuration interaction (FCI) that
does not rest on such decomposition at all. In practice, however, making proper use of the distinction between static and dynamic
correlations still allows one to design algorithms\cite{iCI,iFCI2017b,MBE-FCI2019,i-FCIQMC} that can converge much faster to FCI than those that do not make any use of it.
It is clear that the most effective means for calculating the static and dynamic correlations are diagonalization and perturbation
(including possible resummations, e.g., coupled cluster), respectively.
This explains the efficacy of various multireference perturbation theories\cite{MRPT2Rev1,MRPT2Rev2}. However, the underlying
complete active space (CAS), albeit operationally simple, strongly limits the applicability of such multireference methods for obvious reasons:
(1) The size of CAS grows combinatorially with respect to the numbers of active electrons and active orbitals.
(2) It is by no means trivial to maintain the same CAS when computing potential energy surfaces.
(3) Unless sufficiently large, a given CAS is usually not equally good for all target states.
(4) A large CAS usually contains many intermediate states that are even higher in energy than numerous states in the complimentary space. Treating the former better than the latter is certainly unbalanced.
(5) No matter how large the CAS is, those states having no projections onto the CAS
cannot be captured. The question is then how to generate a variational space that is compact but good enough
for all target states at any geometry. The only way to achieve this
is to introduce some selection procedure that can adapt to the variable static correlation automatically and meanwhile
can be terminated at a stage at which the residual dynamic correlation can well be described by a low-order approach.
This leads naturally to `selected configuration interaction plus second-order perturbation theory' (sCIPT2), a very old idea that can be traced back to
the end of 1960s\cite{Davidson1969,Whitten1969} and the beginning of 1970s\cite{CIPSIa,MRDCIb}. What is common
in these early works is the use of first-order coefficients or second-order energies as an importance measure for selecting configurations, so as to
build up a compact variational space in an iterative manner. The idea was further explored to achieve much improved efficiency
in the subsequent decades\cite{TableCI1980,TableCI1986,TableCI1995,CIPSIb,CIPSIc,Harrison1991,DIESEL-MR-CI,Ruedenberg2001,roth2009importance,CIPSI-DMCa},
especially in recent years\cite{LambdaCI2014,ACI2016,ACI2017,ACI2018,HBCI2016,HBCI2017a,HBCI2017b,ASCI2016,ASCI2017,ASCI2018,ASCI2018PT2,DressedSCI2018,CIPSI-DMCd,CIPSI-DMCe}.
At variance with such deterministic selection, stochastic Monte Carlo\cite{MCCI1998,MCCIPT2,MCCI2017,i-FCIQMCPT2} and machine learning\cite{ML-CI}
types of selection procedures have also been proposed in the past. Note in passing that no particular structure of the wave function
has been assumed in the aforementioned deterministic or stochastic selection schemes, in contrast with those
in terms of a certain structure of the wave function\cite{CCCI1987,CCCI1988,NakatsujiExpCI1991,DDCI1993,DDCI1995,CISD-TQ1996,SORCI2003,Bunge2006selected,RuedenbergDeadwood2009,SeniorityCI2011,RASPT22008,SplitCASPT22011}.
While the latter are potentially more efficient in selecting configurations for specific problems, the former are more general, simply because an assumed structure of the wave function
may not always hold, e.g., for non-energetic properties.

More excitingly, several new deterministic\cite{iCI,iFCI2017b,MBE-FCI2019,ICIGSD2005,DMRGa,DMRGb,IntrinsicScaling1,AdaptiveCC,SparseFCI,knowles2015compressive,ProjCI,TensorFCI2016,iFCI2017a,MBE-FCI2017,MBE-FCI2018,MBE-FCI2019generalized,FCCR2018,RR-FCI},
stochastic\cite{ProMC2008,FCIQMC2009,i-FCIQMC,S-FCIQMC,Heat-BathFCIQMC,FCIQMC-Ten-no,DMQMC2014,DMQMC2015} and stochastic-then-deterministic\cite{CC-FCIQMC2018} algorithms (which do make some selections/truncations but
do not have a final, separate step for dynamic correlation) have recently been introduced
to solve \textit{directly} the FCI problem, among which we think the iterative CI (iCI) approach\cite{iCI} is particularly distinctive:
Born from the (restricted) SDS framework\cite{SDS} for strongly correlated electrons, iCI constructs and diagonalizes a $m N_P\times m N_P$
Hamiltonian matrix at each macro/micro-iteration. Here, $N_P$ is the number of target states, whereas $m=2, 3, 4$ when zero, one or two sets of $N_P$ secondary (buffer) states are used for
the revision (relaxation) of the $N_P$ reference (primary) functions in the presence of $N_P$ first-order (external) functions describing dynamic correlation.
Since the lowest-order realization of the SDS framework, i.e., SDSPT2\cite{SDS,SDSPT2}, performs already very well for prototypical systems of variable near-degeneracies,
it is not surprising that iCI can converge from above to FCI within just a few iterations, even when starting with a very poor initial guess.
Nonetheless, iCI is so far still computationally very expensive, because no truncations of any kind have yet been made.
The way out is to combine iCI with selection of configurations. The resulting much improved algorithm can either be used
as an efficient active space solver to perform multiconfiguration self-consistent field calculations (iCISCF) with large active spaces or
as an effective means (iCIPT2) to approach FCI, where the whole Hilbert space is searched to extract a compact variational space, which is followed by a second-order perturbative treatment of
the complimentary space. The former will be published elsewhere, while the latter is presented here.

The remainder of the paper is organized as follows. The iCI approach\cite{iCI} is first recapitulated in Sec. \ref{SeciCI}.
The Hamiltonian matrix elements over configuration state functions (CSF) are then discussed in detail in Sec. \ref{SecHamiltonian}. The storage and selection of orbital configurations (oCFG) and CSFs are presented in
Sec. \ref{Selection}. Note in passing that
an oCFG $I$ for an $N$-electron system is just a product of $L$ doubly occupied and $N_{\mathrm{o}}= N- 2L=n_{\alpha}+n_{\beta}$ singly occupied spatial orbitals,
which can generate
\begin{eqnarray}
N_{M}^{\mathrm{d}}=C_{N_{\mathrm{o}}}^{n_{\alpha}}=C_{N_{\mathrm{o}}}^{n_{\beta}},\quad N_{\mathrm{o}}=n_{\alpha}+n_{\beta}
\end{eqnarray}
determinants of spin projection $M=(n_{\alpha}-n_{\beta})/2$, which can further form\cite{HelgakerBook}
\begin{eqnarray}
N_{S,M=S}^{\mathrm{c}}=\frac{2S+1}{S_{\mathrm{high}}+S+1}C_{N_{\mathrm{o}}}^{S_{\mathrm{high}}-S},\quad S_{\mathrm{high}}=\frac{1}{2}N_{\mathrm{o}}\label{Ncsf}
\end{eqnarray}
CSFs $\{|I\mu\rangle\}$ of spin projection $M$ and total spin $S=M$.
It can be seen from the ratio $N_{S,M=S}^{\mathrm{c}}/N_{M=S}^{\mathrm{d}}=(2S+1)/(S_{\mathrm{high}}+S+1)$ that,
for oCFGs with a large number of unpaired electrons and a low total spin, the number of CSFs is significantly smaller than that of determinants.
Therefore, it is significantly more advantageous to work with CSFs than with determinants, needless to say that
the correct spin symmetry is always maintained in the former at whatever truncated level.
Sec. \ref{SecPT2} is devoted to an efficient implementation of the state-specific Epstein-Nesbet second-order perturbation theory (PT2)\cite{Epstein,Nesbet}.
To reveal the efficacy of the proposed iCIPT2 approach, some pilot applications
are provided in Sec. \ref{Result}. The account is closed with concluding remarks in Sec. \ref{Conclusion}.
\section{The iCI method}\label{SeciCI}
The \textit{restricted} SDS framework\cite{SDS} for strongly correlated electrons assumes
the following form for the wave function $|\Psi_i\rangle$ of state $i\in[1, N_P]$,
\begin{align}
|\Psi_i\rangle&=\sum_{k=1}^{N_P}|\Psi_k^{(0)}\rangle C_{k,i}+\sum_{k=1}^{N_P}|\Xi_k^{(1)}\rangle C_{k+N_P,i}
+\sum_{k=1}^{N_P}|\Theta_k^{(2)}\rangle C_{k+2N_P,i},\label{SDSCIW}\\
|\Psi_k^{(0)}\rangle&=\sum_{|J\nu\rangle\in P}|J\nu\rangle \bar{C}_{\nu,k}^{J(0)},\\
|\Xi_k^{(1)}\rangle&=Q\frac{1}{E_k^{(0)}-H_0}QH|\Psi_k^{(0)}\rangle=\sum_{|I\mu\rangle\in Q}|I\mu\rangle\bar{C}_{\mu,k}^{I(1)},\quad Q=1-P,\label{1stW}\\
|\Theta_k^{(2)}\rangle&=P_{\mathrm{s}}H|\Xi_k^{(1)}\rangle=\sum_{|J\nu\rangle\in P}|J\nu\rangle\bar{C}_{\nu,k}^{J(2)},\label{2ndW}\\
P&=\sum_{J\nu}^{d_R}|J\nu\rangle\langle J\nu|=P_{\mathrm{m}}+P_{\mathrm{s}},\quad P_{\mathrm{m}}=\sum_{k=1}^{N_P}|\Psi_k^{(0)}\rangle\langle\Psi_k^{(0)}|,
\end{align}
where $\{|\Psi_k^{(0)}\rangle\}_{k=1}^{N_P}$ and $\{|\Xi_k^{(1)}\rangle\}_{k=1}^{N_P}$ are the zeroth-order (primary) and first-order (external) functions, respectively, whereas $\{|\Theta_k^{(2)}\rangle\}_{k=1}^{N_P}$ are the
\textit{not-energy-biased} Lanczos-type (secondary) functions accounting for changes in the static correlation (described by $|\Psi_k^{(0)}\rangle$) due to the inclusion
of dynamic correlation (described by $|\Xi_k^{(1)}\rangle$).
The yet unknown expansion coefficients are to be determined by the generalized secular equation
\begin{eqnarray}
HC=SCE. \label{EigenEQ}
\end{eqnarray}
It can be seen from Eqs. \eqref{1stW} and \eqref{2ndW} that both the external state $|\Xi_k^{(1)}\rangle$
and the secondary state $|\Theta_k^{(2)}\rangle$ are fully contracted and specific to the primary state $|\Psi_k^{(0)}\rangle$.
As such, the dimension of Eq. \eqref{EigenEQ} is only three times the number ($N_P$) of target states, irrespective of the numbers of correlated electrons and orbitals.
The Ansatz \eqref{SDSCIW} is hence a minimal MRCI, more precisely ixc-MRCISD+s (internally and externally contracted multireference configuration interaction with singles and doubles, further augmented with
secondary states) or simply SDSCI in short. The relationships of SDSCI with other wave function-based methods have been scrutinized before\cite{iCI} and are hence not repeated here. Yet, it still deserves to be mentioned that Eq. \eqref{2ndW} is only one of the many possible choices of secondary states\cite{iVI,iVI-TDDFT}.

Although restricted as such, it has recently been demonstrated\cite{DMRG-SDSCI} that SDSCI is a very effective variational method for post-DMRG (density matrix renormalization group) dynamic correlation.
Two extensions of SDSCI have thus far been considered, SDSPT2\cite{SDS,SDSPT2} and iCI\cite{iCI}. The former amounts to replacing the $QHQ$ block of the Hamiltonian matrix in Eq. \eqref{EigenEQ} with $QH_0Q$.
Different from most variants of MRPT2, the CI-like SDSPT2 treats single and multiple states in the same way and is particularly advantageous when a number of states are nearly degenerate, manifesting the effect of the secondary
states in revising the coefficients of the primary states. In contrast, iCI takes the solutions of Eq. \eqref{EigenEQ} as new primary states $\{|\Psi_k^{(0)}\rangle\}_{k=1}^{N_P}$ and repeats the SDS procedure \eqref{SDSCIW} until convergence. It is clear that
each iteration accesses a space that
is higher by two ranks than that of the preceding iteration.
Up to $2M$-tuple excitations (relative to the initial primary
space) can be accessed if $M$ iterations are carried out. Because of
the variational nature, any minor loss of accuracy stemming
from the contractions can be removed by carrying out
some micro-iterations. In other words, by controlling the
numbers of macro- and micro-iterations, iCI will generate a
series of contracted/uncontracted single/multireference
CISD$\cdots$2M, with the resulting energy being physically meaningful
at each level. This feature is particularly warranted for relative energies:
The iterations can be terminated immediately once the desired accuracy has achieved.
Moreover, it has recently been shown that
the micro-iterations of iCI can be generalized to a very effective means
(i.e., iterative vector interaction (iVI)\cite{iVI,iVI-TDDFT})
for partial diagonalization of given matrices. In particular, by combining with the energy-vector following technique,
iVI can directly access interior roots belonging to a predefined window without knowing the number and characters of the roots.
Therefore, iCI has the potential capability to capture any states of many-electron systems without assuming anything in advance.

\section{The Hamiltonian matrix elements}\label{SecHamiltonian}
To make iCI really work for general systems, we have to resolve two major issues: how to select
important CSFs for a target accuracy and how to evaluate efficiently the Hamiltonian matrix elements over randomly
selected CSFs.
The former will be discussed in Sec. \ref{Selection}. For the latter, we adopt the unitary group approach (UGA)\cite{Paldus1980}.
For a better understanding of this approach, we first derive a diagrammatic representation of the spin-free, second-quantized Hamiltonian in Sec. \ref{SecDiagramH}. The expressions for the Hamiltonian matrix elements
are then presented in Sec. \ref{SecMat}, where the upper bounds for the matrix elements over doubly connected oCFGs
are also discussed.
\subsection{Diagrammatic representation of the Hamiltonian}\label{SecDiagramH}
The aim here is to breakdown the spin-free, second-quantized Hamiltonian
\begin{eqnarray}
H&=&\sum_{i,j}h_{ij}E_{ij}+\frac{1}{2}\sum_{i,j,k,l}(ij|kl)e_{ij,kl},\label{sfH}\\
E_{ij}&=&\sum_{\sigma}a_{i\sigma}^\dag a_{j\sigma}=E_{ji}^\dag,\label{EijOp}\\
e_{ij,kl}&=&\sum_{\sigma,\tau} a_{i\sigma}^\dag a_{k\tau}^\dag a_{l\tau} a_{j\sigma}=E_{ij}E_{kl}-\delta_{jk}E_{il}=\{E_{ij}E_{kl}\}\\
&=&\{E_{kl}E_{ij}\}=e_{kl,ij}=e^\dag_{ji,lk}\label{eijkl}
\end{eqnarray}
into a form that is consistent with the diagrams employed in the UGA\cite{Paldus1980} for
the basic coupling coefficients (BCC) between CSFs. A general rule of thumb for this is to reserve
the creation and annihilation characters of indices $(i, k)$ and $(j, l)$, respectively,
when recasting the unrestricted summation into various restricted summations. For instance,
the first, one-body term of $H$ \eqref{sfH} should be decomposed as
\begin{eqnarray}
H_1&=&H_1^0+H_1^1,\\
H_1^0&=&\sum_i h_{ii} E_{ii},\label{H1-0}\\
H_1^1&=&\sum_{i<j} h_{ij}E_{ij}+\sum_{i>j}h_{ij}E_{ij},\label{H1-1}
\end{eqnarray}
where the second term should not be written as $\sum_{i<j}h_{ij}E_{ji}$, so as to merge the two terms together.
The superscripts 0, 1, and 2 in $H_i$ ($i=1,2$) indicate that the terms would
contribute to the Hamiltonian matrix elements over two oCFGs that are related by zero, single and double excitations, respectively.
To breakdown the second, two-body term of $H$ \eqref{sfH}, we notice that
\begin{align}
\sum_{i, j, k, l}&=\sum_{\{i,k\}\cap\{j,l\}\ne \emptyset}+\sum_{\{i,k\}\cap\{j,l\}= \emptyset},\label{H2split}\\
\sum_{\{i,k\}\cap\{j,l\}\ne\emptyset}&=\left(\sum_{i=j=k=l}+\sum_{i=j\ne k=l}+\sum_{i=l\ne k=j}\right)+\left(\sum_{i=j=k\ne l}+\sum_{i=j=l\ne k}+\sum_{k=l=i\ne j}+\sum_{k=l=j\ne i}\right)\nonumber\\
&+\left(\sum_{i=j\ne k\ne l}+\sum_{i=l\ne k\ne j}+\sum_{k=j\ne i\ne l}+\sum_{k=l\ne i\ne j}\right),\label{H2a}\\
\sum_{\{i,k\}\cap\{j,l\}=\emptyset}&=\sum_{i\ne k\ne j\ne l}+\sum_{i=k\ne j\ne l}+\sum_{j=l\ne i\ne k}+\sum_{i=k\ne j=l}.\label{H2b}
\end{align}
By making further use of the particle symmetry of $e_{ij,kl}$ \eqref{eijkl}, the first term of Eq. \eqref{H2split} can readily be decomposed to
\begin{eqnarray}
H_2^0+H_2^1&=&\frac{1}{2}\sum_{\{i,k\}\cap\{j,l\}\ne \emptyset}(ij|kl)e_{ij,kl}, \label{H201def}\\
H_2^0&=&\frac{1}{2}\sum_{i}(ii|ii)e_{ii,ii}+\frac{1}{2}\sum_{i\ne j}\left[(ii|jj)e_{ii,jj}+(ij|ji)e_{ij,ji}\right]\label{H2-0a}\\
&=&\frac{1}{2}\sum_i(ii|ii)E_{ii}(E_{ii}-1)+\sum_{i<j}\left[(ii|jj)E_{ii}E_{jj}+(ij|ji)e_{ij,ji}\right],\label{H2-0b}\\
H_2^1&=&\sum_{i\ne j}\left[(ii|ij)e_{ii,ij}+(ij|jj)e_{ij,jj}\right]+\sum_{i\ne j\ne k}\left[(ij|kk)e_{ij,kk}+(ik|kj)e_{ik,kj}\right]\label{H2-1a}\\
&=&\sum_{i\ne j}\left[(ii|ij)(E_{ii}-1)E_{ij}+(ij|jj)E_{ij}(E_{jj}-1)\right]\nonumber\\
&+&\sum_{i\ne j\ne k}\left[(ij|kk)E_{ij}E_{kk}+(ik|kj)e_{ik,kj}\right].\label{H2-1b}
\end{eqnarray}
The second summation in $H_2^1$ \eqref{H2-1b} reads more explicitly
\begin{eqnarray}
\sum_{i\ne j\ne k}&=&\left(\sum_{i<j<k}+\sum_{j<i<k}+\sum_{k<i<j}+\sum_{k<j<i}\right)+\left(\sum_{i<k<j}+\sum_{j<k<i}\right).\label{ijksum}
\end{eqnarray}
Similarly, the second term of Eq. \eqref{H2split} can be written as
\begin{align}
H_2^2&=\frac{1}{2}\sum_{\{i,k\}\cap\{j,l\}=\emptyset}(ij|kl)e_{ij,kl}\\
&=\frac{1}{2}\sum_{i\ne k\ne j\ne l}(ij|kl)e_{ij,kl}+\frac{1}{2}\sum_{i\ne j\ne k}(ij|kj)e_{ij,kj}+\frac{1}{2}\sum_{i\ne j\ne l}(ij|il)e_{ij,il}+\frac{1}{2}\sum_{i\ne j}(ij|ij)e_{ij,ij}.\label{H2-2}
\end{align}
The first term of $H_2^2$ \eqref{H2-2} can further be written as
\begin{eqnarray}
\frac{1}{2}\sum_{i\ne k\ne j\ne l}(ij|kl)e_{ij,kl}&=&\sum_{i<k}\sum_{j<l}^{\prime}\left[(ij|kl)e_{ij,kl}+(il|kj)e_{il,kj}\right],
\end{eqnarray}
where the summation takes the following form
\begin{eqnarray}
\sum_{i<k}\sum_{j<l}^{\prime}=\sum_{i<k<j<l}+\sum_{j<l<i<k}+\sum_{i<j<k<l}+\sum_{i<j<l<k}+\sum_{j<i<k<l}+\sum_{j<i<l<k}.\label{H2sum-a}
\end{eqnarray}
The second term of Eq. \eqref{H2-2} can further be written as
\begin{align}
\frac{1}{2}\sum_{i\ne j\ne k}(ij|kj)e_{ij,kj}&=\sum_j\sum_{i<k}^\prime(ij|kj)e_{ij,kj},\label{H2-2nd}\\
\sum_j\sum_{i<k}^\prime&=\sum_{i<k<j}+\sum_{i<j<k}+\sum_{j<i<k},\label{H2sum-b}
\end{align}
by observing that interchanging $i$ and $k$ on the left hand side of Eq. \eqref{H2-2nd} gives rise to an identical result.
Similarly, the third term of Eq. \eqref{H2-2} can further be written as
\begin{align}
\frac{1}{2}\sum_{i\ne j\ne l}(ij|il)e_{ij,il}&=\sum_i\sum_{j<l}^\prime(ij|il)e_{ij,il},\label{H2-3rd}\\
\sum_i\sum_{j<l}^\prime&=\sum_{j<l<i}+\sum_{j<i<l}+\sum_{i<j<l}.\label{H2sum-c}
\end{align}
In sum, $H_2^2$ reads
\begin{align}
H_2^2&=\sum_{i<k}\sum_{j<l}^{\prime}\left[(ij|kl)e_{ij,kl}+(il|kj)e_{il,kj}\right]+\sum_j\sum_{i<k}^\prime(ij|kj)e_{ij,kj}\nonumber\\
&+\sum_i\sum_{j<l}^\prime(ij|ik)e_{ij,il}+\frac{1}{2}\sum_{i\ne j}(ij|ij)e_{ij,ij}\label{H2-2f}
\end{align}
in conjunction with Eqs. \eqref{H2sum-a}, \eqref{H2sum-b} and \eqref{H2sum-c} for the first three summations, respectively. Alternatively, $H_2^2$ \eqref{H2-2f} can be written as
\begin{eqnarray}
H_2^2=\sum_{i\le k}\sum_{j\le l}^\prime \left[ 2^{-\delta_{ik}\delta_{jl}}(ij|kl)e_{ij,kl}+(1-\delta_{ik})(1-\delta_{jl})(il|kj)e_{il,kj}\right],\label{EqnDiff2}
\end{eqnarray}
where the prime indicates that $\{j, l\}\cap\{ i, k\} =\emptyset$.


The individual terms of $H$ \eqref{sfH} are summarized in the Table \ref{H-diagram}. The corresponding diagrams are shown in Figs. \ref{Diagrams-0} to \ref{Diagrams-2},
which are drawn with the following conventions:
(1) The enumeration of orbital levels starts with zero and increases from bottom to top.
(2) The left and right vertices (represented by full dots) indicate creation and annihilation operators, respectively, which form a single generator when connected by a non-vertical line.
(3) Products of single generators should always be understood as normal ordered.
For instances, Fig. \ref{Diagrams-1}(a) and 2(b) mean $E_{ij}$ ($i<j$) and $E_{ij}$ ($i>j$), respectively.
The former is a raising generator (characterized
by a generator line going upward from left to right),
whereas the latter is a lowering generator (characterized by
a generator line going downward from left to right);
Fig. \ref{Diagrams-1}(m) means $\{E_{kj}E_{ik}\}=\{E_{ik}E_{kj}\}=e_{ik, kj}$ ($i<j<k$),
which is the exchange counterpart
of the direct generator $e_{ij, kk}=\{E_{kk}E_{ij}\}=\{E_{ij}E_{kk}\}$ ($i<j<k$) shown in Fig. \ref{Diagrams-1}(g).
Note in passing that the labels a$x$, b$x$, c$x$, and d$x$ ($x\in[1,7]$)
for the corresponding diagrams in Figs. \ref{Diagrams-0} to \ref{Diagrams-2} are the same as those in Fig. 6 of Ref. \citenum{Paldus1980}.

When the above diagrams are used to evaluate the BCCs between CSFs,
the level segments can be classified as follows:
\begin{itemize}
\item A: the segment is a terminus and is outside the range of any other generator line.
Upper and lower A-type terminuses of raising (lowering) generators are further denoted by
A$_{\mathrm{R}}$ (A$_{\mathrm{L}}$) and A$^{\mathrm{R}}$ (A$^{\mathrm{L}}$), respectively.
\item B: the segment is a terminus and is within the range of another generator line.
Upper and lower B-type terminuses of raising (lowering) generators are further denoted by
B$_{\mathrm{R}}$ (B$_{\mathrm{L}}$) and B$^{\mathrm{R}}$ (B$^{\mathrm{L}}$), respectively.
\item C$^\prime$: the segment is not a terminus and is within the range of a single generator line.
\item C$^{\prime\prime}$: the segment is not a terminus and is within the range of two generator lines.
\end{itemize}

As a matter of fact, thanks to the conjugacy (bra-ket inversion) relations $\langle I\mu|E_{ij}|J\nu\rangle=\langle I\mu|E_{ij}|J\nu\rangle^*=\langle J\nu|E_{ji}|I\mu\rangle$ and $\langle I\mu|e_{ij,kl}|J\nu\rangle=\langle I\mu|e_{ij,kl}|J\nu\rangle^*=\langle J\nu|e_{ji,lk}|I\mu\rangle$ in the absence of spin-orbit couplings,
only the s2, c$x$ and d$x$ ($x\in[1,7]$) types of diagrams
are needed for the evaluation of matrix elements of the single and double generators in $H_1^1$ \eqref{H1-1}, $H_2^1$ \eqref{H2-1b} and $H_2^2$ \eqref{H2-2f}.
That is, once the BCCs for these types of diagrams are available,
those for the conjugate diagrams can be obtained simply by matrix transpose.
It is also of interest to see that the generators in the s2, c4, c6 and d2 diagrams (required by $H_1^1$ \eqref{H1-1} and $H_2^1$ \eqref{H2-1b}; see Fig. \ref{Diagrams-1})
are subject to $i>j$, while those in the other c$x$ and d$x$ diagrams (required by $H_2^2$ \eqref{H2-2f}; see Fig. \ref{Diagrams-2})
are subject to $k\ge i$ and $k>l\ge j$. Such known conditions facilitate greatly the
determination of specific generators between given oCFG pairs.
It will be shown in Sec. \ref{Selection} that such a choice of diagrams can
be achieved naturally by arranging the oCFGs and CSFs in a particular order.

Given the sequence of segment types $\{Q_r\}$ (see Table \ref{Segments} for the relevant diagrams),
the BCCs can be calculated as\cite{Paldus1980}
\begin{align}
\langle I\mu|E_{ij}|J\nu\rangle&=\Pi_{r\in\Omega_1}W_r=\Pi_{r\in\Omega_1}W(Q_r;\tilde{d}_rd_r, \Delta b_r, b_r),\quad \Delta b_r=b_r-\tilde{b}_r, \label{UGA-1}\\
\langle I\mu|e_{ij,kl}|J\nu\rangle&=\Pi_{r\in\Omega_2^{\mathrm{n}}}W_r\left[\sum_{X=0}^{1}\Pi_{s\in\Omega_2^{\mathrm{o}}}W_s^{(1)}(X)\right]\nonumber\\
&=\Pi_{r\in\Omega_2^{\mathrm{n}}}W(Q_r;\tilde{d}_rd_r, \Delta b_r, b_r)\left[\Pi_{s\in\Omega_2^{\mathrm{o}}}W_s^{(1)}(Q_s; \tilde{d}_sd_s, \Delta b_s=0, X=0)\right.\nonumber\\
&+\left.\Pi_{s\in\Omega_2^{\mathrm{o}}}W_s^{(1)}(Q_s; \tilde{d}_sd_s, \Delta b_s, X=1)\right],\label{UGA-2}
\end{align}
where $d_r$ is the Shavitt step number\cite{Shavitt1977} of level $r$ in the ket CSF $|J\nu\rangle$
($d_r=0$ if level $r$ is not occupied; $d_r=1$ if level $r$ is singly occupied
and spin-up coupled with level $r-1$, i.e., $S_r=S_{r-1}+1/2$; $d_r=2$ if level $r$ is singly occupied
and spin-down coupled with level $r-1$, i.e., $S_r=S_{r-1}-1/2$; $d_r=3$ if level $r$ is doubly occupied),
whereas $b_r$ ($=2S_r=\sum_{p=1}^r[\delta_{1,d_p}-\delta_{2,d_p}]$) is related to the intermediate spin $S_r$ of level $r$
and can directly be calculated from the step number sequence $\mathbf{d}=[d_p]$ characterizing uniquely the ket CSF $|J\nu\rangle$.
The corresponding $d$ and $b$ values in the bra CSF $\langle I\mu|$ are indicated by tildes.
The generator ranges are defined as
$\Omega_1=[\min(i, j), \max(i, j)]$ and $\Omega_2=[\min(i, j), \max(i, j)]\cup[\min(k, l), \max(k, l)]=\Omega_2^{\mathrm{n}}\cup\Omega_2^{\mathrm{o}}$ for the single and double generators,
respectively, with $\Omega_2^{\mathrm{n}}$ and $\Omega_2^{\mathrm{o}}$ being the non-overlapping and overlapping regions of $\Omega_2$, respectively.
Specific segment values $W_r$ and $W^{(Y)}_r$ can be found from Tables III and VII, respectively, in Ref. \citenum{Paldus1980}.
Note in passing that the step numbers for all the (external) levels not in $\Omega_1$/$\Omega_2$
must be identical in the bra and ket CSFs in order for Eq. \eqref{UGA-1}/\eqref{UGA-2}
to be nonzero. Note also that
the BCCs depend only on the structure of oCFG pairs but not on the individual orbitals.
Therefore, it is essential to reutilize them for different oCFG pairs of the same structure (see Sec. \ref{ReuseCoupling}).

\begin{table}[!htp]
\centering
\small
\caption{Diagrammatic representation of the spin-free, second-quantized Hamiltonian \eqref{sfH}. The Einstein summation
convention over repeated indices is assumed. }
\begin{threeparttable}
\begin{tabular}{lllll}\toprule\toprule \multicolumn{1}{l}{Hamiltonian}&\multicolumn{1}{l}{expression}&\multicolumn{1}{l}{range}&\multicolumn{1}{l}{diagram$^a$}&upper bound$^b$\\\toprule
			$H_1^0$ \eqref{H1-0} &$h_{ii}E_{ii}$                     &         &w1\\
			$H_2^0$ \eqref{H2-0b}&$\frac{1}{2}(ii|ii)e_{ii,ii}$      &         &w2 \\
			&$(ii|jj)e_{ii,jj}$                 &$i<j$    &w3 \\
			&$(ij|ji)e_{ij,ji} $                &$i<j$    &b7=c7\\\midrule
			$H_1^1$ \eqref{H1-1} &$h_{ij} E_{ij}$                    &$i<j$    & s1\\
			&                                   &$i>j$    & s2\\
			$H_2^1$ \eqref{H2-1b} &$(ii|ij)e_{ii,ij}$                &$i<j$    & s3\\
			&                                   &$i>j$    & s4 \\
			&$(ij|jj)e_{ij,jj}$                 &$i<j$    & s5 \\
			&                                   &$i>j$    & s6 \\
			&$(ij|kk)e_{ij,kk}$                 &$i<j<k$  & s7 \\
			&                                   &$j<i<k$  & s8\\
			&                                   &$k<i<j$  & s9\\
			&                                   &$k<j<i$  & s10\\
			&                                   &$i<k<j$  & s11\\
			&                                   &$j<k<i$  & s12\\
			&$(ik|kj)e_{ik,kj}$                 &$i<j<k$  & b6\\
			&                                   &$j<i<k$  & c6\\
			&                                   &$k<i<j$  & b4\\
			&                                   &$k<j<i$  & c4\\
			&                                   &$i<k<j$  & a2\\
			&                                   &$j<k<i$  & d2\\\midrule
			$H_2^2$ \eqref{H2-2f}/\eqref{EqnDiff2}&\multirow{6}{*}{$(ij|kl)e_{ij,kl}+(il|kj)e_{il,kj}$}&$i<k<j<l$&a3 + a5&\multirow{6}{*}{$2(|(ij|kl)|+|(il|kj)|)$}\\
			&                                   &$j<l<i<k$&d3 + d5\\
			&                                   &$i<j<k<l$&a1 + b5\\
			&                                   &$i<j<l<k$&c1 + c3\\
			&                                   &$j<i<k<l$&b1 + b3\\
			&                                   &$j<i<l<k$&d1 + c5\\\cline{2-5}
			&                                   &$i<k<j=l$  &a6&\multirow{3}{*}{$2|(ij|kj)|$}\\
			&$(ij|kj)e_{ij,kj}$                 &$i<j=l<k$  &c2\\
			&                                   &$j=l<i<k$  &d4\\\cline{2-5}
			&                                   &$j<l<i=k$  &d6&\multirow{3}{*}{$2|(ij|il)|$}\\
			&$(ij|il)e_{ij,il}$                 &$j<i=k<l$  &b2\\
			&                                   &$i=k<l<j$  &a4\\\cline{2-5}
			&\multirow{2}{*}{$\frac{1}{2}(ij|ij)e_{ij,ij}$}      &$i=k<j=l$    &a7&\multirow{2}{*}{$|(ij|ij)|$}\\
			&                                                    &$j=l<i=k$    &d7\\
			\bottomrule\bottomrule
\end{tabular}
\begin{tablenotes}
\item[a]See Figs. \ref{Diagrams-0}--\ref{Diagrams-2}.
\item[b]The estimated upper bounds $\tilde{H}^{IJ}$ for the matrix elements
$\langle I\mu|H_2^2|J\nu\rangle$, see Sec. \ref{TwoEDiff} and Supporting Information.
\end{tablenotes}
\end{threeparttable}\label{H-diagram}
\end{table}

\clearpage
\newpage
\input{Diagram}
\newpage

\begin{table}[!htp]
    \centering
    \caption{Sequences of segment types required for the evaluation of matrix elements of generators. The left-right order of segment types corresponds to the downward order of levels. The sequences of segment types for the a$x$ and b$x$ diagrams can be obtained from the corresponding ones of d$x$ and c$x$ ($x\in[1,7]$), respectively, by making the $R\leftrightarrow$L inversion.}
    \begin{tabular}{clllc}\toprule\toprule
\multicolumn{1}{l}{diagram}&\multicolumn{1}{l}{generator}&
\multicolumn{1}{l}{range}&\multicolumn{1}{l}{segment sequence}&\multicolumn{1}{l}{remark}\\\toprule
    c7&$e_{ij,ji}^1$&$i<j$&$A_{RL}C''\cdots C''A^{RL}$&Eq. \eqref{H-diag-3}\\ \\
    s2&$E_{ij}$&$i>j$&$A_LC'\cdots C'A^L$&Eq. \eqref{EqnDiff1}\\
    c4&$e^1_{ik,kj}$&$i>j>k$&$A_LC'\cdots C'B_RC''\cdots C'' A^{RL}$& \\
    c6&$e^1_{ik,kj}$&$k>i>j$&$A_{RL}C''\cdots C'' B^RC'\cdots C'A^L$& \\
    d2&$e^1_{ik,kj}$&$i>k>j$&$A_LC'\cdots C'A_L^L C'\cdots C' A^L$& \\ \\
    c1&$e_{ij,kl}$&$k>l>j>i$&$A_LC'\cdots C' A^LA_RC'\cdots C'A^R$&Eq. \eqref{DoubleMat}\\
    c3&$e_{il,kj}$&$k>l>j>i$&$A_LC'\cdots C'B_RC''\cdots C'' B^L C'\cdots C' A^R$& \\
    d1&$e_{ij,kl}$&$k>l>i>j$&$A_LC'\cdots C' A^L A_L C' \cdots C' A^L$& \\
    c5&$e_{il,kj}$&$k>l>i>j$&$A_L C'\cdots C' B_R C'' \cdots C'' B^R C' \cdots C' A^L$& \\
    d3&$e_{ij,kl}$&$k>i>l>j$&$A_L C'\cdots C' B_L C''\cdots C'' B^L C'\cdots C' A^L$& \\
    d5&$e_{il,kj}$&$k>i>l>j$&$A_L C'\cdots C' B_L C''\cdots C'' B^L C'\cdots C' A^L $& \\
    c2&$e_{ij,kj}$&$k>j>i$&$A_LC'\cdots C'A^L_R C' \cdots C'A^R$& \\
    d4&$e_{ij,kj}$&$k>i>j$&$A_LC'\cdots C; B_L C'' \cdots C'' A^{LL}$& \\
    d6&$e_{ij,il}$&$i>l>j$&$A_{LL} C'' \cdots C'' B^L C' \cdots C' A^L$& \\
    d7&$e_{ij,ij}$&$i>j$&$A_{LL}C''\cdots C'' A^{LL}$& \\
                        \bottomrule\bottomrule
    \end{tabular}\label{Segments}
\end{table}

\subsection{Explicit expressions for the Hamiltonian matrix elements}\label{SecMat}
\subsubsection{Zero-electron difference}
For two identical oCFGs, it is the $H_1^0$ \eqref{H1-0} and $H_2^0$ \eqref{H2-0b} parts of $H$ \eqref{sfH} that should be considered, i.e.,
\begin{eqnarray}
\langle I\mu|H_1^0+H_2^0|I\nu\rangle&=&\delta_{\mu\nu}\left[\sum_i h_{ii}n_i^I+\frac{1}{2}\sum_{i}(ii|ii)n_i^I(n_i^I-1)+\sum_{i<j}(ii|jj)n_i^In_j^I\right]\nonumber\\
&&+\sum_{i<j}(ij|ji)\langle I\mu| e_{ij,ji}|I\nu\rangle,\quad n_i^I=\langle I\mu|E_{ii}|I\mu\rangle,\label{H-diag-0}
\end{eqnarray}
where the last, exchange term can be simplified by using Payne's Theorem 1\cite{Payne}, viz.,
\begin{eqnarray}
\langle I\mu|e_{ij,kl}|J\nu\rangle&=&\langle I\mu|e_{ij,kl}^0|J\nu\rangle +\langle I\mu|e_{ij,kl}^1|J\nu\rangle,\\
\langle I\mu|e_{ij,kl}^0|J\nu\rangle&=&-2\langle I\mu|e_{il,kj}^0|J\nu\rangle \mbox{ iff } \max(i, j)\le\min(k, l).
\end{eqnarray}
Here, the superscripts 0 and 1 correspond to the intermediate angular momentum $X=0$ and $X=1$, respectively (see Eq. \eqref{UGA-2}).
More specifically, we have
\begin{eqnarray}
\langle I\mu|e_{ij,ji}|I\nu\rangle&=&\langle I\mu|e_{ij,ji}^0|I\nu\rangle+\langle I\mu|e_{ij,ji}^1|I\nu\rangle,\\
\langle I\mu|e_{ij,ji}^0|I\nu\rangle&=&-\frac{1}{2}\langle I\mu|e_{ii,jj}^0|I\nu\rangle=
\delta_{\mu\nu}(-\frac{1}{2}n_i^I n_j^I+\frac{1}{2}\delta_{ij}n_i^I).
\end{eqnarray}
Therefore, Eq. \eqref{H-diag-0} can be written as
\begin{align}
\langle I\mu|H_1^0+H_2^0|I\nu\rangle&=\delta_{\mu\nu}\left[\sum_i h_{ii}n_i^I+\frac{1}{2}\sum_{i}(ii|ii)n_i^I(n_i^I-1)+\sum_{i<j}n_i^In_j^I[(ii|jj)-\frac{1}{2}(ij|ji)]\right]\nonumber\\
&+\sum_{i<j}(ij|ji)\langle I\mu|e_{ij,ji}^1|I\nu\rangle.\label{H-diag-1}
\end{align}
In terms of the following intermediate quantities
\begin{eqnarray}
f_{ij}&=&h_{ij}+\sum_k \omega_k[(ij|kk)-\frac{1}{2}(ik|kj)],\\\label{Inter1}
\epsilon_i&=&f_{ii}=h_{ii}+\sum_k \omega_k g_{ik},\quad g_{ik}=(ii|kk)-\frac{1}{2}(ik|ki)=g_{ki},\\\label{Inter2}
n_k^I&=&\omega_k+\Delta_k^I,
\end{eqnarray}
where $\{\omega_k\}$ are the occupation numbers of the spatial orbitals of a common reference oCFG,
Eq. \eqref{H-diag-1} can be converted to the following form
\begin{align}
\langle I\mu|H_1^0+H_2^0|I\nu\rangle&=\delta_{\mu\nu}\{\frac{1}{2}\sum_i \omega_i[h_{ii}+\epsilon_i+(\omega_i-2)g_{ii}]+\sum_i\Delta_i^I[\epsilon_i+(\omega_i-1)g_{ii}]\nonumber\\
&+\sum_{i\le j}\Delta_i^Ig_{ij}\Delta_j^I\}+\sum_{i<j}(ij|ji)\langle I\mu|e_{ij,ji}^1|I\nu\rangle,\label{H-diag-3}
\end{align}
which is most suited for implementation. Note in passing that $\langle I\mu|e_{ij,ji}^1|I\nu\rangle$
is nonzero only when both levels $i$ and $j$ are singly occupied.
\subsubsection{One-electron difference}
When oCFG $I$ can be obtained from oCFG $J$ by exciting a single electron, we have
\begin{align}
\langle I\mu |H_1^1 + H_2^1 | J\nu\rangle & =  \langle I\mu |\sum_{i\neq j} \left[h_{ij}E_{ij}+(ii|ij)e_{ii,ij}+(ij|jj)e_{ij,jj}\right]| J\nu\rangle \nonumber\\
&+\langle I\mu |\sum_{i\ne j\neq k} \left[(ij|kk)e_{ij,kk}+(ik|kj)e_{ik,kj}\right]| J\nu\rangle\nonumber\\
&=\sum_{i\neq j}\left[h_{ij}+(n_i^I-1)(ii|ij)+(n_j^J-1)(ij|jj)\right]\langle I\mu|E_{ij}|J\nu\rangle\nonumber\\
&+\sum_{i\ne j\ne k}\left((ij|kk)n_k^J\langle I\mu| E_{ij}|J\nu\rangle+(ik|kj)\langle I\mu|e_{ik,kj}|J\nu\rangle\right).
\label{Single}
\end{align}
In view of Eq. \eqref{ijksum}, we can distinguish two cases for the second compound summation over $k$:
the non-overlapping case (a) $k<\min(i, j)$ or $k>\max(i,j)$, and the overlapping case (b) $\min(i,j)<k<\max(i,j)$.
For the former, Payne's Theorem 1\cite{Payne} can be applied to $\langle I\mu|e_{ik,kj}|J\nu\rangle$, i.e.,
\begin{equation}
\begin{aligned}
&(i j | k k)n_k^J\langle I\mu| E_{i j}|J\nu\rangle+(i k | k j)\langle I\mu| e_{i k, k j}|J\nu\rangle \\
=&(i j | k k)n_k^J\langle I\mu| E_{i j}|J\nu\rangle+(i k | k j)\left[\langle I\mu| e_{i k, k j}^{0}|J\nu\rangle+\langle I\mu| e_{i k, k j}^{1}|J\nu\rangle\right] \\
=&(i j | k k)n_k^J\langle I\mu| E_{i j}|J\nu\rangle+(i k | k j)\left[-\frac{1}{2}\langle I\mu| e_{i j, k k}^{0}|J\nu\rangle+\langle I\mu| e_{i k, k j}^{1}|J\nu\rangle\right] \\
=&n_{k}^J\langle I\mu|E_{i j}|J\nu\rangle\left[(i j | k k)-\frac{1}{2}(i k | k j)\right]+\langle I\mu| e_{i k, k j}^{1}|J\nu\rangle(i k | k j).
\end{aligned}
\label{Inner}
\end{equation}
For the latter, overlapping case we have
\begin{equation}
\begin{aligned}
 &(i j | k k)n_k^J\langle I\mu| E_{i j}| J\nu\rangle+(i k | k j)\langle I\mu| E_{k j} E_{i k}|J\nu\rangle=\\
 &n_{k}^J\langle I\mu|E_{i j}|J\nu\rangle\left[(i j | k k)-\frac{1}{2}(i k | k j)\right] +\left[\frac{1}{2} n_{k}^J\langle I\mu|E_{i j}|J\nu\rangle+\langle I\mu|  E_{k j} E_{i k}|J\nu\rangle\right](ik|kj). \end{aligned}
\label{exter}
\end{equation}
Inserting Eqs. \eqref{Inner} and \eqref{exter} into Eq. \eqref{Single} leads to
\begin{align}
\langle I\mu |H_1^1 + H_2^1 | J\nu\rangle &=\sum_{i\ne j}\left[h_{ij}+(n_i^I-1)(ii|ij)+(n_j^J-1)(ij|jj)\right]\left\langle I\mu|E_{i j}|J\nu\right\rangle\nonumber \\
&+\sum_{i\ne j}\sum_{k\ne i\ne j} n_k^J\left[(ij|kk)-\frac{1}{2}(ik|kj)\right]\left\langle I\mu|E_{i j}|J\nu\right\rangle\nonumber\\
&+\sum_{i\neq j}\sum_{k\in \mbox{ case (a)}}(ik|kj)\langle I\mu| e^1_{ik,kj}|J\nu\rangle\nonumber\\
&+\sum_{i\neq j}\sum_{k\in \mbox{ case (b)}}(ik|kj)\left[\frac{1}{2}n_k^J\left\langle I\mu|E_{i j}|J\nu\right\rangle+\langle I\mu| E_{kj}E_{ik}|J\nu\rangle\right].\label{Single1}
\end{align}
In terms of the following identities
\begin{align}
(n_i^I-1)(ii|ij)+(n_j^J-1)(ij|jj)&=n_i^J[(ij|ii)-\frac{1}{2}(ii|ij)]+n_j^J[(ij|jj)-\frac{1}{2}(ij|jj)]\nonumber\\
&+(n_i^I-\frac{1}{2}n_i^J-1)(ii|ij)+(\frac{1}{2}n_j^J-1)(ij|jj),\\
(n_i^I-\frac{1}{2}n_i^J-1)\left\langle I\mu|E_{i j}|J\nu\right\rangle &= \frac{1}{2}n_i^J\left\langle I\mu|E_{i j}|J\nu\right\rangle,
\end{align}
the first two terms of Eq. \eqref{Single1} can be simplified to
\begin{equation}
\sum_{i\ne j}\left\langle I\mu|E_{i j}|J\nu\right\rangle\left[f_{ij}+\sum_k\Delta_k^J[(ij|kk)-\frac{1}{2}(ik|kj)]+\frac{1}{2}n_i^J(ii|ij)+(\frac{1}{2}n_j^J-1)(ij|jj)\right].
\end{equation}

For the third term of Eq. \eqref{Single1}, the $k$-level segments are $A^{RL}$ and $A_{RL}$ types of terminal segments (see the c4 and c6 diagrams in
Fig. \ref{Diagrams-1} and Table \ref{Segments}).
It can be found from Table VIIa of Ref. \citenum{Paldus1980} that such segment values $W_k^{(1)}$ are all zero if $n_k^I=n_k^J= 0 \text{ or } 2$.
Therefore, $\langle I\mu| e^1_{ik,kj}|J\nu\rangle$ can be nonzero in this case only if $n_k^I=n_k^J=1$.

For the fourth term of Eq. \eqref{Single1}, it is first noticed that the segment values for computing the BCCs
$\left\langle I\mu|E_{i j}|J\nu\right\rangle$ and $\langle I\mu| E_{kj}E_{ik}|J\nu\rangle$
differ only in the $k$-level segments. The latter are $A_L^L$ and $A_R^R$ types of terminal segments
in the d2 and a2 diagrams, respectively (see Fig. \ref{Diagrams-1} and Table \ref{Segments}).
Again, it can be found from Table VIIa of Ref. \citenum{Paldus1980} that such segment values $W_k^{(1)}$ are all zero if $n_k^I=n_k^J=0$.
On the other hand, if $n_k^I=n_k^J=2$, the $k$-level segment value for $\langle I\mu| E_{kj}E_{ik}|J\nu\rangle$ is
either $W^{(1)}(A_L^L;33,\Delta b, b, X)$ or $W^{(1)}(A_R^R;33, \Delta b, b, X)$,
which both equal to -1 for $\Delta b =\pm1$ (with the Yamanouchi-Kotani (YK) phase\cite{Paldus1980}) or 0 for $\Delta b \ne \pm1$,
whereas the $k$-level segment value for $\left\langle I\mu|E_{i j}|J\nu\right\rangle$ is $W(C';33,\Delta b,b)$,
which is 1 for $\Delta b = \pm 1$ (with the YK phase) or 0 for $\Delta b \ne \pm1$ (cf. Table IIIb of Ref. \citenum{Paldus1980}).
Therefore, the sum $\frac{1}{2}n_k^J\left\langle I\mu|E_{i j}|J\nu\right\rangle+\langle I\mu| E_{kj}E_{ik}|J\nu\rangle$ is always zero if $n_k^I=n_k^J=0\text{ or }2$. That is, it
can be nonzero only if $n_k^I=n_k^J=1$.

As such, Eq. \eqref{Single} can, with the YK phase, be simplified to
\begin{equation}
\begin{split}
\langle I\mu|H_1^1+H_2^1|J\nu\rangle&=\sum_{i\ne j}\left\langle I\mu|E_{i j}|J\nu\right\rangle\left[f_{ij}+\sum_k\Delta_k^J[(ij|kk)-\frac{1}{2}(ik|kj)]\right.\\
&+\left.\frac{1}{2}n_i^J(ii|ij)+(\frac{1}{2}n_j^J-1)(ij|jj)\right]\\
&+\sum_{i\neq j}\sum_{k\in\mbox{ exterior open}}(ik|kj)\langle I\mu| e_{ik,kj}^1|J\nu\rangle\\
&+\sum_{i\neq j}\sum_{k\in\mbox{ interior open}}(ik|kj)\left[\frac{1}{2}\left\langle I\mu|E_{i j}|J\nu\right\rangle+\langle I\mu| E_{kj}E_{ik}|J\nu\rangle\right],
\end{split}\label{EqnDiff1}
\end{equation}
where `exterior open' and `interior open' emphasize that level $k$ belongs to
the non-overlapping ($k<\min(i, j)$ or $k>\max(i,j)$) and overlapping ($\min(i,j)<k<\max(i,j)$) cases, respectively, and is singly occupied. To the best of our knowledge, the particular form \eqref{EqnDiff1}
for the Hamiltonian matrix elements over singly excited CSFs has not been documented before in the literature.

\subsubsection{Two-electron difference}\label{TwoEDiff}
When oCFG $I$ can be obtained from oCFG $J$ by exciting two electrons, the Hamiltonian matrix elements read (cf. Eq. \eqref{EqnDiff2} and Table \ref{H-diagram})
\begin{eqnarray}
\langle I\mu|H_2^2|J\nu\rangle=\sum_{i\le k}\sum_{j\le l}^\prime \left[ 2^{-\delta_{ik}\delta_{jl}}(ij|kl)\langle I\mu|e_{ij,kl}|J\nu\rangle+(1-\delta_{ik})(1-\delta_{jl})(il|kj)\langle I\mu|e_{il,kj}|J\nu\rangle\right],\label{DoubleMat}
\end{eqnarray}
where the prime indicates that $\{j, l\}\cap\{ i, k\} =\emptyset$.
The dimension of the matrix $[H_2^2]$ \eqref{DoubleMat} scales as $N_{\mathrm{ref}}n_o^2 n_v^2$, with $N_{\mathrm{ref}}$, $n_o$ and $n_v$ being the
number of CSFs in the primary space and the numbers of occupied and virtual orbitals, respectively.
Enormous efficiency will be gained if it can be estimated
\textit{a priori} that a given oCFG $J$ will interact (through $H_2^2$) strongly with what oCFGs $\{I\}$, such that those unimportant oCFGs are not accessed at all. This goal can be fulfilled by introducing the following inequality
for the general BCCs
\begin{equation}
	\left|\left\langle I\mu|e_{ij,kl}|J\nu\right\rangle\right| \leq 2,\quad \forall \{j,l\}\cap\{i,k\}=\emptyset,\label{Approx}
\end{equation}
which provides upper bounds (denoted as $\tilde{H}^{IJ}$) for the matrix elements $\langle I\mu|H_2^2|J\nu\rangle$ \eqref{DoubleMat} (see Table \ref{H-diagram}).
Note in passing that $\tilde{H}^{IJ}$ depend only on the two-electron integrals, and can hence be sorted
in descendent order and stored in memory from the outset.
The proof of Eq. \eqref{Approx} is given in Supporting Information.

\section{Storage and selection of configurations}\label{Selection}
The variational CSF space $P=\{|I\mu\rangle\}$ for static correlation is to be
selected from the corresponding oCFG space $V_{\mathrm{cfg}}=\{I\}$ that is random in occupation patterns.
To do this efficiently, the oCFGs and CSFs must be sorted and stored properly.
\subsection{Storage of oCFGs and CSFs}
A bitstring representation of the oCFGs is adopted here. Since a spatial orbital can be zero, singly or doubly
occupied, two bits are needed for each orbital. Specifically, $(00)_2$, $(01)_2$ and $(11)_2$ are to be used
for the three kinds of occupancy, respectively, which are manifested
by the number of nonzero bits in each case. An oCFG composed of $N_{\text{orb}}$ spatial orbitals
therefore requires $2\times N_{\text{orb}}$ bits.
However, since CPUs are designed to handle efficiently 64-bit integers, an oCFG will be stored as an array (denoted as \textcolor{blue}{OrbOccBinary}) of 64-bit integers, with the unused space padded with zeros.
The actual storage required for an oCFG is hence $2\times64\times N_{\text{64}}$ bits, with $N_{\text{64}}$ being the number of 64-bit integers, viz.
 \begin{equation}
	 N_{\text{64}} = \left\lfloor\frac{N_{\text{orb}}-1}{32}\right\rfloor+1.
 \end{equation}
Note in passing that the orbitals are indexed from 0 to $N_{\text{orb}}-1$ and go from right to left in the bistring.
The memory requirement for $N_{\mathrm{cfg}}$ oCFGs is just $N_{\mathrm{cfg}}*N_{\text{64}}*8$ bytes.
Apart from the compact storage of oCFGs, the bitstring encoding also allows performing a variety of operations on oCFGs, by taking advantage of the high efficiency of CPUs in bitwise operations on integers.
For instance, given the $2n$-th and $(2n+1)$-th bits of the $m$-th element of \textcolor{blue}{OrbOccBinary}, the index \textcolor{blue}{OrbIndx} for the corresponding orbital can be calculated simply as $32m+n$, while its occupation number can be calculated with Algorithm \ref{AlgorithmOCC}. Likewise, the excitation rank of oCFGs $I$ over $J$ can be determined with Algorithm \ref{AlgorithmRank}. To be compatible with the unitary group approach discussed before (i.e., to stick with the s2, c$x$ and d$x$ ($x\in[1,7]$) types of diagrams),
oCFG $I$ is stored preceding oCFG $J$ (i.e., $I<J$) if and only if $n_p^I>n_p^J$ for $p=\max\{x|n_x^I\neq n_x^J\}$.
The comparison of an oCFG pair can be performed efficiently with Algorithm \ref{AlgorithmOrder}.

\begin{algorithm}
    \caption{Occupation numbers of oCFG in bitstring representation. \textcolor{blue}{BTEST}$(I,n)$ returns TRUE if the $n$-th bit of oCFG $|I\rangle$ is set and FALSE otherwise.}
    \label{AlgorithmOCC}
        \KwIn{OrbOccBinary\_$I$, OrbIndx}
        \KwOut{$n_{\text{OrbIndx}}$}
            $m\gets\lfloor\frac{\text{OrbIndx}}{32}\rfloor$\;
            $n\gets\text{OrbIndx} \mbox{ mod } 32$\;
            \eIf{\textcolor{blue}{BTEST}(OrbOccBinary\_I[m], $2n+1$) = TRUE}
            {
               return $2$ \;
            }
            {
                \eIf{\textcolor{blue}{BTEST}(OrbOccBinary\_I[m], $2n$) = TRUE}
                {
                    return $1$\;
                }
                {
                    return $0$\;
                }
            }
\end{algorithm}

\begin{algorithm}
    \caption{Excitation ranks of oCFGs. \textcolor{blue}{POPCNT}$(I)$ returns the number of non-zero bits for a given integer $I$; \textcolor{blue}{IEOR}$(I,J)$ performs bitwise XOR (exclusive or) logical operation.}
    \label{AlgorithmRank}
        \KwIn{OrbOccBinary\_$I$, OrbOccBinary\_$J$, $N_{64}$}
        \KwOut{Excitation rank}
            Result $\gets$ 0\;
            \For{($i=0;i < N_{64};i = i+1$)}
            {
                Result $\gets$ Result $+$ \textcolor{blue}{POPCNT}(\textcolor{blue}{IEOR}(OrbOccBinary\_$I$[$i$],OrbOccBinary\_$J$[$i$]))\;
            }
            return Result/2\;
\end{algorithm}

\begin{algorithm}
    \caption{Ordering of oCFGs}
    \label{AlgorithmOrder}
        \KwIn{OrbOccBinary\_$I$, OrbOccBinary\_$J$, $N_{64}$}
        \KwOut{1: $I<J$; 0: $I=J$; -1: $I>J$}
            \For{($i= N_{64}-1;i\ge 0;i = i-1$)}
            {
                \If{OrbOccBinary\_$I$[$i$]$\neq$ OrbOccBinary\_$J$[$i$]}
                {
                    \eIf{OrbOccBinary\_$I$[$i$]$>$ OrbOccBinary\_$J$[$i$]}
                    {
                        return $1$\;
                    }
                    {
                    return $-1$\;
                    }
                }
            }
            return 0\;
\end{algorithm}

An oCFG $I$ can generate $N^I_{S,S}$ CSFs $\{|I\mu\rangle\}_{\mu=0}^{N_{S,S}^I-1}$ of total spin $S$ (cf. Eq. \eqref{Ncsf}) via, e.g., the genealogical coupling scheme\cite{HelgakerBook}.
Such CSFs are characterized uniquely by the Shavitt step number sequences $\{(d_0^{\mu}d_1^{\mu}\cdots)\}|_{\mu=0}^{N_{S,S}^{I-1}}$, which can be arranged in a lexical (dictionary) order (i.e., $|I\mu\rangle$ precedes $|I\nu\rangle$ if and only if $d_p^{\mu}<d_p^{\nu}$ for $p=\min\{x|d_x^{\mu}\neq n_x^{\nu}\}$), so as to fix the relative ordering $\mu$ uniquely. A 32-bit integer is required to store the relative ordering of CSFs $\{|I\mu\rangle\}$.
Therefore, the memory requirement for $\tilde{N}_{\mathrm{csf}}$ CSFs selected from
$N_{\mathrm{cfg}}$ oCFGs amounts to $N_{\mathrm{cfg}}*N_{\text{64}}*8+\tilde{N}_{\mathrm{csf}}*4$ bytes. In the largest calculation of benzene ($N_{\mathrm{cfg}}=878834$, $\tilde{N}_{\mathrm{csf}}=1381841$),
the memory for this is only 32 Mb.
A given list of CSFs will be sorted first by the oCFGs to which they belong and then by their relative ordering.

\subsection{Reutilization of BCCs}\label{ReuseCoupling}
As discussed before, the segment values for common unoccupied or common doubly occupied orbitals in the bra and ket oCFGs are all one under the YK phase, if the BCC does not vanish. Therefore, only singly occupied orbitals or orbitals with different occupation numbers in the bra and ket oCFGs need to be considered explicitly when evaluating the BCCs. This leads to
seven occupation patterns, which can be ordered by the code numbers shown in Table \ref{ROT}.
An oCFG pair ($I,J$) can hence be characterized by a `reduced occupation table' (ROT) consisting of two number sequences, \textcolor{blue}{ROT\_Orb} and \textcolor{blue}{ROT\_Code}.
The former records the orbital sequence whereas the latter records
the corresponding code sequence, after deleting the common unoccupied and common doubly occupied orbitals in both cases.
Every code sequence (e.g., (021430) for the example shown in Table \ref{CoeffExam}) will be converted to an array of 64-bit integers (3 bits for each code) and stored on a node of a red-black tree for bilinear search.

The BCCs $\langle I\mu|E_{ij}|J\nu\rangle$ and $\langle I\mu|e_{ij,kl}|J\nu\rangle$ ($\mu\in[0,N_{S,S}^I-1]$, $\nu\in[0, N_{S,S}^J-1]$, $i,j,k,l\in[0,N_{\text{orb}}-1]$)
depend only on the structure of oCFG pair $(I, J)$ but not on the individual orbitals $\{i, j, k, l\}$. Therefore,
they can be rewritten as $\langle I\mu|E_{\bar{i}\bar{j}}|J\nu\rangle$ and $\langle I\mu|e_{\bar{i}\bar{j},\bar{k}\bar{l}}|J\nu\rangle$, respectively, in terms of
the reduced orbital indices $\{\bar{i}, \bar{j}, \bar{k},\bar{l}\}\in[0, LenROT-1]$, which have one-to-one correspondence with the orbital indices $\{i, j, k, l\}$ recorded in \textcolor{blue}{ROT\_Orb}
(see the example in Table \ref{CoeffExam}). Here, $LenROT$ is the length of the code sequence
(NB: $LenROT\le 2 + N_o$ ($LenROT\le 4 + N_o$) for singly (doubly) connected oCFG pairs, with $N_o$ being the number of common singly occupied orbitals).
Different oCFG pairs sharing the same \textcolor{blue}{ROT\_Code} have the same
BCCs. That is, once available, the BCCs can be reused many times.
For instance, the matrix element $\langle I\mu|H_2^2|J\nu\rangle$ for the example in Table \ref{CoeffExam} read
\begin{align}
\langle I\mu|H_2^2|J\nu\rangle&=(32|65)\langle I\mu|e_{32,65}|J\nu\rangle+(35|62)\langle I\mu|e_{35,62}|J\nu\rangle\\
&=(32|65)\langle I\mu|e_{\bar{2}\bar{1},\bar{4}\bar{3}}|J\nu\rangle+(35|62)\langle I\mu|e_{\bar{2}\bar{3},\bar{4}\bar{1}}|J\nu\rangle.
\end{align}
If another oCFG pair $(I',J')$ shares the same \textcolor{blue}{ROT\_Code} but different \textcolor{blue}{ROT\_Orb} $(i_0',i_1',\cdots,i_5')$, the matrix element $\langle I'\mu|H_2^2|J'\nu\rangle$ can be calculated as
\begin{align}
\langle I'\mu|H_2^2|J'\nu\rangle&=(i_2'i_1'|i_4'i_3')\langle I'\mu|e_{\bar{2}\bar{1},\bar{4}\bar{3}}|J'\nu\rangle+(i_2'i_3'|i_4'i_1')\langle I'\mu|e_{\bar{2}\bar{3},\bar{4}\bar{1}}|J'\nu\rangle\\
&=(i_2'i_1'|i_4'i_3')\langle I\mu|e_{\bar{2}\bar{1},\bar{4}\bar{3}}|J\nu\rangle+(i_2'i_3'|i_4'i_1')\langle I\mu|e_{\bar{2}\bar{3},\bar{4}\bar{1}}|J\nu\rangle.
\end{align}

It deserves to be stressed that it is precisely the use of ROT that renders the present
`tabulated orbital configuration based unitary group approach' (TOC-UGA) distinct from the table CI method\cite{TableCI1980,TableCI1986,TableCI1995}: No expensive line-up permutations are needed here.
Instead, given a code sequence \textcolor{blue}{ROT\_Code}, the BCCs can directly be calculated by using Eqs. \eqref{UGA-1} and \eqref{UGA-2}, with the following information:
If oCFG $I$ arises from $J$ by exciting one electron from orbital $j$ to orbital $i$, $\bar{j}$ would then correspond to code 2 or 4, while $\bar{i}$ would correspond to code 1 or 3 because
of the relations $n_j^I=n_j^J-1$ and $n_i^I=n_i^J+1$. Likewise, if oCFG $I$ arises from $J$ by exciting two electrons from orbitals $j$ and $l$ $(\ge j)$ to orbitals $i$ and $k$ $(\ge i)$ [NB: $\{i,k\}\cap\{j,l\}=\emptyset$],
$\bar{j}$ and $\bar{l}$ $(\ge \bar{j})$
would correspond to code 2, 4 or 6, while $\bar{i}$ and $\bar{k}$ $(\ge \bar{i})$ would correspond to code 1, 3 or 5 because of the relations $n_j^I=n_j^J-1$, $n_l^I=n_l^J-1$, $n_i^I=n_i^J+1$ and $n_k^I=n_k^J+1$. Codes 6 and 5 further imply $\bar{j}=\bar{l}$ and $\bar{i}=\bar{k}$, respectively. If $\bar{i}<\bar{j}$ in $E_{\bar{i}\bar{j}}$ or $\bar{k}<\bar{l}$ in $e_{\bar{i}\bar{j},\bar{k}\bar{l}}$, a bra-ket inversion should be invoked when calculating
the BCCs in terms of the diagrams documented in Table \ref{Segments}.

\begin{table}[!htp]
\centering
\caption{Code numbers for the occupation patterns of oCFG pairs}
\begin{tabular}{p{1.5cm}p{1cm}<{\centering}p{1cm}<{\centering}p{1cm}<{\centering}p{1cm}
<{\centering}p{1cm}<{\centering}p{1cm}<{\centering}p{1cm}<{\centering}p{1cm}
<{\centering}p{1cm}<{\centering}}\toprule
 		code   &0&1&2&3&4&5&6&$\textcolor{red}{\times}$&$\textcolor{red}{\times}$\\\toprule
 		bra occ&1&1&0&2&1&2&0&\textcolor{red}{0}&\textcolor{red}{2}\\
 		ket occ&1&0&1&1&2&0&2&\textcolor{red}{0}&\textcolor{red}{2}\\\bottomrule
\end{tabular}\label{ROT}
\end{table}

 \begin{table}[!htp]
 	\centering
 	\caption{Illustration on the reutilization of basic coupling coefficients}	\begin{tabular}{p{4cm}p{1cm}<{\centering}p{1cm}<{\centering}p{1cm}<{\centering}p{1cm}<{\centering}p{1cm}<{\centering}p{1cm}<{\centering}p{1cm}<{\centering}p{1cm}<{\centering}p{1cm}<{\centering}}\toprule
 		OrbIndx&0&1&2&3&4&5&6&7\\\midrule
        bra occ&0&1&0&1&2&1&2&1\\
 		ket occ&0&1&1&0&2&2&1&1\\\midrule
        ROT\_Orb&$\textcolor{red}{\times}$&1&2&3&$\textcolor{red}{\times}$&5&6&7\\
        Reduced indices&$\textcolor{red}{\times}$&$\bar{0}$&$\bar{1}$&$\bar{2}$&$\textcolor{red}{\times}$&$\bar{3}$&$\bar{4}$&$\bar{5}$\\
 		ROT\_Code&$\textcolor{red}{\times}$&0&2&1&$\textcolor{red}{\times}$&4&3&0\\\bottomrule
 	\end{tabular}
 \label{CoeffExam}
 \end{table}

\subsection{Connections between oCFGs}

To evaluate the Hamiltonian matrix elements efficiently, the single and double connections between randomly selected oCFGs must be established. While this can be done in a number of ways, we consider here the residue-based sorting algorithm\cite{ASCI2018}.
%

The $n$-th order residues are defined as those oCFGs that can be generated by removing $n$ electrons in all possible ways from a reference oCFG. The $n$-th order residues of an oCFG space $V_{\mathrm{cfg}}$ are the union of the $n$-th order residues of the oCFGs in $V_{\mathrm{cfg}}$. Only first- and second-order residues are needed in the present work.
They are stored in memory in structure arrays \textcolor{blue}{R$_1$} and \textcolor{blue}{R$_2$}, respectively, where every element consists of an array (\textcolor{blue}{OrbOccBinary}) of 64-bit integers recording a residue (in the same way as recording an oCFG), an array (\textcolor{blue}{CfgIndx}) of 32-bit integers recording the index of the oCFG from which the residue is generated, and an array (\textcolor{blue}{OrbIndx}) of 16-bit integers recording the indices of the orbitals from which the electrons are removed (NB: if two electrons are removed from the same orbital, the orbital will be recorded twice).
Once all the first (second) order residues have been generated, sort the structure array \textcolor{blue}{R$_1$} (\textcolor{blue}{R$_2$}) by residues:
a unique residue is recorded only once; its parent oCFGs and corresponding orbitals are put into arrays \textcolor{blue}{CfgIndx} and \textcolor{blue}{OrbIndx}, respectively.
The memory requirement is modest. Assuming that the second-order residues of $N_{\mathrm{cfg}}$ oCFGs of $N$ electrons and $N_{\mathrm{orb}}$ orbitals are all distinct,
the upper limit for the memory requirement can be estimated to be $M_2^R=N_{\mathrm{cfg}}\times C_{N}^2\times\left[(\lfloor\frac{N_{\mathrm{orb}}-1}{32}+1\rfloor)*\frac{64}{8}+\frac{32}{8}+\frac{16}{8}*2\right]$ bytes.
In the largest calculation of benzene ($N=30$, $N_{\mathrm{orb}}=108$, $N_{\mathrm{cfg}}=878834$),
the actual memory is 6.5Gb, which is much smaller than the upper limit $M_2^R=15.3$ Gb. The reason for this is twofold:
(a) Different oCFGs may generate the same residues. (b) The number of residues generated from an oCFG lies between
$C_{\frac{N}{2}}^2+\frac{N}{2}$ (in case every orbital is doubly occupied) and $C_{N}^2$.

The sorted residues will be used repeatedly in the evaluation of Hamiltonian matrix elements,
selection of oCFGs and perturbative treatment of dynamic correlation. It is just that they should be updated when the space
$V_{\mathrm{cfg}}$ is modified.

\subsection{Hamiltonian construction}
\begin{figure}
\centering
{\resizebox{0.5\textwidth}{!}{\includegraphics{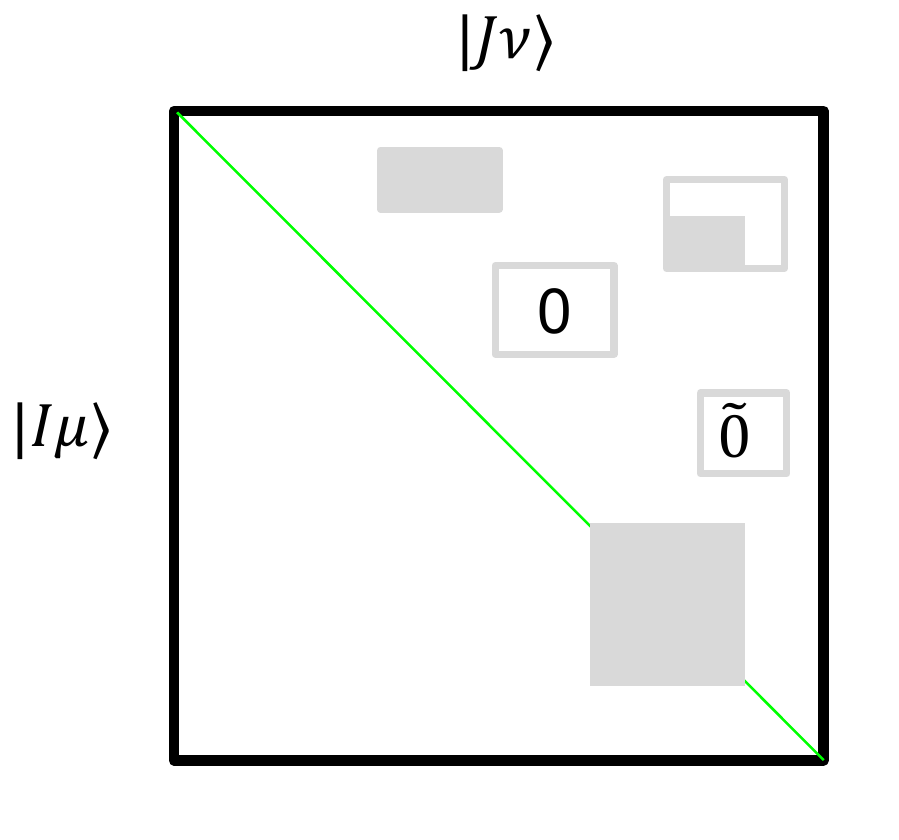}}}
\caption{Schematic illustration of the structure of the symmetric Hamiltonian matrix $[H^{IJ}_{\mu\nu}]$}
\label{Hstructure}
\end{figure}

Given the sorted residue array \textcolor{blue}{R$_2$} of the oCFG space $V_{\mathrm{cfg}}$, the Hamiltonian matrix elements over the CSFs in space $P$ (selected from $V_{\mathrm{cfg}}$) can readily be calculated, since
the oCFGs that generate the same second-order residue are either doubly or singly connected. The algorithm goes as follows:
\begin{enumerate}
    \item Loop over oCFG $I\in V_{\mathrm{cfg}}$ and calculate the diagonal-block elements $\langle I\mu|H_1^0+H_2^0|I\nu\rangle$, with $|I\mu\rangle$ and $|I\nu\rangle$ belonging to $P$.
    \item Loop over all the elements in \textcolor{blue}{R$_2$}:
    \begin{itemize}
    \item Loop over unique oCFG pairs $(I, J)$ ($I<J$) generating the same residue: If they are doubly connected, calculate the matrix elements $\langle I\mu|H_2^2|J\nu\rangle$, with $|I\mu\rangle$ and $|J\nu\rangle$ belonging to $P$. Otherwise, store $(I, J)$ in array \textcolor{blue}{SINGLE}. Note in passing that the double generators $e_{ij,kl}$ ($k\ge i$, $k>l\ge j$) can readily be fixed here
        by making use of the orbital indices $j$ and $l$ recorded in \textcolor{blue}{OrbIndx}.
    \end{itemize}
    \item Remove duplicates in \textcolor{blue}{SINGLE} and calculate the matrix elements $\langle I\mu|H_1^1+H_2^1|J\nu\rangle$, with $|I\mu\rangle$ and $|J\nu\rangle$ belonging to $P$.
     The single generators $E_{ij}$ ($i>j$) can be set up by deleting the common orbital indices of oCFGs $I$ and $J$ recorded in \textcolor{blue}{OrbIndx}.
\end{enumerate}
Note in passing that when the space $P$ is modified to $P^\prime$, the Hamiltonian matrix
for the unaltered part $P_0$ of $P$ need not be recalculated (NB: $P_0$ is spanned by those oCFGs $\{I\}$
common in $P$ and $P^\prime$ but excluding those whose CSFs are not identical due to selection). This is important when $P^\prime$ differs from $P$ only marginally. Overall, the Hamiltonian matrix in block form $[H^{IJ}]$ (i.e., $H^{IJ}_{\mu\nu}=\langle I\mu|H|J\nu\rangle$) is very sparse (see Fig. \ref{Hstructure}):
some blocks are strictly zero or nearly zero; some blocks are full rectangular while some are not full rectangular after the selection of individual CSFs;
some diagonal blocks can be of very large size if the oCFGs have many singly occupied orbitals. Therefore, the matrix must be stored in a suitable sparse form (see Appendix \ref{SparseH}),
so as to facilitate the use of sparse matrix-vector multiplications.

\subsection{Selection of oCFGs and CSFs}
The aim of selection is to find, iteratively, a better variational CSF space $P^{(k+1)}$ for the expansion of the wave function $|\Psi^{(k+1)}\rangle$ by feeding in
the wave function $|\Psi^{(k)}\rangle=\sum_{ |I\mu\rangle\in P^{(k)}}C^{I(k)}_{\mu}|I\mu\rangle$ of the previous iteration $k$.
The selection of important CSFs consists of two steps, ranking and pruning. In the ranking step, proper rank values $\{A^I_{\mu}\}$ for the
CSFs $\{|I\mu\rangle\}$ outside of $P^{(k)}$ will be evaluated. Those CSFs with rank values larger than
the ranking-threshold $C_{\mathrm{q}}$ are then used to extend $P^{(k)}$ to $\bar{P}^{(k+1)}$.
After constructing and diagonalizing the Hamiltonian matrix in $\bar{P}^{(k+1)}$, one obtains an improved wave function
$|\bar{\Psi}^{(k+1)}\rangle=\sum_{|I\mu\rangle\in\bar{P}^{(k+1)}} \bar{C}^{I(k+1)}_{\mu}|I\mu\rangle$ with energy $\bar{E}^{(k+1)}$.
In the pruning step, those CSFs in $\bar{P}^{(k+1)}$ but with coefficients smaller in absolute value than the pruning-threshold $C_{\mathrm{min}}$ are discarded, so as to reduce $\bar{P}^{(k+1)}$ to $P^{(k+1)}$.
In principle, the Hamiltonian matrix in the reduced space $P^{(k+1)}$ should be reconstructed and diagonalized
to obtain $|\Psi^{(k+1)}\rangle=\sum_{|I\mu\rangle\in P^{(k+1)}} C^{I(k+1)}_{\mu}|I\mu\rangle$ with energy $E^{(k+1)}$.
However, this is not necessary for a sufficiently small $C_{\mathrm{min}}$: $\{C^{I(k+1)}_{\mu}\}$ and $E^{(k+1)}$ can simply be set to the corresponding $\{\bar{C}^{I(k+1)}_{\mu}\}$ and $\bar{E}^{(k+1)}$, respectively.

\begin{figure}
\centering
{\resizebox{0.8\textwidth}{!}{\includegraphics{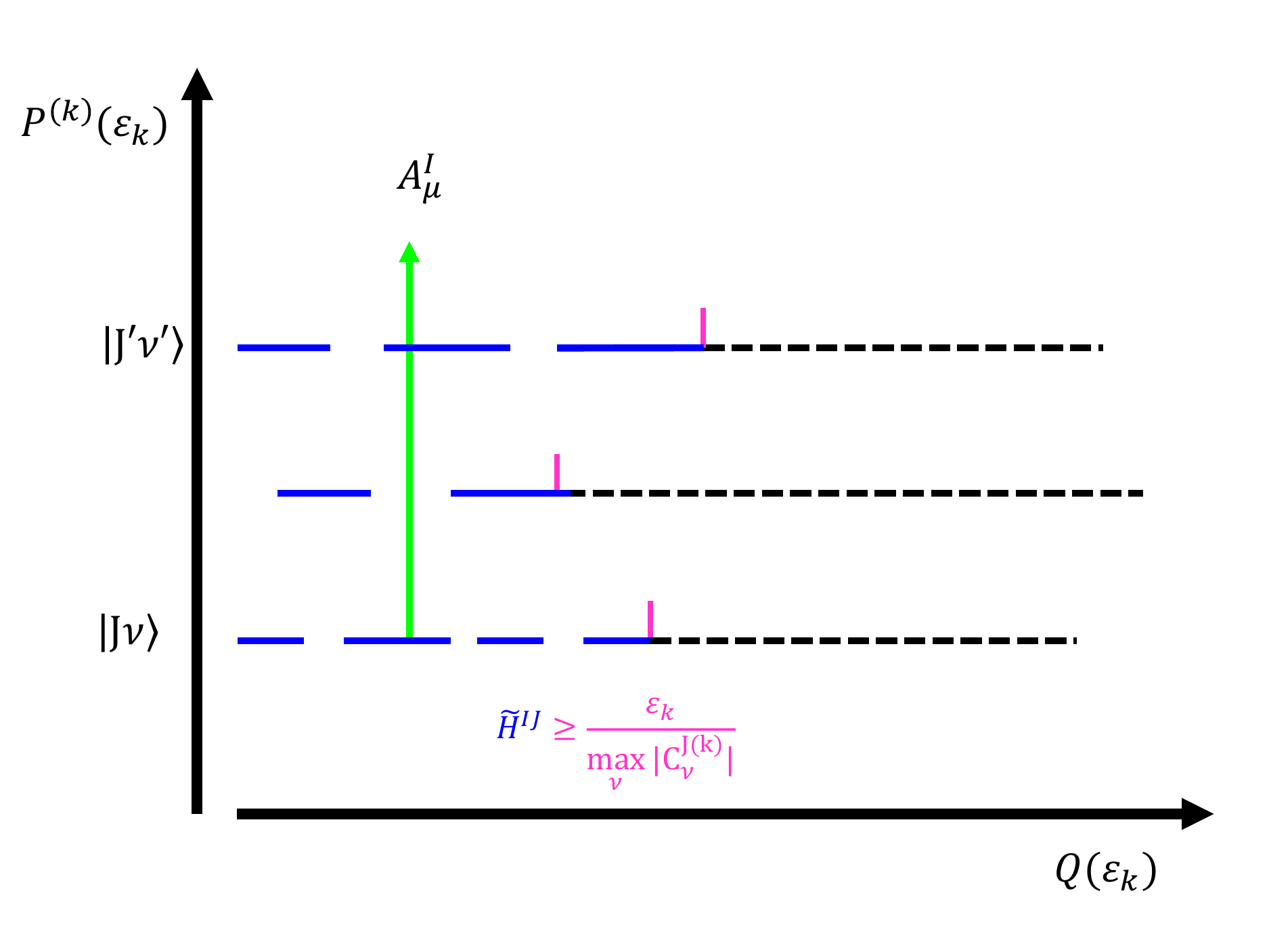}}}
\caption{Illustration of the iCI ranking \eqref{iCIrank}: At each macro-iteration $k$,
doubly excited oCFGs $\{I\}$ from an oCFG $J\in P^{(k)}$ are retained only if they satisfy $\tilde{H}^{IJ}\ge \varepsilon_k/\max_{\nu}|C_{\nu}^{J(k)}|$ (broken blue lines);
those unimportant oCFGs are not accessed at all (dash black lines); some oCFGs of $P^{(k)}$ do not
contribute to $A_{\mu}^I$ due to the restriction \eqref{iCI-J} (blanks between blue lines). }
\label{iCIselection-int}
\end{figure}

As for the ranking, an obvious choice\cite{CIPSIa} is the (pruned\cite{ASCI2016}) first-order coefficient in absolute value,
\begin{equation}
A^I_{\mu}(\mathrm{ASCI}) = \left|\frac{\langle I\mu|H|\Psi^{(k)}\rangle}{E^{(k)}-H_{\mu\mu}^{II}}\right|=\left|\frac{\sum_{|J\nu\rangle\in P^{(k)}}H^{IJ}_{\mu\nu}C^{J(k)}_{\nu}}{E^{(k)}-H_{\mu\mu}^{II}}\right|\ge C_{\mathrm{q}}. \label{ASCIrank}
\end{equation}
Albeit robust, Eq. \eqref{ASCIrank} is computationally too expensive (due to the summation in the numerator) for the purpose of ranking.
It was therefore simplified to
\begin{equation}
A^I_{\mu}(\mathrm{HCI})=\max_{|J\nu\rangle\in P^{(k)}}|H^{IJ}_{\mu\nu}C^{J(k)}_{\nu}|
\stackrel{\mathrm{CSF}\rightarrow\mathrm{Det}}\longrightarrow \max_{\nu\in P^{(k)}} |H_{\mu\nu}C_{\nu}^{(k)}|\ge\epsilon_{\mathrm{q}}\label{HBCIrank}
\end{equation}
in the heat-bath CI approach (HCI)\cite{HBCI2016}, which works with determinants though.
Since the Hamiltonian matrix elements $H_{\mu\nu}$ over doubly connected determinants $|\Phi_{\mu}\rangle$ and $|\Phi_{\nu}\rangle$ depend only on the two-electron integrals (the absolute values of
which can be sorted in descendent order and stored in memory from the outset), the ranking can be made extremely
efficient: Those determinants $\{|\Phi_{\mu}\rangle\}$ that are doubly excited from $|\Phi_{\nu}\rangle\in P^{(k)}$ are never accessed if $|H_{\mu\nu}|<\epsilon_{\mathrm{q}}/|C_{\nu}^{(k)}|$\cite{HBCASSCF}. However, those determinants of high energies may become unimportant even though they satisfy condition \eqref{HBCIrank}.
Therefore, the variational space determined by the integral-driven selection \eqref{HBCIrank} is usually larger than that determined
by the coefficient-driven selection \eqref{ASCIrank}, particularly when large basis sets are used (NB: this problem
can partly be resolved by introducing an approximate denominator to condition \eqref{HBCIrank}, see the Supporting Information of Ref.\cite{HBCI-Cr2}).
The recent development\cite{ASCI2018} of the adaptive sampling CI (ASCI) approach\cite{ASCI2016} reveals that
sorting based algorithms can render the evaluation of $A_{\mu}^I$ \eqref{ASCIrank} very fast. Here,
the integral- and coefficient-driven algorithms are combined for the selection of doubly excited CSFs, viz.,
\begin{equation}
A^I_{\mu}(\mathrm{iCI})= \left|\frac{\sum_{J\in P^{(k)}}^{(\varepsilon_k)}
(\sum_{\nu,|J\nu\rangle\in P^{(k)}}H^{IJ}_{\mu\nu}C^{J(k)}_{\nu})}{E^{(k)}-H_{\mu\mu}^{II}}\right|\ge C_{\mathrm{q}}, \label{iCIrank}
\end{equation}
where the summation over oCFG $J\in P^{(k)}$ is subject to the following condition
\begin{equation}
\max_{\nu}|\tilde{H}^{IJ}C_{\nu}^{J(k)}|\ge\varepsilon_k=\frac{1}{2}\varepsilon_{k-1},\quad J\in P^{(k)}.\label{iCI-J}
\end{equation}
Here, $\tilde{H}^{IJ}$ are the estimated upper bounds for the two-body Hamiltonian matrix elements $\langle I\mu|H_2^2|J\nu\rangle$ (see Table \ref{H-diagram}). Specifically,
like condition \eqref{HBCIrank}, those oCFGs $\{I\}$ that are doubly excited from oCFG $J$ in $P^{(k)}$ are never accessed if $\tilde{H}^{IJ}<\varepsilon_k/\max_{\nu} |C_{\nu}^{J(k)}|$ (see Fig. \ref{iCIselection-int}).
After this integral-driven screening, the individual CSFs $\{|I\mu\rangle\}$ of the selected oCFGs $\{I\}$ are further selected using their (approximate) first-order coefficients \eqref{iCIrank}
(NB: the approximation arises from the restriction \eqref{iCI-J} on the summation over $J$ in the numerator), just
like condition \eqref{ASCIrank}. Moreover, as indicated by the relation $\varepsilon_k=\frac{1}{2}\varepsilon_{k-1}$, the integral-threshold $\varepsilon_k$
is to be reduced by a factor of two with each iteration, meaning that the external oCFG space $Q$ is accessed incrementally (i.e., $Q(\varepsilon_k)$).
Since the calculation of $A^I_{\mu}(\mathrm{iCI})$ \eqref{iCIrank} becomes increasingly more expensive as the integral-threshold $\varepsilon_k$ gets reduced,
we decide to repeat the selections for a given integral-threshold $\varepsilon_k$ until the following condition is fulfilled
\begin{equation}
S=\frac{|P^{(k)}_j\cap P^{(k)}_{j-1}|}{|P^{(k)}_j\cup P^{(k)}_{j-1}|}> S_{\mathrm{p}},
\end{equation}
where the numerator (denominator) is the number of CSFs in the intersection (union) of the $P^{(k)}_j(\epsilon_k)$ and $P^{(k)}_{j-1}(\epsilon_k)$
spaces between two adjacent micro-iterations (designated by subscript $j$). Upon convergence at $j=t$, the space $P^{(k)}_t(\epsilon_k)$ is diagonalized via the iVI approach\cite{iVI,iVI-TDDFT}
to obtain $|\Psi^{(k+1)}\rangle$ and $E^{(k+1)}$. Before going to the next macro-iteration, Algorithm \ref{AlgorithmNO} is invoked to decide whether the rotation of orbitals and update of $\epsilon_k$
need to be performed. The algorithm also decides when to terminate the selection.

\begin{algorithm}
    \caption{Determination of rotation of orbitals and update of $\varepsilon_k$.
    Return \textbf{Yes} to terminate the selection or \textbf{No} to go next iteration.
    The condition for the convergence of NOs is as follows.
    The NOs $\{\phi_p^{(k+1)}\}$ and $\{\phi_p^{(k)}\}$ are related by $\phi_p^{(k+1)}=\sum_{q}c_{qp}\phi_q^{(k)}$.
    If $\exists q \text{ s.t. } |C_{qp}|\geq 0.99$, $\phi_p^{(k+1)}$ is considered to be converged (the same as $\phi_q^{(k)}$).
    If the sum of the occupation numbers of the converged NOs deviates from the total number of correlated electrons by less than 0.05,
    $\{\phi_p^{(k+1)}\}$ are considered to be the same as $\{\phi_p^{(k)}\}$.}
   \label{AlgorithmNO}
    \eIf {Orbital rotation is allowed}
    {
        \eIf{$|E^{(k+1)}-E^{(k)}|\le 10\Delta_{\mathrm{e}}$}
        {
            Construct and diagonalize the one-particle reduced density matrix to generate natural orbitals (NO) $\{\phi_p^{(k+1)}\}$,
            followed by integral transformation and rediagonalization of $P^{(k)}_t(\varepsilon_k)$. \\
            \eIf{$\{\phi_p^{(k+1)}\}$ are converged}
            {$\varepsilon_{k+1}=\varepsilon_k/2$ and switch off orbital rotation.}
            {$\varepsilon_{k+1}=\varepsilon_k$.}
        }
        {
            $\varepsilon_{k+1}=\varepsilon_k/2$.
        }
        return \textbf{No}.
    }
    {
        \eIf{$|E^{(k+1)}-E^{(k)}|\le \Delta_{\mathrm{e}}$}
        {
            return \textbf{Yes};
        }
        {
            $\varepsilon_{k+1}=\varepsilon_k/2$ , return \textbf{No}.
        }
    }

\end{algorithm}

\begin{figure}
\centering
\includegraphics[width=\textwidth]{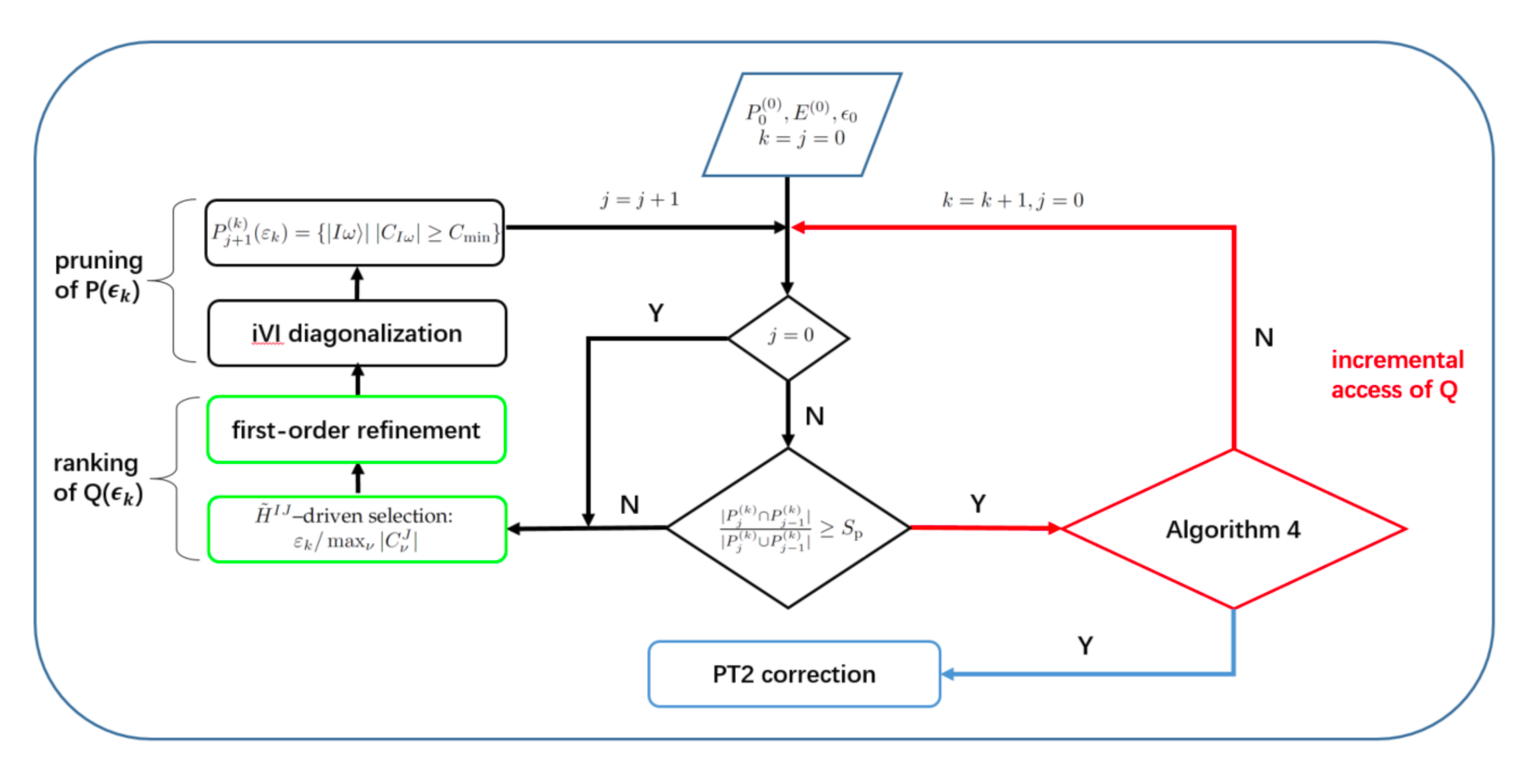}
\caption{Flowchart for iCI selection (Scheme I)}
\label{iCIselection1}
\end{figure}

To illustrate the above algorithm more clearly, a flowchart is provided in Fig. \ref{iCIselection1}. Simply put,
given an integral-threshold $\varepsilon$ to confine the accessible external oCFG space $Q(\varepsilon)$ (macro-iteration), the variational CSF space $P(\varepsilon)$ is updated iteratively until convergence (micro-iteration). Repeat this with
a reduced $\varepsilon$ until no new NOs are needed. As for the thresholds, the dynamically adjusted integral-threshold
$\varepsilon$ affects only the number of iterations; both $S_{\mathrm{p}}$
and $\Delta_{\mathrm{e}}$ can be set to conservative values (e.g., 0.95 and $3\times10^{-4}$, respectively);
the ranking-threshold $C_{\mathrm{q}}$ can simply be chosen to be the same as the
coefficient pruning-threshold $C_{\mathrm{min}}$. Therefore, only $C_{\mathrm{min}}$ is truly a free parameter, which controls the size of the variational space and hence the final accuracy. It also deserves to be mentioned
that the integral-threshold $\epsilon$ is usually larger than $C_{\mathrm{min}}/10$ upon termination of
the selection.

\begin{figure}
\centering
\includegraphics[width=\textwidth]{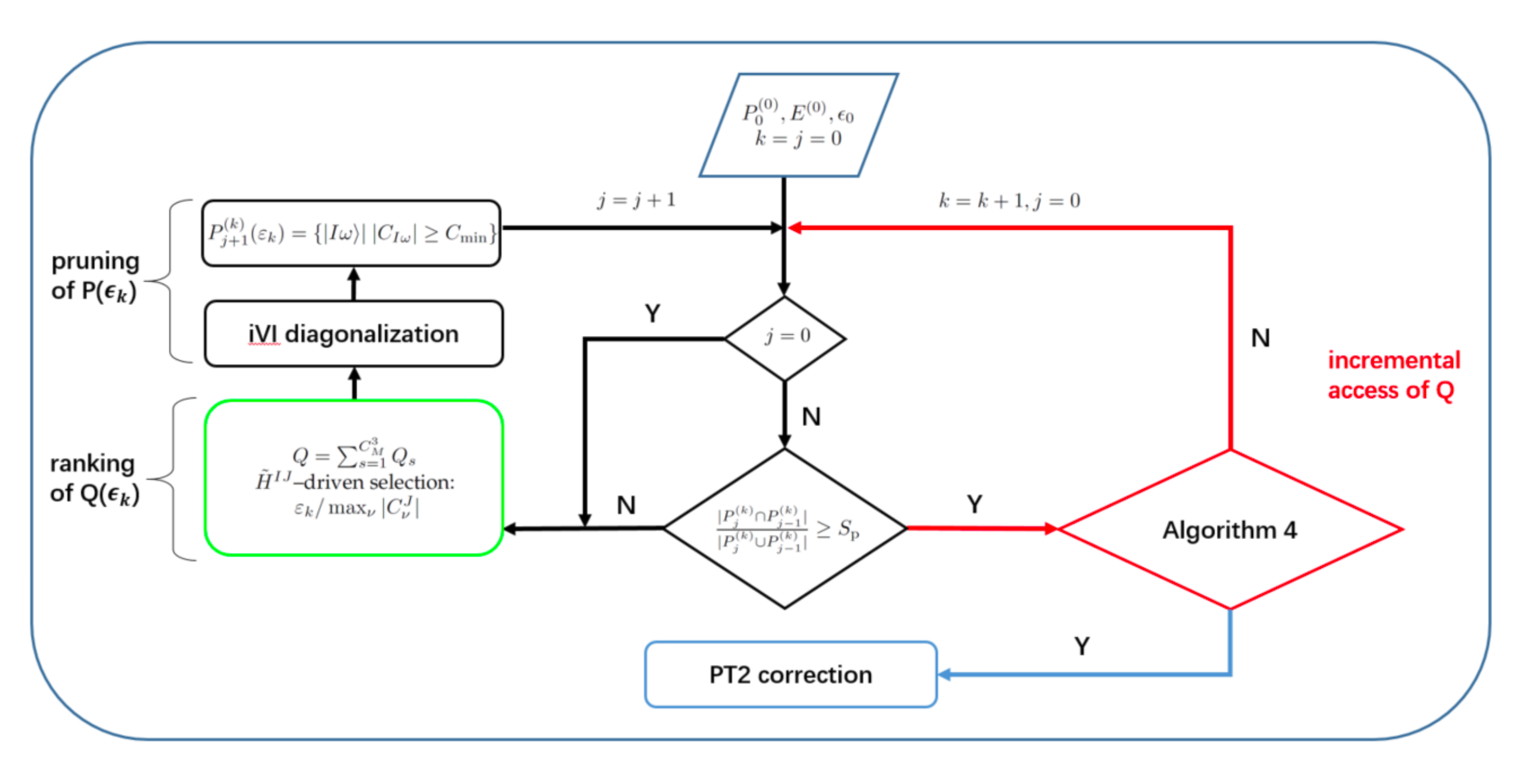}
\caption{Flowchart for iCI selection (Scheme II)}
\label{iCIselection2}
\end{figure}

The above algorithm is truly efficient. However, there may exist a memory bottleneck even if the external space
is accessed only incrementally. This bottleneck can be resolved by decomposing the external space into
disjoint subsets $\{Q_s\}$ (see Sec. \ref{SecPT2}), such that only one subspace is accessed at a time, with no communications between different subspaces.
This leads to a modified algorithm shown in Fig. \ref{iCIselection2}.
The major difference from the previous algorithm lies in that the oCFGs $\{I\}\in Q_s$ have to be selected individually according to condition \eqref{iCI-J}. That is, a doubly excited oCFG is first generated but is then dumped if it does not fulfill condition \eqref{iCI-J}.


\section{Constraint-based Epstein-Nesbet PT2}\label{SecPT2}
Having determined a high-quality yet compact variational CSF space $P$ and hence the zeroth-order eigenpairs $\{E_i^{(0)},|\Psi_i^{(0)}\rangle\}$,
we are now ready to account for the dynamic correlation via, e.g., the state-specific Epstein-Nesbet second-order perturbation theory (PT2)\cite{Epstein,Nesbet}
\begin{align}
E_{c,i}^{(2)} = \sum_{|I\mu\rangle\in Q}\frac{\left|\langle I\mu|H|\Psi_i^{(0)}\rangle\right|^2}{E_i^{(0)}-H_{\mu\mu}^{II}}=\sum_{|I\mu\rangle\in Q} \frac{\left|\sum_{|J\nu\rangle\in P}H^{IJ}_{\mu\nu}C^J_{\nu,i}\right|^2}{E_i^{(0)}-H_{\mu\mu}^{II}}.\label{ENPT2}
\end{align}

One problem associated with expression \eqref{ENPT2} lies in that it is memory intensive:
The number of external CSFs $\{|I\mu\rangle\}$ scales as $N_{\mathrm{csf}} n_o^2 n_v^2$, with $N_{\mathrm{csf}}$ being the number of CSFs $\{|J\nu\rangle\}$ in $P$.
To reduce the memory requirement,
we adopt the constraint PT2 algorithm proposed recently by Tubman et al.\cite{ASCI2018PT2}. The basic idea
is to decompose the external space $Q$ into disjoint subsets $\{Q_s\}$. For instance, the triplet constraints give rise to
$C_{N_{\mathrm{orb}}}^3$ subsets, each of which is characterized by a particular set of
three numbers $(p, q, r)$ ($p>q>r$) that specify the three highest occupied spatial orbitals in the oCFGs. That is,
all the orbitals are unoccupied in the oCFGs if they are higher than $r$ but different from $p$ and $q$, i.e.,
 $(0, \cdots, 0, n_p, 0, \cdots, 0, n_q, 0, \cdots, 0, n_r)$ in which $n_p, n_q, n_r\in[1,2]$.
Such subspaces can readily be generated by adding one and two electrons into the first- and second-order residues of the $P$ space, respectively, with the following rules:
Given a triplet $(p, q, r)$ that defines, say, subspace $Q_s$, a valid first-order residue can have at most one zero,
whereas a valid second-order residue can have at most two zeros in the occupation numbers $(m_p, m_q, m_r)$.
If present, they must be filled with at least one electron. If singly occupied, $p/q/r$ can still accept one electron.
Except for these on-site cases, only those orbitals lower than $r$ can accept electrons (if any).
More importantly, valid residues, whether first- or second-order,
are located \textit{continuously} in different segments of the sorted residue array \textcolor{blue}{OrbOccBinary}.
The number of such segments is at most $3\times 3\times 3-1=26$ $(m_p, m_q, m_r \in[0,2]\backslash m_p=m_q=m_r=0)$.
The head and tail of every segment can readily be found by using the bisection method.
For a first-order residue with electron-depletion orbital $j$ recorded in array \textcolor{blue}{OrbIndx}, the orbital occupied by the added electron corresponds to $i$ in $E_{ij}$.
Note that the case of $i=j$ is also allowed here, for CSF $|J\mu\rangle$ may also contribute to the perturbation to another CSF $|J\nu\rangle$ from the same oCFG $J$.
For a second-order residue with electron-depletion orbitals $j$ and $l$ ($\ge j$) recorded in array \textcolor{blue}{OrbIndx}, the two orbitals occupied by the added two electrons corresponds to $i$ and $k$ $(\ge i)$ in $e_{ij,kl}$, after eliminating the zero- and one-body excitations [i.e., $\{ i, k\}\cap\{j, l\} \ne \emptyset$; cf. Eq. \eqref{H201def}]. If $i<j$ in $E_{ij}$ or $k<l$ in $e_{ij,kl}$, a bra-ket inversion should be invoked when calculating
the BCCs in terms of the diagrams documented in Table \ref{Segments}.

Being disjoint, the subspaces $\{Q_s\}$ can be embarrassingly parallelized, viz.
\begin{align}
E_{c,i}^{(2)} &= \sum_{s=1}^{C_{N_{\mathrm{orb}}}^3} \sum_{|I\mu\rangle\in Q_s} \frac{\left|\langle I\mu|H|\Psi_i^{(0)}\rangle\right|^2}{E_i^{(0)}-H_{\mu\mu}^{II}}.\label{ENPT2Q}
\end{align}
Obviously, the same technique can be applied to the evaluation of $A_{\mu}^I$ \eqref{ASCIrank}/\eqref{iCIrank} as well.

Another problem associated with Eq. \eqref{ENPT2} and also with Eq. \eqref{ENPT2Q}
lies in that some of the singly and doubly excited CSFs from space $P$ are already present in $P$ and it is very expensive to check this
(which is not the case for the ranking \eqref{ASCIrank}/\eqref{HBCIrank}/\eqref{iCIrank}, because the number of important CSFs to be added to space $P$ is not very many,
such that the double check of duplications in $P$ is very cheap). This issue can be resolved\cite{ASCI2018PT2} by precomputing
the PT2-like energy $\bar{E}_{c,i}^{(2)}$ for such singly and doubly excited CSFs spanning $\bar{P}\in P$, which is finally subtracted from the PT2 energy $\tilde{E}_{c,i}^{(2)}$ for all the singly and doubly excited CSFs spanning
$\tilde{Q}=Q\cup\bar{P}$:
\begin{align}
E_{c,i}^{(2)} &=\tilde{E}_{c,i}^{(2)}-\bar{E}_{c,i}^{(2)},\label{ENPT2diff}\\
\tilde{E}_{c,i}^{(2)}&= \sum_{s=1}^{C_{N_{\mathrm{orb}}}^3} \sum_{|I\mu\rangle\in \tilde{Q}_s} \frac{\left|\sum_{|J\nu\rangle\in P, |J\nu\rangle\ne |I\mu\rangle}H_{\mu\nu}^{IJ}C_{\nu,i}^J\right|^2}{E_i^{(0)}-H_{\mu\mu}^{II}},\quad \tilde{Q}_s=Q_s\cup\bar{P},\label{ENPT2Qt}\\
\bar{E}_{c,i}^{(2)}&=\sum_{|I\mu\rangle\in \bar{P}} \frac{\left|\langle I\mu|H|\Psi_i^{(0)}\rangle-C_{\mu,i}^I H^{II}_{\mu\mu}\right|^2}{E_i^{(0)}-H_{\mu\mu}^{II}}\label{ENPT2Pa}\\
                   &=\sum_{|I\mu\rangle\in \bar{P}} (C_{\mu,i}^I)^2 (E_i^{(0)}-H^{II}_{\mu\mu}),\label{ENPT2P}
\end{align}
where use of the relation $\langle I\mu|H|\Psi_i^{(0)}\rangle=C_{\mu,i}^I E_i^{(0)}$ has been made when going from Eq. \eqref{ENPT2Pa} to Eq. \eqref{ENPT2P}.
The negative term in the numerator of Eq. \eqref{ENPT2Pa} arises from the fact that the diagonal terms (zero-body excitations) have been excluded in Eq. \eqref{ENPT2Qt}.
This particular arrangement\cite{ASCI2018PT2} simplifies the evaluation of $\bar{E}_{c,i}^{(2)}$ via Eq. \eqref{ENPT2P}.
Finally, if wanted, Eq. \eqref{ENPT2Qt} can be approximated as
\begin{align}
\tilde{E}_{c,i}^{(2)}(\tau)&=\sum_{s=1}^{C_{N_{\mathrm{orb}}}^3} \sum_{I\in \tilde{Q}_s}^{(\tau)}\left[\sum_{\mu, |I\mu\rangle\in \tilde{Q}_k} \frac{\left|\sum_{|J\nu\rangle\in P, |J\nu\rangle\ne |I\mu\rangle}H_{\mu\nu}^{IJ}C_{\nu,i}^J\right|^2}{E_i^{(0)}-H_{\mu\mu}^{II}}\right],
\end{align}
where the summation over doubly excited oCFGs $\{I\}$ is subject to the following condition
\begin{equation}
\max_{\nu}|\tilde{H}^{IJ}C_{\nu,i}^J|\ge\tau,
\end{equation}
with $\tau$ being a very conservative threshold for truncating the first-order interacting space $Q=\sum^{\oplus}_s Q_s$. Note that
this truncation does not affect those oCFGs $\{I\}$ belonging to space $\bar{P}\in P$, because $\tau$ is orders of magnitude smaller than $\varepsilon$ employed for the selection of $P$ (cf. \eqref{iCI-J}).
Therefore, contamination of $\bar{P}$ to $E_{c,i}^{(2)}(\tau)=\tilde{E}_{c,i}^{(2)}(\tau)-\bar{E}_{c,i}^{(2)}$ will not arise. It is expected that the difference $E_{c,i}^{(2)}(\tau=0)-E_{c,i}^{(2)}(\tau)$ is only weakly dependent on the coefficient
pruning-threshold $C_{\mathrm{min}}$, i.e., $E_{c,i}^{(2)}(C^S_{\mathrm{min}},\tau=0)-E_{c,i}^{(2)}(C^S_{\mathrm{min}},\tau)\simeq E_{c,i}^{(2)}(C^L_{\mathrm{min}},\tau=0)-E_{c,i}^{(2)}(C^L_{\mathrm{min}},\tau)$ should hold for $C_{\mathrm{min}}^S<C_{\mathrm{min}}^L$. When this has been reached, we can get $E_{c,i}^{(2)}(C^S_{\mathrm{min}},\tau=0)$ from $E_{c,i}^{(2)}(C^L_{\mathrm{min}},\tau=0)+\left[E_{c,i}^{(2)}(C^S_{\mathrm{min}},\tau)-E_{c,i}^{(2)}(C^L_{\mathrm{min}},\tau)\right]$,
so as to reduce the computational costs.

\section{Pilot applications}\label{Result}
\subsection{\ce{C2}}
The proposed iCIPT2 approach is first applied to carbon dimer, a system that is known to
have strong multireference characters even at the equilibrium distance. The calculations start with HF orbitals but which are rotated to natural orbitals (NO) during the selection procedure (see Algorithm \ref{AlgorithmNO}).
Although $D_{2h}$ instead of $D_{\infty h}$ symmetry is used, the terms of the states can be assigned correctly by considering simultaneously
the degeneracy of the calculated energies and the correspondences between the $D_{2h}$ and $D_{\infty h}$ irreducible representations.

\begin{figure}[!htp]
	\centering
	\includegraphics[width=\textwidth]{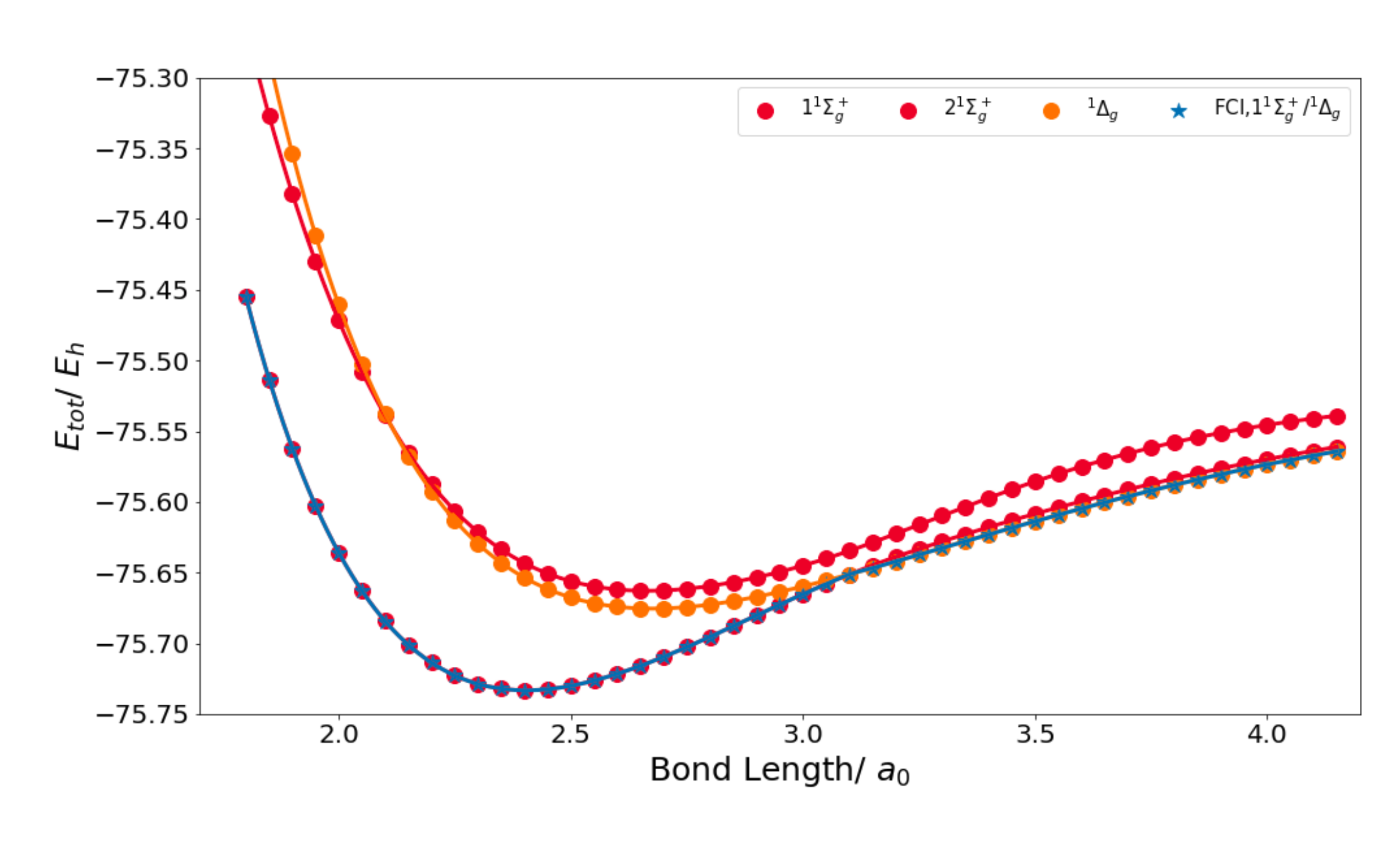}
	\caption{Potential energy curves for the lowest three states of \ce{C2} by all-electron iCIPT2/cc-pVDZ ($C_{\text{min}}=10^{-4}$).
The all-electron FCI/cc-pVDZ results are taken from Ref.\cite{MPS-LCC}. }\label{C2curve}
\end{figure}

\begin{figure}[!htp]
	\centering
	\includegraphics[width=\textwidth]{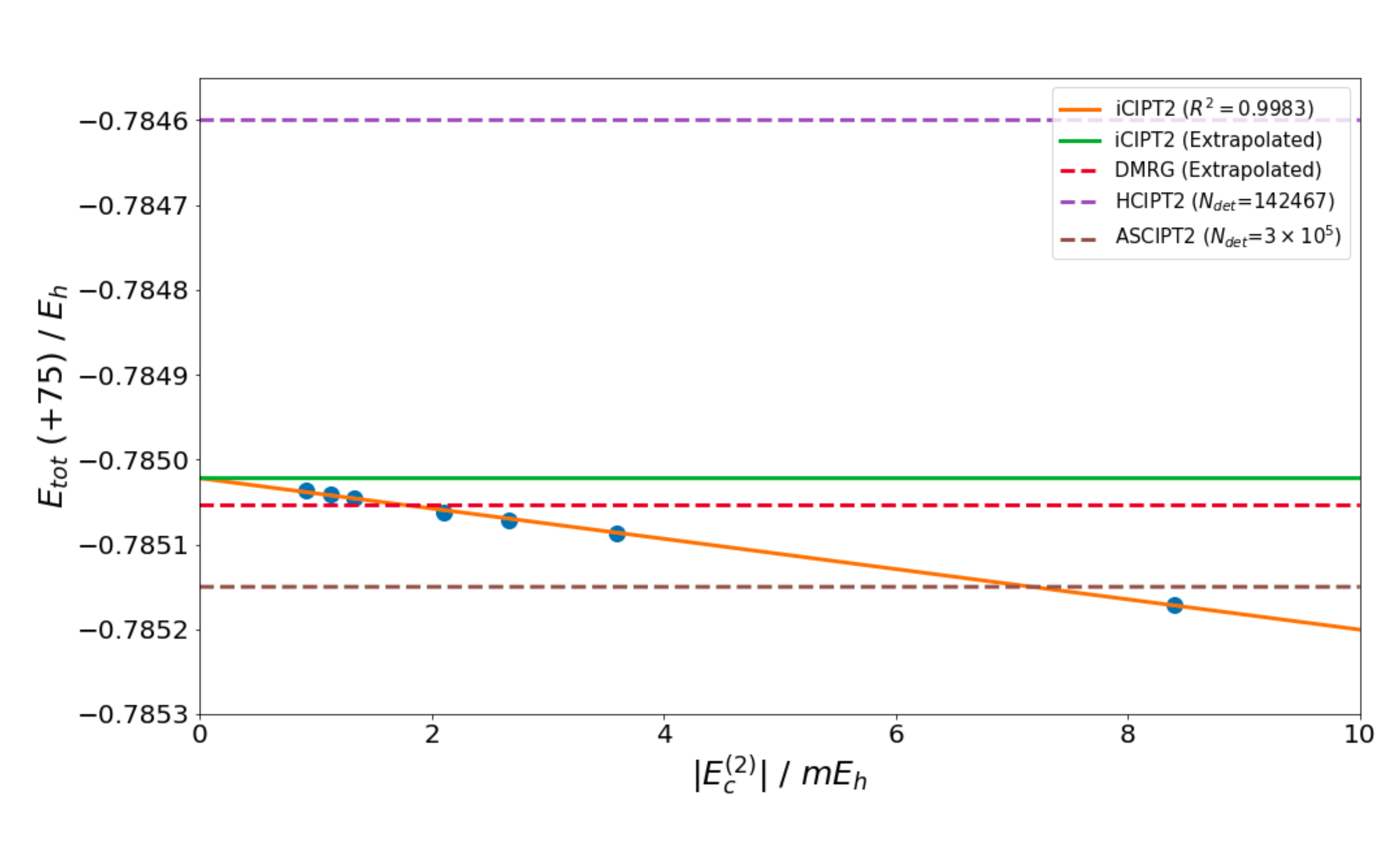}
	\caption{Comparison of the frozen-core ground state energies of \ce{C2} at $R=1.24253$ \AA~by different methods with cc-pVTZ.
Blue dots refer to different $C_{\mathrm{min}}$ values in iCIPT2.
The extrapolated values are -75.785022(3), -75.785054, -75.78515 and -75.7846(3) $\mathrm{E_h}$ for iCIPT2, DMRG\cite{DMRG2015}, ASCIPT2\cite{ASCI2018PT2} and HCIPT2\cite{HBCI2017c}, respectively.}\label{C2TZ}
\end{figure}

\begin{figure}[!htp]
	\centering
	\includegraphics[width=\textwidth]{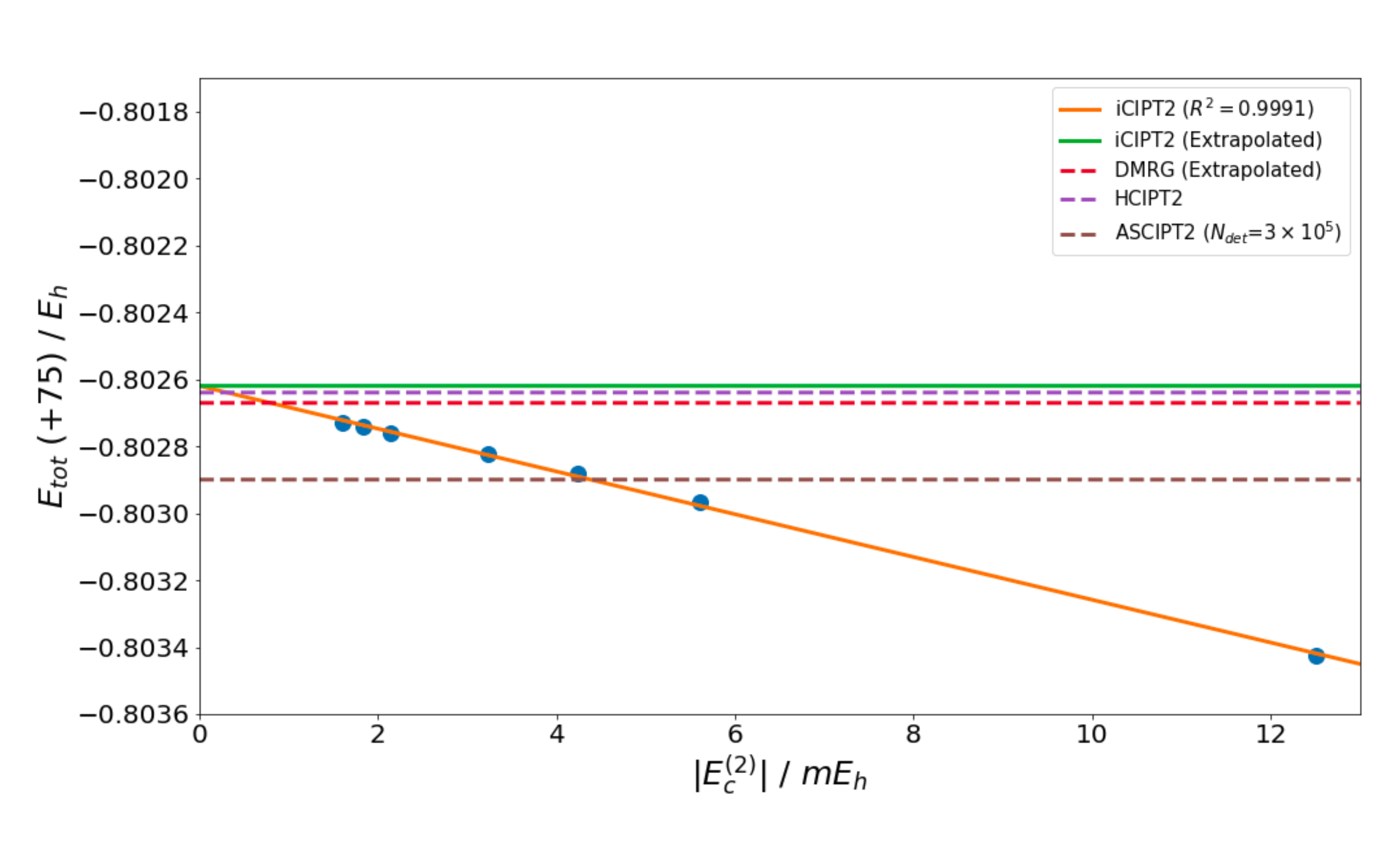}
	\caption{Comparison of the frozen-core ground state energies of \ce{C2} at $R=1.24253$ \AA~by different methods with cc-pVQZ.
Blue dots refer to different $C_{\mathrm{min}}$ values in iCIPT2.
The extrapolated values are -75.802620(13), -75.802671, -75.80264 and -75.80290 $\mathrm{E_h}$ for iCIPT2, DMRG\cite{DMRG2015}, ASCIPT2\cite{ASCI2018PT2} and HCIPT2\cite{HBCI2017b}, respectively.}\label{C2QZ}
\end{figure}

\begin{figure}[!htp]
	\centering
	\includegraphics[width=\textwidth]{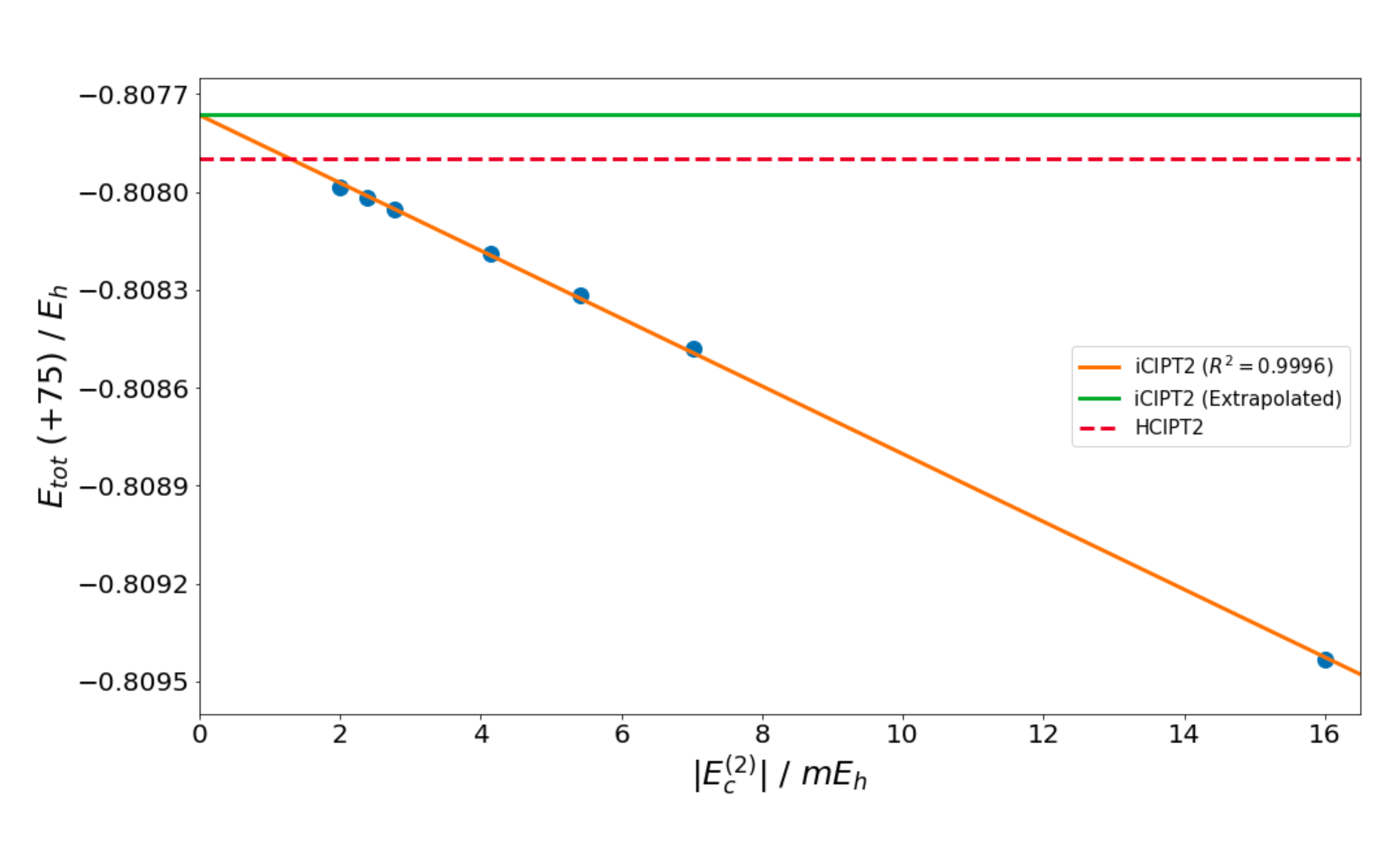}
	\caption{Comparison of the frozen-core ground state energies of \ce{C2} at $R=1.24253$ \AA~by different methods with cc-pV5Z.
Blue dots refer to different $C_{\mathrm{min}}$ values in iCIPT2.
The extrapolated values are -75.807764(18) and -75.80790(3) a.u. for iCIPT2 and HCIPT2\cite{HBCI2017b}, respectively.}\label{C25Z}
\end{figure}

To compare directly with previous results, we first report the frozen-core ground state energies of \ce{C2} calculated
at the equilibrium distance (1.24253 \AA) with the cc-pVXZ (X = D, T, Q, 5) basis sets\cite{Dunning1989}. It is seen from Table \ref{C2energies} and more clearly from Figs. \ref{C2TZ} to \ref{C25Z} that
the iCIPT2 energies converge steadily as the coefficient pruning-threshold $C_{\mathrm{min}}$ decreases, and become full agreement (within $0.1$ $\mathrm{mE_h}$) with the corresponding DMRG\cite{DMRG2015}, ASCIPT2\cite{ASCI2018PT2}
and HCIPT2\cite{HBCI2017c,HBCI2017b} results
when $C_{\mathrm{min}}=10^{-4}$, $10^{-4}$, $0.3\times 10^{-4}$ and $0.25\times 10^{-4}$ are used for the DZ, TZ, QZ and 5Z bases, respectively.
Since the convergence is so smooth, one can stop at any desired accuracy. Encouraged by this,
we further report the iCIPT2 excitation energies for the 8 lowest excited states of \ce{C2} at their equilibrium distances, which are known experimentally\cite{CarbonExp}.
It is seen from Table \ref{C2Excitation} that the maximum and mean absolute deviations from the experimental values\cite{CarbonExp} are only 0.02 eV and 0.01 eV, respectively, by
the all-electron calculations with both the QZ and 5Z bases. The frozen-core approximation introduces noticeable errors (ca. 0.02 eV on average).
Another issue related to all selection-based schemes lies in whether a smooth potential energy surface can be produced. This is indeed the case for iCIPT2. As can be seen from Fig. \ref{C2curve} (and Supporting Information),
the all-electron FCI/DZ potential energy curve\cite{MPS-LCC} for the ground state of \ce{C2} is nicely reproduced by iCIPT2/DZ with $C_{\mathrm{min}}=10^{-4}$:
the deviation of iCIPT2 from FCI is not larger than 0.1 $\mathrm{mE_h}$ for the whole range of the FCI curve and the crossing between $^1\Sigma_g^+$ and  $^1\Delta_g$ at 3.10 bohrs
is also reproduced. For comparison, the matrix product state-based linearized coupled-cluster (MPS-LCC) method\cite{MPS-LCC} deviates from FCI by 0.1--4.8 $\mathrm{mE_h}$
and exhibits a discontinuity at 3.10 bohrs. It also deserves to be mentioned that the iCIPT2 energy difference between the noninteracting C$\cdots$C and two carbon atoms is at most 1 $\mathrm{mE_h}$
for all the bases considered here, manifesting that iCIPT2 is nearly size consistent.


\begin{table}[!htp]
	\tiny
	\caption{Frozen-core iCIPT2/cc-pVXZ (X = D, T, Q, 5) calculations of the ground state of \ce{C2} at $R=1.24253$ \AA.
   $C_{\mathrm{min}}$: threshold for pruning configuration state functions (CSF) in the variational (var) space;
  $N_{\mathrm{cfg}}$: number of orbital configurations in the variational space; $N_{\mathrm{csf}}$: number of CSFs
  corresponding to $N_{\mathrm{cfg}}$; PT2: state-specific Epstein-Nesbet second-order perturbation theory.}
    \begin{threeparttable}
	\centering
	\begin{tabular}{ccrrrrllrrrrrrrrrrrrrrrrrrrrrrrr}
    \toprule
		&&&&&&\multicolumn{2}{c}{energy ($\mathrm{E_h}$)}
		&\multicolumn{3}{c}{wall time (second)}\\\cline{7-8}\cline{9-11}
		basis
		&\multicolumn{1}{c}{$\text{C}_{\text{min}}/10^{-4}$}
		&\multicolumn{1}{c}{$N_{\text{cfg}}$}
		&\multicolumn{1}{c}{$N_{\text{csf}}$}
		&\multicolumn{1}{c}{$\tilde{N}_{\text{csf}}$\tnote{a}}
		&\multicolumn{1}{c}{$\tilde{N}_{\text{det}}$\tnote{b}}
		&\multicolumn{1}{c}{var}
		&\multicolumn{1}{c}{total}
		&\multicolumn{1}{c}{var}
		&\multicolumn{1}{c}{PT2}
		&\multicolumn{1}{c}{total}\\\toprule
		DZ
		  &3.0        &4025 &11771 &6662 &22046 &-75.725235&-75.728478 &2.0 &0.4&2.4\\
		  &1.0        &10858&36043 &20009&36043 &-75.727359&-75.728514 &5.8 &0.9&6.7\\
		  &0.7        &14840&51247 &28405&101193&-75.727736&-75.728518 &9.3 &1.2&10.5\\
		  &0.5        &19850&71054 &39323&142051&-75.727989&-75.728520 &13.4&1.6&15.0\\
		  &0.3        &30023&113084&62566&230920&-75.728232&-75.728521 &23.3&2.2&25.5\\
		  &DMRG\tnote{d}       &     &      &     &      &          &-75.728556  &    &   &\\
		  &ASCIPT2\tnote{e}&&&&300000&-75.72836&-75.72855&&&\\
          &HCIPT2\tnote{f}&&&&28566&-75.7217&-75.7286(2)&&&\\
		\midrule
		TZ
		  &3.0        &7619 &21222 &11407 &36670 &-75.776765&-75.785171 &9  &3 &12\\
		  &1.0        &23951&77471 &39342 &135058&-75.781493&-75.785087 &25 &9 &34\\
		  &0.7        &34708&117741&59187 &207007&-75.782402&-75.785071 &39 &13&52\\
		  &0.5        &46186&162497&82205 &292205&-75.782952&-75.785063 &54 &17&71\\
		  &0.3        &78289&291552&145729&529906&-75.783710&-78.785045 &104&29&133\\
          &0.25       &94256 &358507&178041&652141&-75.783912&-75.785041&129&33&162\\
		  &0.2        &117839&459603&227080&839058&-75.784120&-75.785036&171&42&213\\
          &0.0\tnote{c}        &     &      &     &      &          &  -75.785022(3)       &    &   &  \\
		  &DMRG\tnote{d}       &     &      &      &      &          &-75.785054 &   &  &\\
		  &ASCIPT2\tnote{e}&&&&300000&-75.78196&-75.78515&&&\\
          &HCIPT2\tnote{f}&&&&28566&-75.7738&-75.7846(3)&&&\\
          \midrule
		QZ
		  &3.0        &9131  &24350 &13147 &41483 &-75.790913&-75.803425 &51 &14 &65\\
		  &1.0        &29750 &91695 &46530 &156565&-75.797365&-75.802968 &77 &41 &118\\
		  &0.7        &44534 &144199&71648 &245693&-75.798640&-75.802882 &110&60 &170\\
		  &0.5        &64498 &218116&106986&373386&-75.799585&-75.802824 &169&85 &254\\
		  &0.3        &112453&405609&194548&695920&-75.800617&-75.802760 &338&147&485\\
          &0.25       &136467&503417&239787&864600&-75.800903&-75.802743&416&171&587\\
		  &0.2        &164647&615634&295452&1073145&-75.801133&-75.802730&484&209&693\\
          &0.0\tnote{c}        &     &      &     &      &          &  -75.802620(13)       &    &   &  \\
		  &DMRG\tnote{d}       &      &      &      &      &          &-75.802671 &   &   &\\
          &ASCIPT2\tnote{e}&&&&300000&-75.79807&-75.80290&&&\\
          &HCIPT2\tnote{g}&&&&&&-75.80264&&&\\
          \bottomrule
        5Z
		&3.0 &9512  &25145 &13563 &42624 &-75.793434&-75.809434&273&46 &319\\
		&1.0 &33806 &101439&51317 &170618&-75.801447&-75.808479&396&132&528\\
		&0.7 &49422 &154744&77314 &261569&-75.802895&-75.808318&452&189&641\\
		&0.5 &71918 &235234&115843&399168&-75.804043&-75.808190&590&275&865\\
		&0.3 &126783&442037&212308&749953&-75.805291&-75.808055&940&467&1407\\
		&0.25&154888&551770&262589&934869&-75.805629&-75.808019&1032&595&1627\\
		&0.2 &197617&721564&340880&1225120&-75.805987&-75.807985&1305&726&2031\\
		&0.0\tnote{c}&&&&&&-75.807764(18)&&&\\
		&HCIPT2\tnote{g}&&&&&&-75.80790(3)&&&\\
    \bottomrule
	\end{tabular}
\begin{tablenotes}
\item[a]Number of CSFs after selection, among which about 0.5\% have coefficients slightly smaller in absolute values than $C_{\mathrm{min}}$ (which is due to a final diagonalization).
\item[b]Estimated number of determinants according to the expression $\sum_I\frac{\tilde{N}_{\mathrm{csf}}^I}{N_{\mathrm{csf}}^I}N_{\mathrm{det}}^I$, with $N_{\mathrm{det}}^I$, $N_{\mathrm{csf}}^I$ and $\tilde{N}_{\mathrm{csf}}^I$ being the numbers of determinants, CSFs and selected CSFs of oCFG $I$, respectively.
\item[c]Extrapolated value by linear fit of the $E_{\mathrm{total}}$ vs. $|E_c^{(2)}|$ plot.
\item[d]Ref.\cite{DMRG2015}.
\item[e]Ref.\cite{ASCI2018PT2}.
\item[f]Ref.\cite{HBCI2017c}.
\item[g]Ref.\cite{HBCI2017b}.
\end{tablenotes}
    \end{threeparttable}\label{C2energies}
\end{table}

\begin{table}[!htp]
	\caption{iCIPT2 adiabatic excitation energies of \ce{C2} with
$C_{\mathrm{min}}$ being $0.3\times 10^{-4}$ and $0.5\times 10^{-4}$ for cc-pVQZ and cc-pV5Z, respectively. FZ: frozen-core; AE: all-electron;
MAX: maximum absolute error; MAE: mean absolute error. }
\begin{threeparttable}
	\centering
	\begin{tabular}{clcccccccccccccccccccccc}\toprule
		&&\multicolumn{6}{c}{adiabatic excitation energy (eV)}\\\cline{3-8}
		\multicolumn{1}{c}{state}
		&\multicolumn{1}{c}{$R_{\text{eq}}$ (\AA)\tnote{a}}
		&\multicolumn{1}{c}{$\Delta E_{\text{FC,QZ}}$}
		&\multicolumn{1}{c}{$\Delta E_{\text{AE,QZ}}$}
		&\multicolumn{1}{c}{$\Delta E_{\text{FC,5Z}}$}
		&\multicolumn{1}{c}{$\Delta E_{\text{AE,5Z}}$}
		&\multicolumn{1}{c}{$\Delta E_{\text{FC,5Z}}$\tnote{b}}
		&\multicolumn{1}{c}{Experiment\tnote{a}}\\\midrule
		$X^1\Sigma_g^+$        &1.24253&0.00&0.00&0.00&0.00&0.00&0.00\\
		$a^3\Pi_u$             &1.312  &0.07&0.09&0.07&0.09&0.07&0.09\\
		$b^3\Sigma_g^-$        &1.369  &0.77&0.81&0.77&0.81&0.78&0.80\\
		$A^1\Pi_u$             &1.318  &1.03&1.05&1.03&1.05&1.03&1.04\\
		$c^3\Sigma_u^+$        &1.208  &1.17&1.14&1.16&1.12&1.16&1.13\\
		$B^1\Delta_g$          &1.385  &1.48&1.52&1.48&1.52&1.49&1.50\\
		$B^{\prime1}\Sigma_g^+$&1.377  &1.89&1.93&1.90&1.93&1.90&1.91\\
		$d^3\Pi_g$             &1.266  &2.51&2.49&2.50&2.47&2.50&2.48\\
		$C^1\Pi_g$             &1.255  &4.30&4.27&4.28&4.25&4.29&4.25\\\midrule
        \multicolumn{2}{c}{MAX}&0.05&0.02&0.03&0.02&0.04&\\
        \multicolumn{2}{c}{MAE}&0.03&0.01&0.02&0.01&0.02&\\
        \bottomrule
	\end{tabular}
\begin{tablenotes}
\item[a]Ref.\cite{CarbonExp}.
\item[b]HCIPT2\cite{HBCI2017b}.
\end{tablenotes}
\end{threeparttable}\label{C2Excitation}
\end{table}


\newpage

\subsection{\ce{O2}}
\begin{figure}[!htp]
	\centering
	\includegraphics[width=\textwidth]{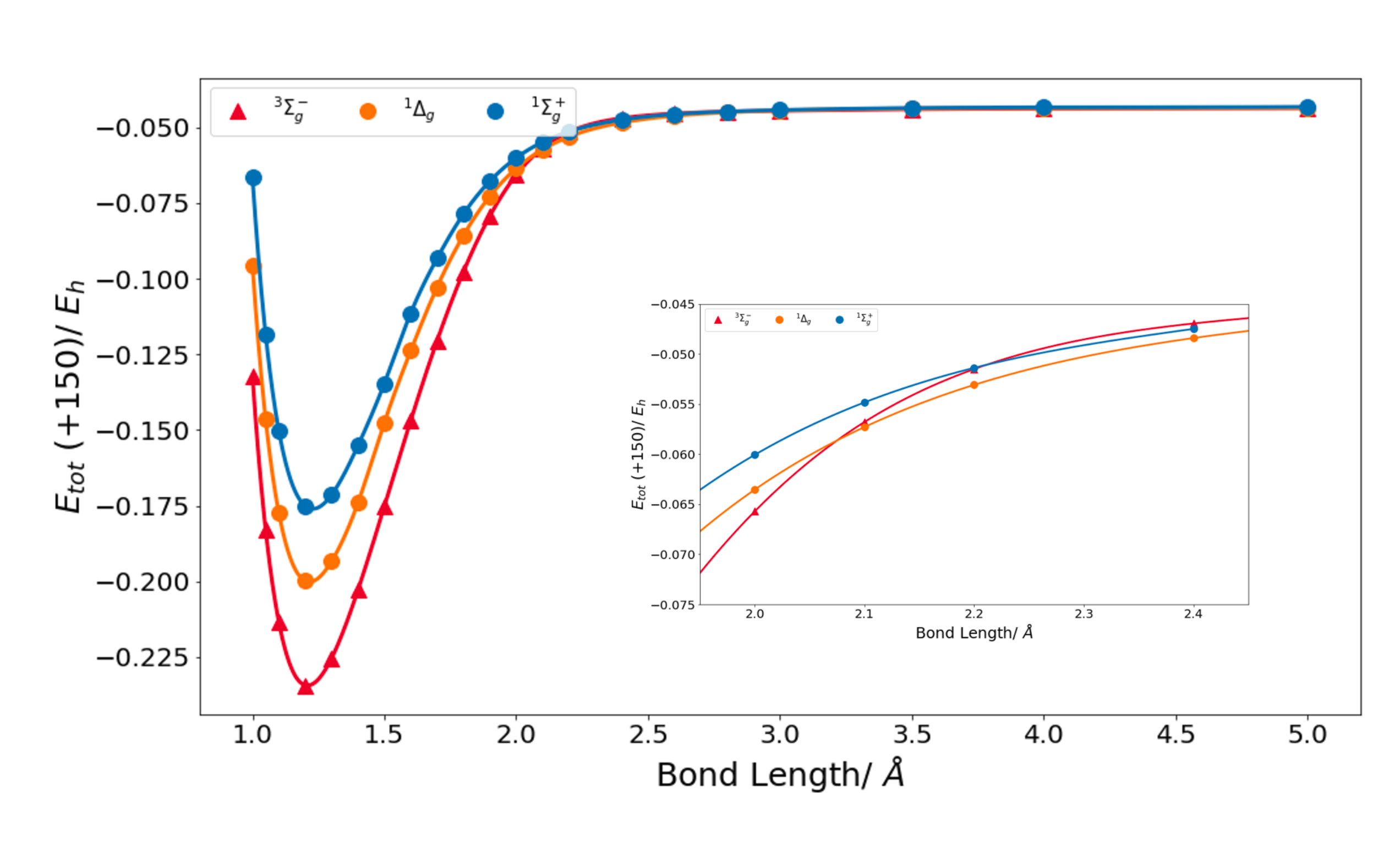}
	\caption{Potential energy curves of the $^3\Sigma_g^-$, $^1\Delta_g$ and $^1\Sigma_g^+$ states of oxygen by iCIPT2/cc-pVQZ. }\label{O2Curve}
\end{figure}

One particularly good feature of the CSF-based iCIPT2 lies in that it can describe the ground and excited states of arbitrary open-shell systems. To show this, we calculate the potential energy curves of
the lowest three states (i.e., $^3\Sigma_g^-$, $^1\Delta_g$ and $^1\Sigma_g^+$) of \ce{O2}, which are most relevant to chemical, biological and photocatalytic
reactions. The potential energy curves (see Fig. \ref{O2Curve}) are obtained by cubic spline interpolations of the pointwise energies (cf. Supporting Information) calculated by iCIPT2/cc-pVQZ with $C_{\text{min}}=0.5\times 10^{-4}$.
It can be seen from the inset of Fig. \ref{O2Curve} that the $^3\Sigma_g^-$ state is crossed by $^1\Delta_g$ and $^1\Sigma_g^+$ around 2.07 and 2.21 \AA, respectively. Some spectroscopic constants of \ce{O2} are presented in Table \ref{ConstantOxygen}.
The equilibrium bond length, harmonic vibrational frequency and dissociation energy (after correcting for zero-point energy) of the ground state $^3\Sigma_g^-$ are 1.2085 \text{\AA}, 1573.9 cm$^{-1}$ and 5.0936 eV,
respectively, which are very close to the corresponding experimental values (1.2075 \text{\AA}, 1580 cm$^{-1}$ and 5.080 eV)\cite{Herzberg1950Molecular}.
The adiabatic excitation energies of $^1\Delta_g$ and $^1\Sigma_g^+$ are
0.93 and 1.59 eV, respectively, which are again very close to the corresponding experimental values (0.95 and 1.61 eV)\cite{Atkins2009Shriver}.

\begin{table}[!htp]
    \centering
    \small
    \caption{Spectroscopic constants of the $^3\Sigma_g^-$, $^1\Delta_g$ and $^1\Sigma_g^+$ states of \ce{O2}. $\omega_e$: harmonic vibrational frequency;
    ZPE: zero-point energy; $D_e$: dissociation energy after correcting ZPE; $\Delta E_{\mathrm{v}}$/$\Delta E_{\mathrm{a}}$: vertical/adiabatic excitation energy. }
    \begin{tabular}{cccccccccc}\toprule
    state
    &$R_e/$\AA
    &$\omega_e/\mathrm{cm^{-1}}$
    &$\omega_eX_e/\mathrm{cm^{-1}}$
    &$\omega_eY_e/\mathrm{cm^{-1}}$
    &\multicolumn{1}{c}{$\omega_eZ_e/\mathrm{cm^{-1}}$}
    &ZPE/eV
    &$D_e$/eV
    &$\Delta E_{\mathrm{v}}$/eV
    &$\Delta E_{\mathrm{a}}$/eV\\\toprule
    $^3\Sigma_g^-$&1.2085&1573.9&15.866&3.0327& 0.4847&0.0976&5.0936&0.00&0.00\\
    $^1\Delta_g$  &1.2164&1503.3&21.256&1.4019&-0.0609&0.0932&4.1616&0.93&0.93\\
    $^1\Sigma_g^+$&1.2291&1429.4&19.870&2.5417& 0.3715&0.0886&3.5054&1.60&1.59\\\bottomrule
    \end{tabular}\label{ConstantOxygen}
\end{table}


\subsection{\ce{Cr2}}
\begin{figure}[!htp]
	\centering
	\includegraphics[width=\textwidth]{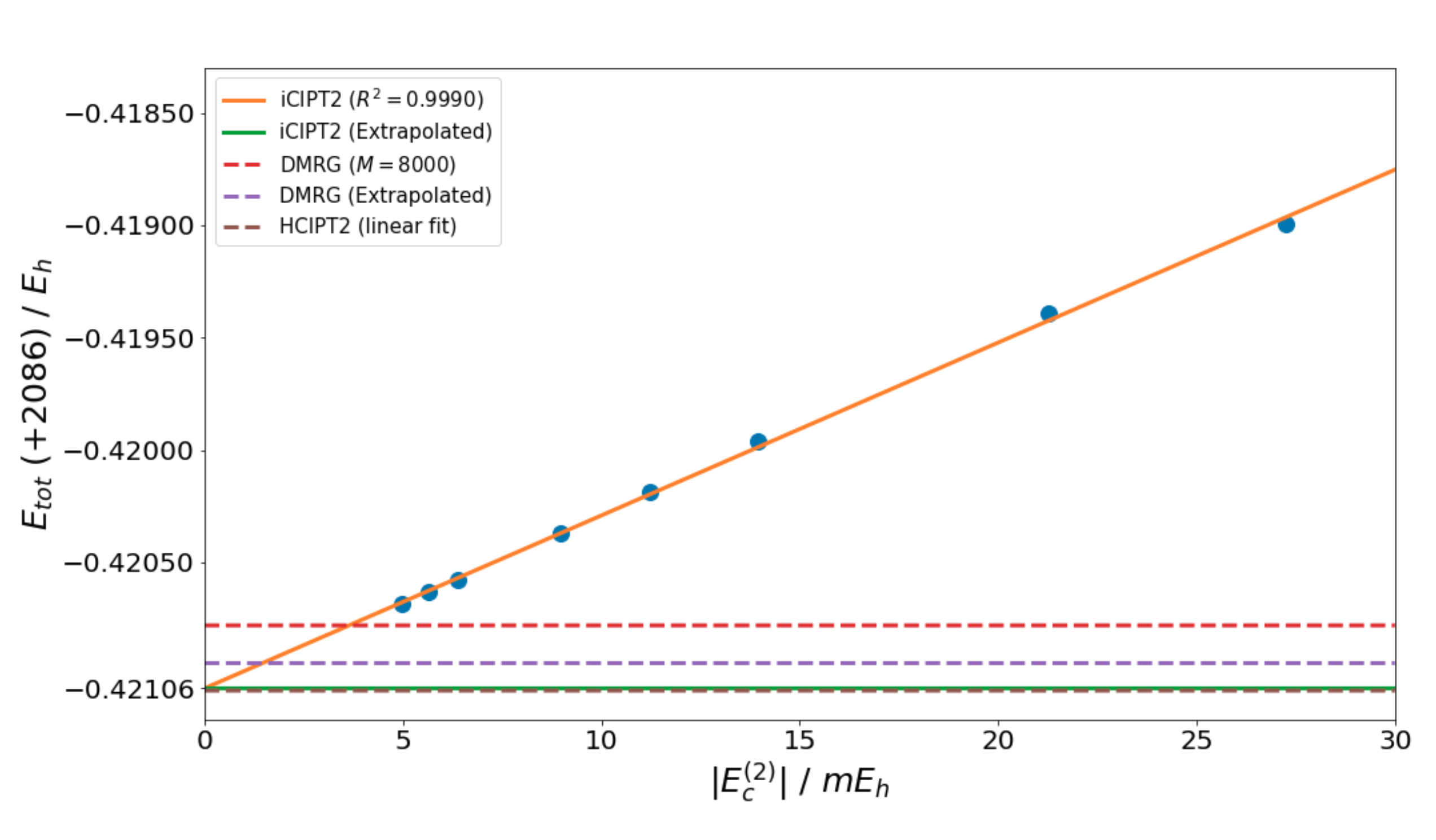}
	\caption{Comparison of the Hartree-Fock core ground state energies of \ce{Cr2} at $R=1.5$ \AA~by different methods with the Ahlrichs SV basis.
		Blue dots refer to different $C_{\mathrm{min}}$ values in iCIPT2.
		The extrapolated values are -2086.421056(35), -2086.420948, -2086.42107 $\mathrm{E_h}$ for iCIPT2, DMRG\cite{DMRG2015} and HCIPT2\cite{HBCI-Cr2}, respectively.}\label{Cr24}
\end{figure}

\begin{figure}[!htp]
	\centering
	\includegraphics[width=\textwidth]{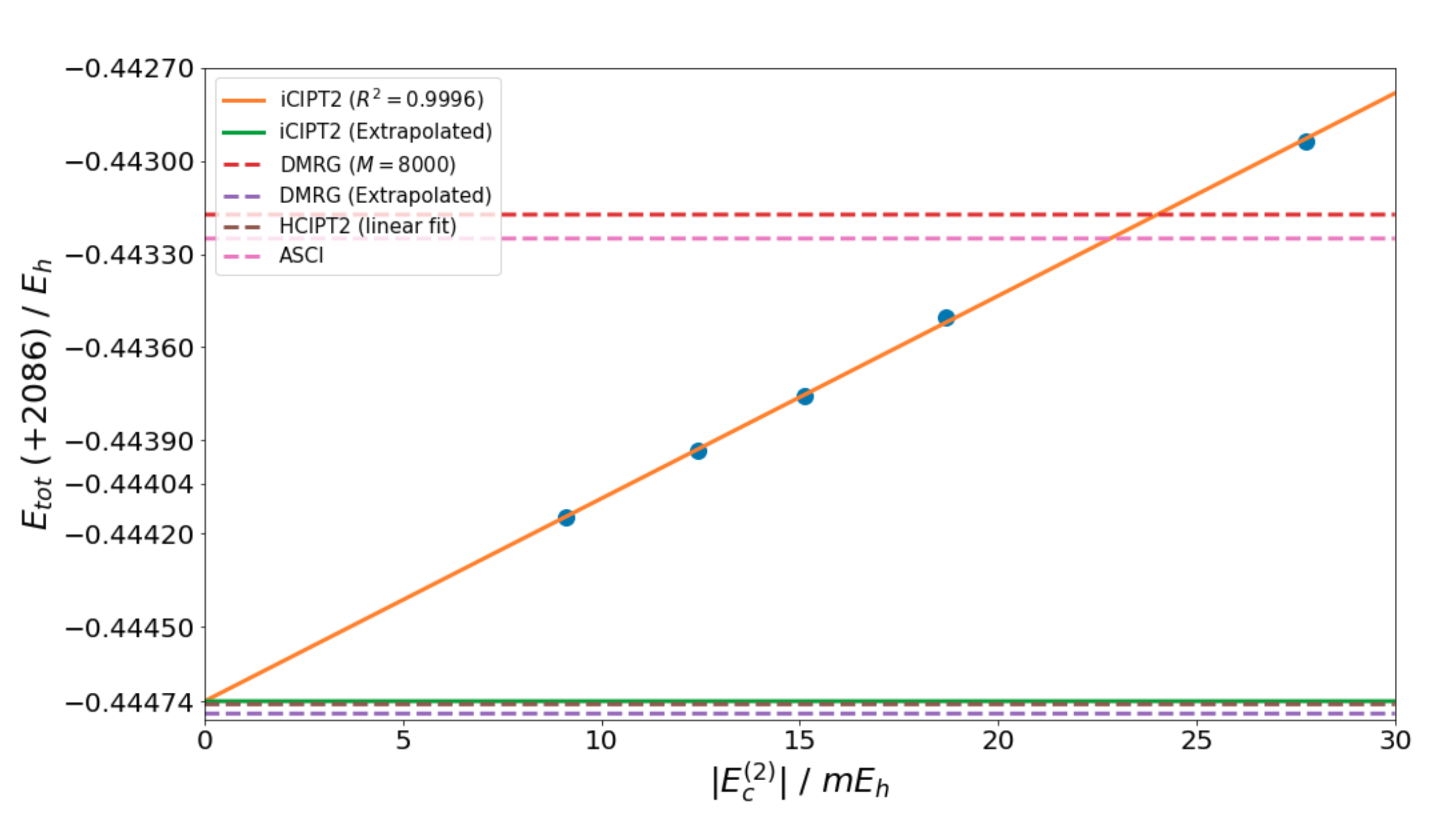}
	\caption{Comparison of the all-electron ground state energies of \ce{Cr2} at $R=1.5$ \AA~by different methods with the Ahlrichs SV basis.
Blue dots refer to different $C_{\mathrm{min}}$ values in iCIPT2.
The extrapolated values are -2086.444740(41), -2086.44478(32), -2086.44475 and -2086.44325 $\mathrm{E_h}$ for iCIPT2, DMRG\cite{DMRG2015}, HCIPT2\cite{HBCI-Cr2} and ASCI\cite{ASCI2016} respectively. }\label{Cr48}
\end{figure}

Compared with \ce{C2} and \ce{O2}, \ce{Cr2} is a more stringent test system for quantum chemical methods due to strong entanglement between the static and dynamic components of correlation. The frozen-core (24e, 30o) and all-electron (48e, 42o) FCI spaces consist of $1.17\times10^{14}$ ($9.35\times10^{14}$) and $1.42\times10^{21}$ ($1.56\times10^{22}$) CSFs (determinants), respectively. In order to compare directly with the previous results\cite{FCIQMC2014,DMRG2015,ASCI2016,HBCI-Cr2},
the ground state energy of \ce{Cr2} at an interatomic distance of 1.5 \AA~ is calculated with the Ahlrichs SV basis\cite{AhlrichsSV}. 
Noticing that the calculated energy may be dependent on how the core orbitals are defined, both the Hartree-Fock and CAS(12e, 12o) cores
are considered in the frozen-core calculations. The remaining orbitals
are natural orbitals that are optimized during the selection procedure (see again Algorithm \ref{AlgorithmNO}).
The results are presented in Table \ref{Cr2FC}. For a better illustration, the results 
by the Hartree-Fock core and all-electron calculations are further plotted in Figs. \ref{Cr24} and \ref{Cr48}, respectively. 
Apart from the excellent agreement between the extrapolated values by all the methods quoted here, only the following
points need to be pointed out: (1) The all-electron iCIPT2 calculations with $C_{\mathrm{min}}$ smaller than $0.3\times 10^{-4}$ cannot be performed due to shortage of memory space. Nevertheless, the calculated data is enough for a high quality linear fit ($R^2=0.9996$), yielding an extrapolated energy  (-2086.44474 $\mathrm{E_h}$) that almost coincides with the extrapolated DMRG (-2086.44478 $\mathrm{E_h}$)\cite{DMRG2015} and HBCIPT2 (-2086.44475 $\mathrm{E_h}$)\cite{HBCI-Cr2} values; (2) It is a bit surprising that the all-electron DMRG (m = 8000) energy matches the iCIPT2 one with $C_{\mathrm{min}}$ around $1.5\times 10^{-4}$, instead of some value below $0.2\times 10^{-4}$ (cf. the Hartree-Fock core calculations); (3)
Although the wall times are reported here, they cannot be compared with those by the closely related ASCIPT2\cite{ASCI2018} and HBCIPT\cite{HBCI-Cr2} methods,
simply because the parallel calculations were performed with different numbers of CPU cores of different computers.

\begin{figure}[!htp]
	\centering
	\includegraphics[width=\textwidth]{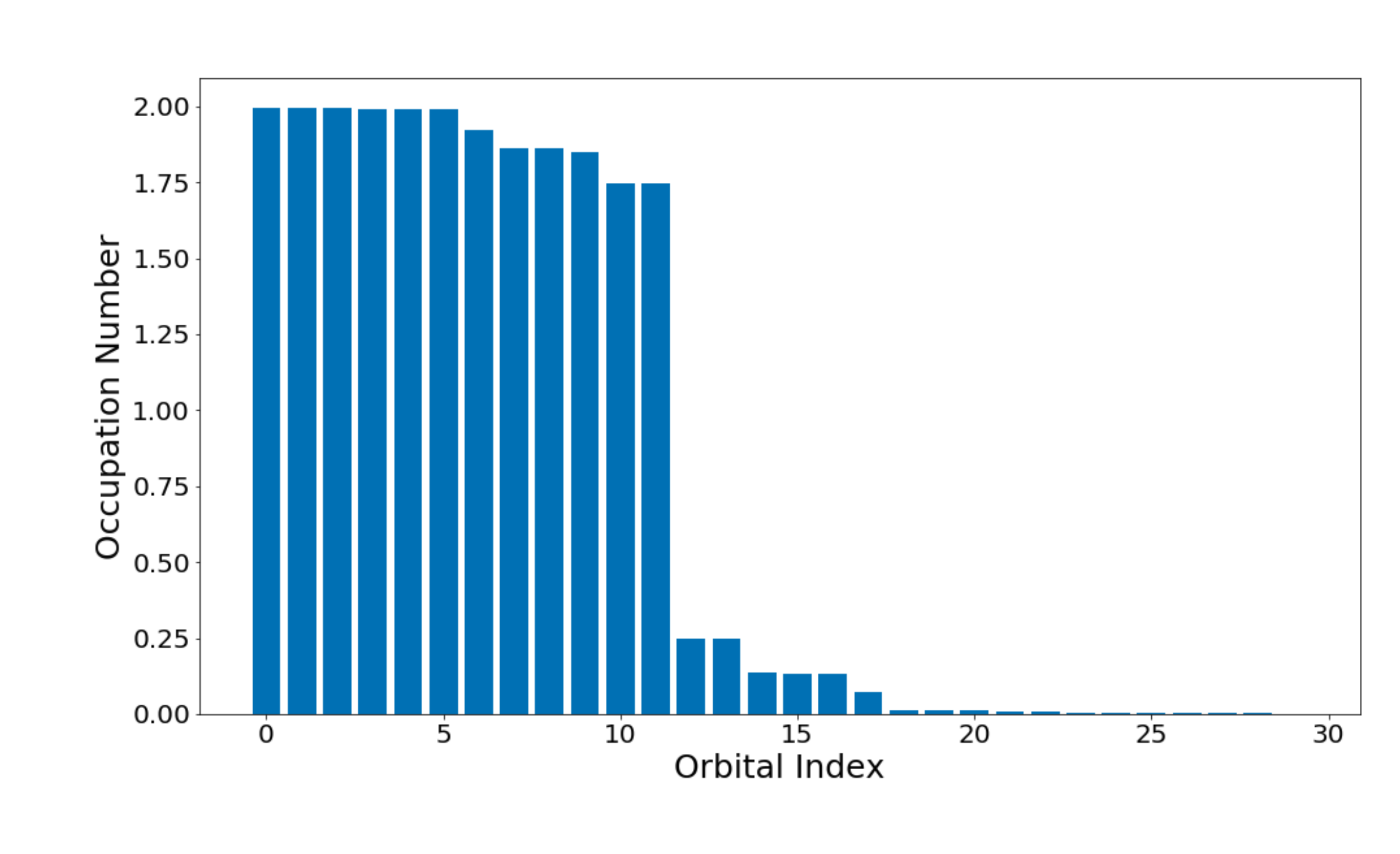}
	\caption{Occupation numbers of natural orbitals ($C_{\text{min}}=0.2\times10^{-4}$) for Cr$_2$ at $R = 1.5$ \AA~[Hartree Fock core, (24e, 30o)].}
    \label{CrOcc24}
\end{figure}

We further analyze the variational wave functions. As can be seen from Tables \ref{Cr2FZRank} and \ref{Cr2FZSeniority}, the highest excitation rank (relative to the leading, closed-shell oCFG) and the highest seniority number of the oCFGs in the variational space amount to 10 and 12, respectively, in the frozen-core calculations. It can also be seen from Fig. \ref{CrOcc24} that essentially all the NOs are sampled by the selection procedure. Although not documented here, the same hold also for the all-electron calculations.
All these establish that \ce{Cr2} is beyond the capability of single-reference methods.


\begin{table}[!htp]
	\tiny
	\caption{iCIPT2/SV calculations of Cr$_2$ at $R = 1.5$ \AA. For additional explanations see Table \ref{C2energies}.}
	\begin{threeparttable}
		\centering
		\begin{tabular}{ccrrrrllrrr}\toprule
			&
			&
			&
			&
			&\multicolumn{2}{c}{energy ($\mathrm{E_h}$)}
			&\multicolumn{3}{c}{wall time (sec)}\\\cline{6-7}\cline{8-10}
			\multicolumn{1}{c}{$\text{C}_{\text{min}}/10^{-4}$}
			&\multicolumn{1}{c}{$N_{\text{cfg}}$}
			&\multicolumn{1}{c}{$N_{\text{csf}}$}
			&\multicolumn{1}{c}{$\tilde{N}_{\text{csf}}$}
			&\multicolumn{1}{c}{$\tilde{N}_{\text{det}}$}
			&\multicolumn{1}{c}{var}
			&\multicolumn{1}{c}{total}
			&\multicolumn{1}{c}{var}
			&\multicolumn{1}{c}{PT2}
			&\multicolumn{1}{c}{total}\\\toprule
            \multicolumn{10}{c}{Hartree-Fock core, (24e, 30o)} \\
			3.0 &19442 &98889  &32416  &121239 &-2086.391741&-2086.418993&66  &16  &82\\
			2.0 &31763 &176002 &56882  &219688 &-2086.398124&-2086.419391&99  &28  &127\\
			1.0 &72982 &486500 &143981 &583862 &-2086.406023&-2086.419964&243 &65  &308\\
			0.7 &109678&791934 &227610 &944209 &-2086.408963&-2086.420184&312 &100 &412\\
			0.5 &164704&1282820&359831 &1524930&-2086.411391&-2086.420369&534 &171 &705\\
			0.3 &299454&2629089&703541 &3072786&-2086.414197&-2086.420576&961 &496 &1457\\
			0.25&371511&3422901&892099 &3935360&-2086.414997&-2086.420631&1236&723 &1959\\
			0.2 &457771&4452704&1147403&5131066&-2086.415723&-2086.420683&1452&1026&2478\\
			0.0\tnote{a} &&    &       &       &            &-2086.421056(35)&&&\\
		    DMRG (m=8000)\tnote{b}&      &       &      &       &            &-2086.420780&    &   &\\
	        DMRG (m=$\infty$)\tnote{b}&      &       &      &       &            &-2086.420948&    &   &\\
			HCIPT2\tnote{c} &&&&       &         &-2086.420934 (5) &&&\\
			HCIPT2\tnote{d} &&&&       &         &-2086.42107   &&&\\
			FCIQMC\tnote{e} &&&&       &         &-2086.4212(3)   &&&\\ \\
           \multicolumn{10}{c}{CAS(12e, 12o) core, (24e, 30)} \\
		  3.0&19474&99466&32333&120826&-2086.392182&-2086.419290&52&16&68\\
		  2.0&31713&176201&56766&219417&-2086.398555&-2086.419751&77&28&105\\
		  1.0&72811&486435&143675&582778&-2086.406445&-2086.420352&170&67&237\\
		  0.7&112885&822216&232969&966906&-2086.409557&-2086.420565&293&109&402\\
		  0.5&163864&1276103&358252&1518671&-2086.411808&-2086.420755&435&175&610\\
		  0.3&298404&2618279&701672&3064718&-2086.414594&-2086.420960&868&504&1372\\
		  0.25&370538&3412674&890956&3930293&-2086.415389&-2086.421014&1122&746&1868\\
		  0.2&458347&4477991&1148115&5133858&-2086.416130&-2086.421076&1262&1020&2282\\
          0.0\tnote{a} &      &       &       &        &            &-2086.421470(16)&&&\\
          HCIPT2\tnote{c} &&&&       &         &-2086.421385(5) &&&\\
          HCIPT2\tnote{d} &&&&       &         &-2086.42152     &&& \\ \\
                     \multicolumn{10}{c}{All-electron, (48e, 42o)} \\
		  2.0&34765&187681&60650&232203&-2086.415171&-2086.442935&205&125&330\\
		  1.0&81270&522899&156280&628276&-2086.424817&-2086.443505&433&394&827\\
		  0.7&128132&900081&256897&1055972&-2086.428629&-2086.443755&761&662&1423\\
		  0.5&188398&1420214&399815&1679269&-2086.431489&-2086.443933&1871&1115&2663\\
		  0.3&349927&2972292&796412&3447559&-2086.435048&-2086.444149&2121&2309&4430\\
          0.0\tnote{a} &      &       &       &        &            &-2086.444740(41)&&&\\
          DMRG (m=8000)\tnote{b}&      &       &      &       &            &-2086.443173&    &    &\\
		  DMRG (m=$\infty$)\tnote{b}&      &       &      &       &            &-2086.44478(32)&    &    &\\
          HCIPT2\tnote{c}&&&& & &-2086.444586(10)&&&\\
          HCIPT2\tnote{d}&&&& & &-2086.44475&&&\\
          ASCI\tnote{f}  &&&& & &-2086.44325&&&\\
          
			\bottomrule
		\end{tabular}
		\begin{tablenotes}
			\item[a]Extrapolated value by linear fit of the $E_{\mathrm{total}}$ vs. $|E_c^{(2)}|$ plot ($R^2>0.999$).
			\item[b]Ref.\cite{DMRG2015}.
			\item[c]Ref.\cite{HBCI-Cr2} (weighted quadratic fit).
			\item[d]Ref.\cite{HBCI-Cr2} (linear fit).
			\item[e]Ref.\cite{FCIQMC2014}.
            \item[f]Ref.\cite{ASCI2016}.
		\end{tablenotes}
	\end{threeparttable}\label{Cr2FC}
\end{table}


\begin{table}[!htp]
	\small
	\caption{Excitation ranks of oCFGs (relative to the leading, closed-shell oCFG) in the variational space ($C_{\text{min}}=0.2\times10^{-4}$) for \ce{Cr2} at $R = 1.5$ \AA~[Hartree Fock core, (24e, 30o)]. For additional explanations see Table \ref{C2energies}.}
	\begin{threeparttable}
		\centering
		\begin{tabular}{crrrr}\toprule
			\multicolumn{1}{c}{Excitation rank}
			&\multicolumn{1}{c}{$N_{\text{cfg}}$}
			&\multicolumn{1}{c}{$\tilde{N}_{\text{csf}}$}
			&\multicolumn{1}{c}{Contribution}
			&\multicolumn{1}{c}{}
			\\\toprule
			0   &1     &1     &59.6262\%&59.6262\%\\
			1   &20    &20    & 0.0720\%&59.6982\%\\
			2   &1663  &2709  &29.3529\%&89.0511\%\\
			3   &24218 &52509 & 0.8260\%&89.8771\%\\
			4   &115938&291396& 7.6426\%&97.5197\%\\
			5   &93636 &236271& 0.5409\%&98.0606\%\\
			6   &120545&307891& 1.4567\%&99.5173\%\\
			7   &48520 &128955& 0.1965\%&99.7138\%\\
			8   &35856 &87461 & 0.2197\%&99.9335\%\\
			9   &11539 &28595 & 0.0407\%&99.9742\%\\
			>=10&5835  &11595 & 0.0258\%&100.0000\%\\\bottomrule
		\end{tabular}
	\end{threeparttable}\label{Cr2FZRank}
\end{table}

\begin{table}[!htp]
	\centering
	\small
	\caption{Seniority numbers of oCFGs in the variational space ($C_{\text{min}}=0.2\times10^{-4}$) for \ce{Cr2} at $R = 1.5$ \AA~[Hartree Fock core, (24e, 30o)]. For additional explanations see Table \ref{C2energies}.}
	\begin{tabular}{crrrrr}\toprule
		\multicolumn{1}{c}{Seniority}
		&\multicolumn{1}{c}{$N_{\text{cfg}}$}
		&\multicolumn{1}{c}{$N_{\text{csf}}$}
		&\multicolumn{1}{c}{$\tilde{N}_{\text{csf}}$}
		&\multicolumn{1}{c}{Contribution}
		&\multicolumn{1}{c}{}
		\\\toprule
		0 &10146 &10146  &10146 &72.5872\%&72.5872\%\\
		2 &19669 &19669  &19669 & 1.1228\%&73.7100\%\\
		4 &131839&263678 &190184&23.2162\%&96.9262\%\\
		6 &142933&714665 &318859& 0.9821\%&97.9083\%\\
		8 &123231&1725234&448527& 2.0114\%&99.9197\%\\
		10&24986 &1049412&127341& 0.0621\%&99.9818\%\\
		12&4919  &649308 &32503 & 0.0182\%&100.0000\%\\
		14&48    &20592  &174   & 0.0000\%&100.0000\%\\\bottomrule
	\end{tabular}\label{Cr2FZSeniority}
\end{table}	

\subsection{\ce{C6H6}}
\begin{figure}[!htp]
	\centering
	\includegraphics[width=\textwidth]{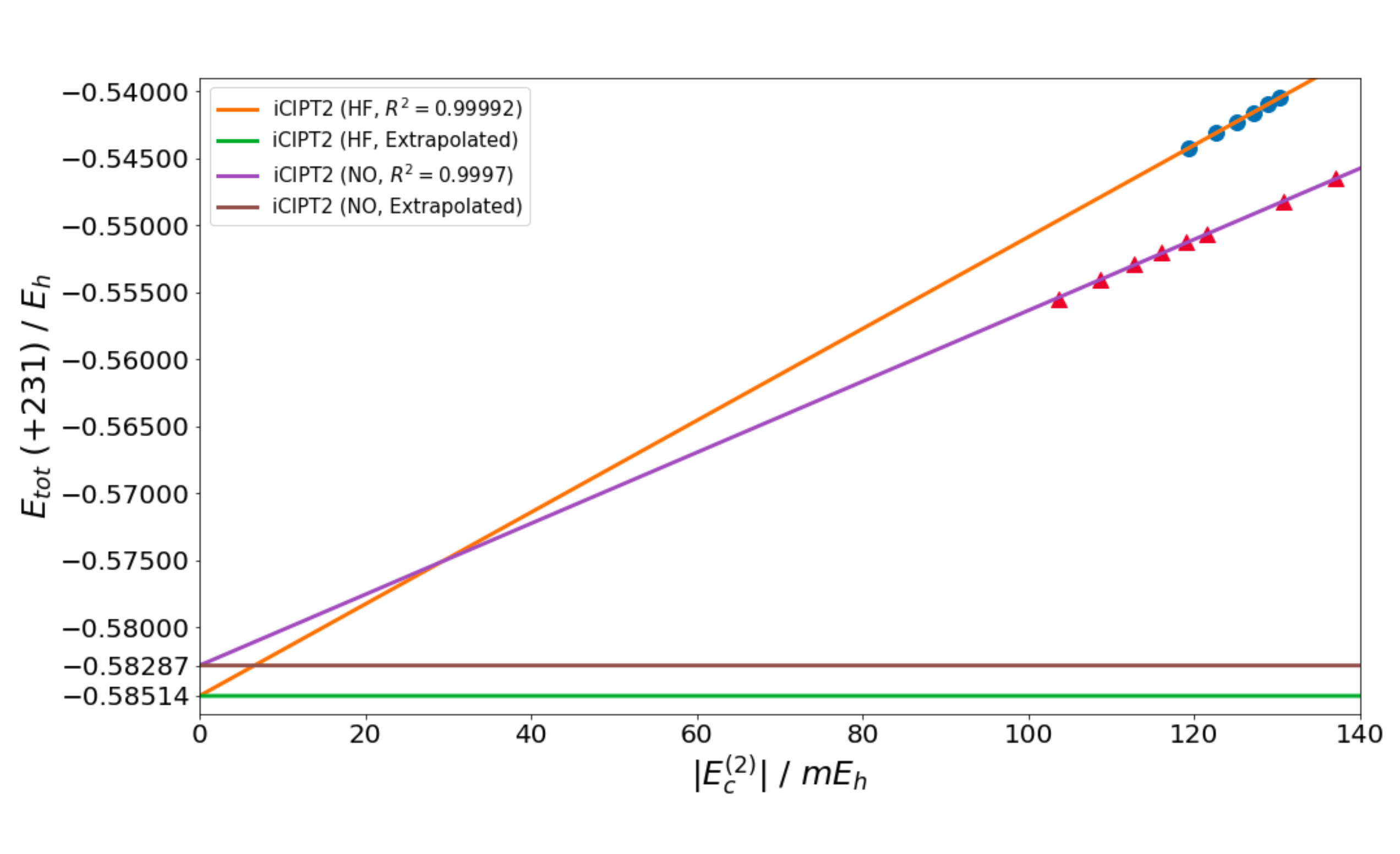}
	\caption{Extrapolation of the correlation energy of benzene.}\label{BenzeneEn}
\end{figure}

As a final example, the frozen-core ground state energy of benzene at equilibrium\cite{BenzeneGeom} is calculated with iCIPT2/cc-pVDZ.
Although benzene is not really a strongly correlated system, the (30e, 108o) space consists of $6.95\times10^{33}$ CSFs or $9.5\times10^{34}$ determinants
and therefore represents a great challenge if one wants to obtain nearly FCI result. The calculations employ $D_{2h}$ symmetry.
The results with HF and natural orbitals are presented in Table \ref{BenHFNO}. A linear extrapolation of the calculated correlation energies is further plotted in Fig. \ref{BenzeneEn}. Several points deserve to be mentioned here:
\begin{enumerate}[(1)]
\item The largest calculation (with $C_{\mathrm{min}}=5\times 10^{-5}$) involves $1.00\times 10^6$  ($1.38\times 10^6$) CSFs in the variational
space and $8.69\times10^{11}$ ($1.38\times 10^{12}$) CSFs in the first-order interacting space and took 1215.75 (2225.85) minutes of wall time
on a single node with two 2.60 GHz Intel Xeon E5-2640 v3 processors, when HF (natural) orbitals are used.
This gives rise to -822.4 (-833.3) $\mathrm{mE_h}$ for the overall correlation energy, which is 95.3\% (96.8\%) of
the extrapolated value $-863.3$ (-861.1 $\mathrm{mE_h}$).
\item The extrapolated correlation energies using natural and HF orbitals differ by only 2.2 $\mathrm{mE_h}$, indicating that the present iCIPT2 calculations
are essentially converged.
\item The atomization energies of benzene are 55.307 and 55.245 eV for the HF and natural orbital based calculations, respectively.
\item It is very much surprising that the majority of the CSFs in the variational space are open shells, e.g.,
$1.6 \times 10^4$ closed-shell CSFs, $2.7\times 10^4$ CSFs with 2 unpaired electrons,
$4.1\times 10^5$ CSFs with 4 unpaired electrons, $3.6\times 10^5$ CSFs with 6 unpaired electrons, and $5.7\times 10^5$ CSFs with 8 unpaired electrons in the case of $C_{\mathrm{min}}=5\times10^{-5}$ with natural orbitals.

\end{enumerate}

\begin{table}[!htp]
	\small
	\caption{Calculated correlation energies $E_c$ ($=E_c^{\mathrm{var}}+E_c^{(2)}$) for benzene at equilibrium (D$_{2h}$ symmetry; cc-pVDZ; (30e, 108o); $E_{\mathrm{HF}}=-230.721819$ $\mathrm{E_h}$). }
\begin{threeparttable}
	\centering

	\begin{tabular}{cclrclr}\toprule
    $C_{\text{min}}$
    &\multicolumn{3}{c}{HF orbitals}
    &\multicolumn{3}{c}{Natural orbitals}\\\cline{2-7}
    &\multicolumn{1}{c}{$E_c^{\mathrm{var}}/\mathrm{mE_h}$}
    &\multicolumn{1}{c}{$E_c/\mathrm{mE_h}$}
    &\multicolumn{1}{c}{$T$$/s$\tnote{a}}
    &\multicolumn{1}{c}{$E_c^{\mathrm{var}}/\mathrm{mE_h}$}
    &\multicolumn{1}{c}{$E_c/\mathrm{mE_h}$}
    &\multicolumn{1}{c}{$T$$/s$\tnote{a}}\\\toprule
10.0 & -589.336 & -802.909 &301   & -612.591 & -816.724&379\\

 5.0 & -649.922 & -810.127 &837   & -656.729 & -819.366&669\\

 3.0 & -667.714 & -812.884 &1309  & -676.147 & -822.384&1481\\

 2.0 & -675.959 & -814.997 &2363  & -687.627 & -824.699&2794\\

 1.5 & -681.083 & -816.564 &3036  & -695.623 & -826.403&5066\\

 1.0 & -688.358 & -818.664 &9875  & -707.243 & -828.807&15206\\

 0.9 & -690.363 & -819.194 &12712 & -710.440 & -829.460&22335\\

 0.8 & -692.695 & -819.797 &17447 & -714.110 & -830.223&38399\\

 0.7 & -695.422 & -820.487 &26334 & -718.441 & -831.135&49238\\

 0.6 & -698.767 & -821.330 &42053 & -723.616 & -832.289&77344\\

 0.5 & -703.066 & -822.427 &72945 & -729.983 & -833.690&133551\\
 0.0 &          &-863.32(54)\tnote{b}&& &-861.05(51)\tnote{c}&\\\bottomrule
	\end{tabular}
\begin{tablenotes}
\item[a] (1) CPU: 2.60 GHz Intel Xeon E5-2640 v3$\times$2, 16 cores; (2) memory: 128 Gb;
(3) parallelization: OpenMP, 16 threads.
\item[b]Extrapolated value by linear fit ($E_c=0.34257\times|E_c^{(2)}|-863.32$ with $R^2=0.99992$) of the $E_{\mathrm{c}}$ vs. $|E_{\mathrm{c}}-E_{\mathrm{c}}^{\mathrm{var}}|$ with $C_{\text{min}}=\{1.0,0.9,\cdots,0.6,0.5\}\times 10^{-4}$. 
\item[c]Extrapolated value by linear fit ($E_c=0.26498\times|E_c^{(2)}|-861.05$ with $R^2=0.9997$)  of the $E_{\mathrm{c}}$ vs. $|E_{\mathrm{c}}-E_{\mathrm{c}}^{\mathrm{var}}|$ with $C_{\text{min}}=\{3.0,2.0,\cdots,0.6,0.5\}\times 10^{-4}$.
\end{tablenotes}
\end{threeparttable}\label{BenHFNO}
\end{table}	

\section{Conclusions and outlook}\label{Conclusion}
A very efficient and robust approach, iCIPT2, has been developed for systems of strongly correlated electrons. It combines well-defined
algorithms for the selection of important oCFGs/CSFs over the whole Hilbert space with a very efficient implementation of Epstein-Nesbet second-order perturbation theory. Effectively only one parameter, the coefficient pruning-threshold $C_{\mathrm{min}}$, is invoked to control the size of the variational space that has no CSFs with coefficients smaller in absolute value than $C_{\mathrm{min}}$. Another salient feature of iCIPT2 lies in
that the external space is split into disjoint subsets, each of which is accessed only incrementally during the selection procedure. As has been demonstrated, iCIPT2 can describe very well not only the ground states but also the excited states of closed- and open-shell systems.
In line with the previous findings\cite{HBCI2017b}, once the total-PT2 correlation energy plot has reached a linear domain, the extrapolated total
correlation energy is very close to the FCI value. Nonetheless, one should be aware that,
even in the linear regime, the slope of the total-PT2 correlation energy plot depends, albeit weakly, on what MRPT2
and what orbitals are used (cf. Fig. \ref{BenzeneEn}). Such dependence may result in an uncertainty of a few millihartrees. While such accuracy
is sufficient for most applications, one may consider to go beyond MRPT2 by formulating, e.g., a size-consistent incomplete model space LCC\cite{IMSCPL1986,GMSPT21986,IMSHeff1989}.

\section*{Supporting Information}

The Supporting Information is available free of charge, including the proof of inequality \eqref{Approx} as well as the
pointwise energies of \ce{C2} and \ce{O2}.

\begin{acknowledgement}
The research of this work was supported by NSFC (Grant Nos. 21033001 and 21973054)
and the North Dakota University System.
\end{acknowledgement}

\newpage

\appendix

\section{Data structure of the Hamiltonian matrix}\label{SparseH}

\subsection{Diagonal blocks}
Both the first-order coefficient \eqref{iCIrank} and second-order energy \eqref{ENPT2Q} require the
diagonal matrix elements $H^{II}_{\mu\mu}$ $(\mu\in[0, N_{S,S}^I-1])$, the evaluation of which is not cheap due to summations over both orbitals and orbital pairs. However,
the calculation can be simplified by introducing a common reference oCFG with occupation numbers $\{\omega_k\}$ for the spatial orbitals.
As shown by Eq. \eqref{H-diag-3}, apart from the first, constant term in the curly brackets, the remaining two terms therein depend only on
the differential occupations $\{\Delta_i^I\}$, whereas both $i$ and $j$ in
$\langle I\mu|e^1_{ij,ji}|I\mu\rangle$ are singly occupied. As discussed in Sec. \ref{ReuseCoupling}, after introducing the reduced spatial
orbital indices $\{\bar{i}, \bar{j}, \bar{k}, \bar{l}\}$, oCFGs with the same number of singly occupied orbitals share the same BCCs $\langle I\mu|e^1_{\bar{i}\bar{j},\bar{j}\bar{i}}|I\mu\rangle$.
Moreover, the off-diagonal elements $H^{II}_{\mu\nu}$ within a diagonal block also depend only on the BCCs $\langle I\mu|e^1_{\bar{i}\bar{j},\bar{j}\bar{i}}|I\nu\rangle$.

To store $\langle I\mu|e^1_{\bar{i}\bar{j},\bar{j}\bar{i}}|I\mu\rangle$, we define the compound index $(ij)=\frac{\bar{j}(\bar{j}-1)}{2}+\bar{i}$ and a matrix $\textcolor{blue}{Ccf\_Dg}$ with elements
$\textcolor{blue}{Ccf\_Dg}[\mu][(ij)]=\langle I\mu|e^1_{\bar{i}\bar{j},\bar{j}\bar{i}}|I\mu\rangle$. The integrals in the summation $\sum_{i<j}(ij|ji)\langle I\mu|e^1_{\bar{i}\bar{j},\bar{j}\bar{i}}|I\mu\rangle$
are stored in an array $\textcolor{blue}{Int}$ with elements $\textcolor{blue}{Int}[(ij)]=(ij|ji)$. This way, the summation over $i$ and $j$ can be viewed as the inner product of vectors $\textcolor{blue}{Ccf\_Dg}[\mu][*]$
and $\textcolor{blue}{Int}[*]$. The calculation of all the diagonal elements within a diagonal block can then be viewed as the matrix-vector product of matrix $\textcolor{blue}{Ccf\_Dg}[*][*]$ and
vector $\textcolor{blue}{Int}[*]$, as shown in Algorithm \ref{AlgorithmDiagonal}.

\begin{algorithm}
    \caption{Calculation of $H_{\mu\mu}^{II}$. $N_{S,S}^I$ is the number of CSFs with spin $S$.}
    \label{AlgorithmDiagonal}
        (Precalculation) Given the reference oCFG $\Omega$ with occupation numbers $\{\omega_k\}$, calculate $C=\frac{1}{2}\sum_i\omega_i[h_{ii}\varepsilon_i
        +(\omega_i-2)g_{ii}]$, $B_i=\varepsilon_i+(\omega_i-1)g_{ii}$ and $g_{ij}$, with $\varepsilon_i$ and $g_{ij}$ defined in Eqs. \eqref{Inter1} and \eqref{Inter2}.

        Input oCFG $I$ and store nonzero differential occupations (relative to $\Omega$) in an array $\{(i,\Delta_i^I)\}$.

        Calculate $C'= C+\sum_i\Delta_i^IB_i+\sum_{i\leq j}\Delta_i^Ig_{ij}\Delta_j^I$.

        Get molecular integrals $\textcolor{blue}{Int}[(ij)]=(ij|ji)$

        \For{($\mu=0;\mu< N_{S,S}^I;\mu=\mu+1$)}
        {
            $H_{\mu\mu}^{II} = \sum_{ij}\textcolor{blue}{Ccf\_Dg}[\mu][(ij)]\times\textcolor{blue}{Int}[(ij)]$;

            $H_{\mu\mu}^{II} = H_{\mu\mu}^{II}+C'$.
        }
        return 0\;
\end{algorithm}

For the off-diagonal matrix elements $H^{II}_{\mu\nu}$ $(\mu\ne\nu)$, we define the compound index $(\mu\nu)=\mu\times N^I_{S,S}+\nu$ and a sparse matrix $\textcolor{blue}{Ccf\_Off}[(\mu\nu)][(ij)]=\langle I\mu|e^1_{\bar{i}\bar{j},\bar{j}\bar{i}}|I\nu\rangle$ that can be stored
in CSR format. Since every off-diagonal element $H^{II}_{\mu\nu}$ can be viewed as the inner product of the sparse vector $\textcolor{blue}{Ccf\_Off}[(\mu\nu)][*]$ and
the dense vector $\textcolor{blue}{Int}[*]$, all the off-diagonal elements together
can be viewed as the product of the sparse matrix $\textcolor{blue}{Ccf\_Off}[*][*]$ and the dense vector $\textcolor{blue}{Int}[*]$.

\subsection{Singly excited blocks}
Without loss of generality we assume oCFG $I$ is generated from oCFG $J$ by moving one electron from spatial orbital $j$ to $i$ with $j<i$. By virtue of the
reduced orbital indices, Eq. \eqref{EqnDiff1} becomes
\begin{align}
\langle I\mu|H^1_1+H^1_2|J\nu\rangle&=\langle I\mu|E_{\bar{i}\bar{j}}|J\nu\rangle\left[f_{ij}+\sum_k\Delta_k^j[(ij|kk)-\frac{1}{2}(ik|kj)]\right.\nonumber\\
&+\left.\frac{1}{2}n_i^J(ii|ij)+(\frac{1}{2}n_j^J-1)(ij|jj)\right]\nonumber\\
&+\sum_{k\in\text{exterior open}}(ik|kj)\langle I\mu|e^1_{\bar{i}\bar{k},\bar{k}\bar{j}}|J\nu\rangle\nonumber\\
&+\sum_{k\in\text{interior open}}(ik|kj)\left[\frac{1}{2}\langle I\mu|E_{\bar{i}\bar{j}}|J\nu\rangle+\langle I\mu|E_{\bar{k}\bar{j}}E_{\bar{i}\bar{k}}|J\nu\rangle\right].\label{1eDiffBlock}
\end{align}
Assuming that the oCFG pair $(I,J)$ corresponds to a ROT with length $LenROT$, Eq. \eqref{1eDiffBlock} can be converted to the following form
\begin{equation}
\langle I\mu|H^1_1+H^1_2|J\nu\rangle = \sum_{k=0}^{LenROT}\textcolor{blue}{Ccf}[(\mu\nu)][k]\times\textcolor{blue}{Int}[k],
\end{equation}
where
\begin{equation}
\begin{split}
\textcolor{blue}{Ccf}[(\mu\nu)][\bar{k}+1]=
\left\{
\begin{split}
\langle I\mu|E_{\bar{i}\bar{j}}|J\nu\rangle&,\text{ }\bar{k}=-1,\\
\langle I\mu|e^1_{\bar{i}\bar{k},\bar{k}\bar{j}}|J\nu\rangle&,\text{ }\bar{k}<\bar{j},\bar{k}<\bar{i},\\
\frac{1}{2}\langle I\mu|E_{\bar{i}\bar{j}}|J\nu\rangle+\langle I\mu|E_{\bar{k}\bar{j}}E_{\bar{i}\bar{k}}|J\nu\rangle&,\text{ }\bar{j}<\bar{k}<\bar{i},\\
0&,\text{ }\bar{k}=\bar{i},\bar{j},\\
\end{split}
\right.
\end{split}
\end{equation}
\begin{equation}
\begin{split}
\textcolor{blue}{Int}[\bar{k}+1]=
\left\{
\begin{split}
f_{ij}+\sum_k\Delta_k^j[(ij|kk)-\frac{1}{2}(ik|kj)]+\frac{1}{2}n_i^J(ii|ij)+(\frac{1}{2}n_j^J-1)(ij|jj),\text{ } \bar{k}=-1,\\
(ik|kj),\text{ } \bar{k}\in[0, LenROT-1].\\
\end{split}
\right.
\end{split}
\end{equation}


\subsection{Doubly excited blocks}
To expedite the discussion, we assume that oCFG $I$ is generated from $J$ by moving two electrons from $j$ and $l$ to $i$ and $k$, subject to
the conditions $j\le l$, $i\le k$, and $\{j, l\}\cap\{ i, k\} =\emptyset$.
Eq. \eqref{DoubleMat} is then reduced to
\begin{equation}
\langle I\mu|H^2_2|J\nu\rangle=2^{-\delta_{ik}\delta_{jl}}(ij|kl)\langle I\mu|e_{\bar{i}\bar{j},\bar{k}\bar{l}}|J\nu\rangle+(1-\delta_{ik})(1-\delta_{jl})(il|kj)\langle I\mu|e_{\bar{i}\bar{l},\bar{k}\bar{j}}|J\nu\rangle,\label{DoubleMatSparse}
\end{equation}
the first and second terms of which are direct and exchange, respectively. This matrix is very sparse and can be stored in CSR format consisting of three arrays, $\textcolor{blue}{Row}$,
$\textcolor{blue}{Col}$ and $\textcolor{blue}{Mat}$. $\textcolor{blue}{Row}$ and
$\textcolor{blue}{Col}$ are only relevant to ROT $(I,J)$ and can be calculated once for all. To construct $\textcolor{blue}{Mat}$,
the BCCs should be stored in an appropriate way. For the case of $j<l$ and $i<k$, four arrays, $\textcolor{blue}{Drct}$, $\textcolor{blue}{DrctLoc}$, $\textcolor{blue}{Xchng}$ and $\textcolor{blue}{XchngLoc}$, are needed.
$\textcolor{blue}{Drct}$ with length $nDrct$ stores the direct type of BCCs, while $\textcolor{blue}{Xchng}$ with length $nXchng$ stores the exchange type of BCCs.
The other two arrays record to which matrix element of $\textcolor{blue}{Mat}$ the BCC will contribute. For instance, if $\exists p, q$ s.t. $\textcolor{blue}{DrctLoc}[p]=\textcolor{blue}{XchngLoc}[q]=r$,
then
\begin{equation}
\textcolor{blue}{Mat}[r]=\textcolor{blue}{Drct}[p]\times(ij|kl)+\textcolor{blue}{Xchng}[q]\times(il|kj).
\end{equation}
The whole matrix elements can be calculated according to Algorithm \ref{AlgorithmDoubleD}.
As for the case of $i=k$ or $j=l$, the exchange term in Eq. \eqref{DoubleMatSparse} is absent, such that the simpler Algorithm \ref{AlgorithmDoubleE} can be used.

\begin{algorithm}
    \caption{Calculation of $\langle I\mu|H^2_2|J\nu\rangle$ \eqref{DoubleMatSparse} for $i<k$, $j< l$ and $\{j, l\}\cap\{ i, k\} =\emptyset$.}
\label{AlgorithmDoubleD}
        Read \textcolor{blue}{Row, Col, Drct, DrctLoc, Xchng, XchngLoc};

        Get molecular integrals $(ij|kl),(il|kj)$;

        \For{($r=0;r<\text{nDrct};r=r+1$)}
        {
            $\textcolor{blue}{Mat}[\textcolor{blue}{DrctLoc}[r]]+=\textcolor{blue}{Drct}[r]\times(ij|kl)$;
        }

        \For{($r=0;r<\text{nXchng};r=r+1$)}
        {
            $\textcolor{blue}{Mat}[\textcolor{blue}{XchngLoc}[r]]+=\textcolor{blue}{Xchng}[r]\times(il|kj)$;
        }

        return $[\textcolor{blue}{Row}, \textcolor{blue}{Col}, \textcolor{blue}{Mat}]$\;
\end{algorithm}

\begin{algorithm}
    \caption{Calculation of $\langle I\mu|H^2_2|J\nu\rangle$ \eqref{DoubleMatSparse} for $i=k$ or $j=l$ and $\{j, l\}\cap\{ i, k\} =\emptyset$.}
\label{AlgorithmDoubleE}
        Read \textcolor{blue}{Row, Col, Drct};

        Get molecular integrals $(ij|kl)$;

        \For{($r=0;r<\text{nDrct};r=r+1$)}
        {
            $\textcolor{blue}{Mat}[r]+=\textcolor{blue}{Drct}[r]\times(ij|kl)$;
        }

        return $[\textcolor{blue}{Row}, \textcolor{blue}{Col}, \textcolor{blue}{Mat}]$\;
\end{algorithm}

\clearpage
\newpage
\bibliography{iCI}


\end{document}

%% file: Diagram.tex

\begin{figure*}[!htp]
	\centering
	\begin{tabular}{cccccc}
		\begin{tikzpicture}[scale = 0.40,thick]
		\draw (0,0)--(0,5);
		\draw (3,0)--(3,5);
		\draw[fill] (3,2.5) circle [radius=0.1];
		\draw[fill] (0,2.5) circle [radius=0.1];
		\draw (3,2.5) -- (0,2.5);
		\node at (-0.5,2.5) {\Large $i$};
		\node at (3.5,2.5) {\Large $i$};
		\end{tikzpicture}
		&
		\begin{tikzpicture}[scale = 0.40,thick]
		\draw (0,0)--(0,5);
		\draw (3,0)--(3,5);
		\draw[fill] (3,2.5) circle [radius=0.1];
		\draw[fill] (0,2.5) circle [radius=0.1];
		\draw (3,2.5) arc [start angle = 0,end angle = 180,radius = 1.5];
		\draw (3,2.5) arc [start angle = 360,end angle = 180,radius = 1.5];
		\node at (-0.5,2.5) {\Large $i$};
		\node at (3.5,2.5) {\Large $i$};
		\end{tikzpicture}
		&
        \begin{tikzpicture}[scale = 0.40,thick]
		\draw (0,0)--(0,5);
		\draw (3,0)--(3,5);
		\draw[fill] (3,1) circle [radius=0.1];
		\draw[fill] (3,4) circle [radius=0.1];
		\draw[fill] (0,1) circle [radius=0.1];
		\draw[fill] (0,4) circle [radius=0.1];
		\draw (3,1) -- (0,1);
		\draw (3,4) -- (0,4);
		\node at (-0.5,1) {\Large $i$};
		\node at (3.5,1) {\Large $i$};
		\node at (-0.5,4) {\Large $j$};
		\node at (3.5,4) {\Large $j$};
		\end{tikzpicture}
		&
		\begin{tikzpicture}[scale = 0.40,thick]
		\draw (0,0)--(0,5);
		\draw (3,0)--(3,5);
		\draw[fill] (3,1) circle [radius=0.1];
		\draw[fill] (3,4) circle [radius=0.1];
		\draw[fill] (0,1) circle [radius=0.1];
		\draw[fill] (0,4) circle [radius=0.1];
		\draw (3,1) -- (0,4);
		\draw (3,4) -- (0,1);
		\node at (-0.5,1) {\Large $i$};
		\node at (3.5,1) {\Large $i$};
		\node at (-0.5,4) {\Large $j$};
		\node at (3.5,4) {\Large $j$};
		\end{tikzpicture}
		\\
		(a) w1& (b) w2 & (c) w3 & (d) b7=c7\\
	\end{tabular}
	\caption{Diagrammatic representation of $H_1^0$ (a) and $H_2^0$ (b-d)}
	\label{Diagrams-0}
\end{figure*}

\begin{figure*}[!htp]
	\centering
	\begin{tabular}{cccccc}
		\begin{tikzpicture}[scale = 0.40,thick]
		\draw (0,0)--(0,5);
		\draw (3,0)--(3,5);
		\draw[fill] (3,4) circle [radius=0.1];
		\draw[fill] (0,1) circle [radius=0.1];
		\draw (3,4) -- (0,1);
		\node at (-0.5,1) {\Large $i$};
		\node at (3.5,4) {\Large $j$};
		\end{tikzpicture}
		&
		\begin{tikzpicture}[scale = 0.40,thick]
		\draw (0,0)--(0,5);
		\draw (3,0)--(3,5);
		\draw[fill] (3,1) circle [radius=0.1];
		\draw[fill] (0,4) circle [radius=0.1];
		\draw (3,1) -- (0,4);
		\node at (3.5,1) {\Large $j$};
		\node at (-0.5,4) {\Large $i$};
		\end{tikzpicture}
		&
		\begin{tikzpicture}[scale = 0.40,thick]
		\draw (0,0)--(0,5);
		\draw (3,0)--(3,5);
		\draw[fill] (3,4) circle [radius=0.1];
		\draw[fill] (3,1) circle [radius=0.1];
		\draw[fill] (0,1) circle [radius=0.1];
		\draw (3,4) -- (0,1);
		\draw (3,1) -- (0,1);
		\node at (-0.5,1) {\Large $i$};
		\node at (3.5,1) {\Large $i$};
		\node at (3.5,4) {\Large $j$};
		\end{tikzpicture}
		&
		\begin{tikzpicture}[scale = 0.40,thick]
		\draw (0,0)--(0,5);
		\draw (3,0)--(3,5);
		\draw[fill] (3,1) circle [radius=0.1];
		\draw[fill] (0,4) circle [radius=0.1];
		\draw[fill] (3,4) circle [radius=0.1];
		\draw (3,1) -- (0,4);
		\draw (3,4) -- (0,4);
		\node at (3.5,1) {\Large $j$};
		\node at (-0.5,4) {\Large $i$};
		\node at (3.5,4) {\Large $i$};
		\end{tikzpicture}
		&
		\begin{tikzpicture}[scale = 0.40,thick]
		\draw (0,0)--(0,5);
		\draw (3,0)--(3,5);
		\draw[fill] (3,4) circle [radius=0.1];
		\draw[fill] (0,1) circle [radius=0.1];
		\draw[fill] (0,4) circle [radius=0.1];
		\draw (3,4) -- (0,1);
		\draw (3,4) -- (0,4);
		\node at (-0.5,1) {\Large $i$};
		\node at (3.5,4) {\Large $j$};
		\node at (-0.5,4) {\Large $j$};
		\end{tikzpicture}
		&
		\begin{tikzpicture}[scale = 0.40,thick]
		\draw (0,0)--(0,5);
		\draw (3,0)--(3,5);
		\draw[fill] (3,1) circle [radius=0.1];
		\draw[fill] (0,4) circle [radius=0.1];
		\draw[fill] (0,1) circle [radius=0.1];
		\draw (3,1) -- (0,4);
		\draw (3,1) -- (0,1);
		\node at (3.5,1) {\Large $j$};
		\node at (-0.5,4) {\Large $i$};
		\node at (-0.5,1) {\Large $j$};
		\end{tikzpicture}
		\\
		(a) s1& (b) s2& (c) s3 & (d) s4 &(e) s5 & (f) s6\\ \\
		\begin{tikzpicture}[scale = 0.40,thick]
		\draw (0,0)--(0,5);
		\draw (3,0)--(3,5);
		\draw[fill] (3,4) circle [radius=0.1];
		\draw[fill] (0,4) circle [radius=0.1];
		\draw[fill] (3,2.5) circle [radius=0.1];
		\draw[fill] (0,1) circle [radius=0.1];
		\draw (3,2.5) -- (0,1);
		\draw (3,4) -- (0,4);
		\node at (-0.5,4) {\Large $k$};
		\node at (3.5,2.5) {\Large $j$};
		\node at (-0.5,1) {\Large $i$};
		\node at (3.5,4) {\Large $k$};
		\end{tikzpicture}
        &
		\begin{tikzpicture}[scale = 0.40,thick]
		\draw (0,0)--(0,5);
		\draw (3,0)--(3,5);
		\draw[fill] (3,4) circle [radius=0.1];
		\draw[fill] (0,4) circle [radius=0.1];
		\draw[fill] (3,1) circle [radius=0.1];
		\draw[fill] (0,2.5) circle [radius=0.1];
		\draw (3,1) -- (0,2.5);
		\draw (3,4) -- (0,4);
		\node at (-0.5,4) {\Large $k$};
		\node at (3.5,1) {\Large $j$};
		\node at (-0.5,2.5) {\Large $i$};
		\node at (3.5,4) {\Large $k$};
		\end{tikzpicture}
        &
		\begin{tikzpicture}[scale = 0.40,thick]
		\draw (0,0)--(0,5);
		\draw (3,0)--(3,5);
		\draw[fill] (3,1) circle [radius=0.1];
		\draw[fill] (3,4) circle [radius=0.1];
		\draw[fill] (0,1) circle [radius=0.1];
		\draw[fill] (0,2.5) circle [radius=0.1];
		\draw (3,1) -- (0,1);
		\draw (3,4) -- (0,2.5);
		\node at (-0.5,2.5) {\Large $i$};
		\node at (3.5,1) {\Large $k$};
		\node at (-0.5,1) {\Large $k$};
		\node at (3.5,4) {\Large $j$};
		\end{tikzpicture}
		&
		\begin{tikzpicture}[scale = 0.40,thick]
		\draw (0,0)--(0,5);
		\draw (3,0)--(3,5);
		\draw[fill] (3,1) circle [radius=0.1];
		\draw[fill] (3,2.5) circle [radius=0.1];
		\draw[fill] (0,1) circle [radius=0.1];
		\draw[fill] (0,1) circle [radius=0.1];
		\draw (3,1) -- (0,1);
		\draw (3,2.5) -- (0,4);
		\node at (-0.5,4) {\Large $i$};
		\node at (3.5,1) {\Large $k$};
		\node at (-0.5,1) {\Large $k$};
		\node at (3.5,2.5) {\Large $j$};
		\end{tikzpicture}
		&
		\begin{tikzpicture}[scale = 0.40,thick]
		\draw (0,0)--(0,5);
		\draw (3,0)--(3,5);
		\draw[fill] (3,2.5) circle [radius=0.1];
		\draw[fill] (3,4) circle [radius=0.1];
		\draw[fill] (0,1) circle [radius=0.1];
		\draw[fill] (0,2.5) circle [radius=0.1];
		\draw (3,2.5) -- (0,2.5);
		\draw (3,4) -- (0,1);
		\node at (-0.5,1) {\Large $i$};
		\node at (3.5,2.5) {\Large $k$};
		\node at (-0.5,2.5) {\Large $k$};
		\node at (3.5,4) {\Large $j$};
		\end{tikzpicture}
		&
		\begin{tikzpicture}[scale = 0.40,thick]
		\draw (0,0)--(0,5);
		\draw (3,0)--(3,5);
		\draw[fill] (3,2.5) circle [radius=0.1];
		\draw[fill] (3,1) circle [radius=0.1];
		\draw[fill] (0,4) circle [radius=0.1];
		\draw[fill] (0,2.5) circle [radius=0.1];
		\draw (3,2.5) -- (0,2.5);
		\draw (3,1) -- (0,4);
		\node at (-0.5,4) {\Large $i$};
		\node at (3.5,2.5) {\Large $k$};
		\node at (-0.5,2.5) {\Large $k$};
		\node at (3.5,1) {\Large $j$};
		\end{tikzpicture}\\
		(g) s7& (h) s8& (i) s9& (j) s10& (k) s11& (l) s12\\ \\
		\begin{tikzpicture}[scale = 0.4,thick]
		\draw (0,0)--(0,5);
		\draw (3,0)--(3,5);
		\draw[fill] (3,4) circle [radius=0.1];
		\draw[fill] (3,2.5) circle [radius=0.1];
		\draw[fill] (0,1) circle [radius=0.1];
		\draw[fill] (0,4) circle [radius=0.1];
		\draw (3,2.5) -- (0,4);
		\draw (3,4) -- (0,1);
		\node at (-0.5,4) {\Large $k$};
		\node at (3.5,4) {\Large $k$};
		\node at (-0.5,1) {\Large $i$};
		\node at (3.5,2.5) {\Large $j$};
		\end{tikzpicture}
		&
        \begin{tikzpicture}[scale = 0.4,thick]
		\draw (0,0)--(0,5);
		\draw (3,0)--(3,5);
		\draw[fill] (3,4) circle [radius=0.1];
		\draw[fill] (3,1) circle [radius=0.1];
		\draw[fill] (0,2.5) circle [radius=0.1];
		\draw[fill] (0,4) circle [radius=0.1];
		\draw (3,1) -- (0,4);
		\draw (3,4) -- (0,2.5);
		\node at (-0.5,4) {\Large $k$};
		\node at (3.5,4) {\Large $k$};
		\node at (-0.5,2.5) {\Large $i$};
		\node at (3.5,1) {\Large $j$};
		\end{tikzpicture}
		&
        \begin{tikzpicture}[scale = 0.4,thick]
		\draw (0,0)--(0,5);
		\draw (3,0)--(3,5);
		\draw[fill] (3,1) circle [radius=0.1];
		\draw[fill] (3,4) circle [radius=0.1];
		\draw[fill] (0,1) circle [radius=0.1];
		\draw[fill] (0,3) circle [radius=0.1];
		\draw (3,1) -- (0,3);
		\draw (3,4) -- (0,1);
		\node at (-0.5,1) {\Large $k$};
		\node at (3.5,1) {\Large $k$};
		\node at (-0.5,3) {\Large $i$};
		\node at (3.5,4) {\Large $j$};
		\end{tikzpicture}
		&
        \begin{tikzpicture}[scale = 0.4,thick]
		\draw (0,0)--(0,5);
		\draw (3,0)--(3,5);
		\draw[fill] (3,1) circle [radius=0.1];
		\draw[fill] (3,3) circle [radius=0.1];
		\draw[fill] (0,1) circle [radius=0.1];
		\draw[fill] (0,4) circle [radius=0.1];
		\draw (3,1) -- (0,4);
		\draw (3,3) -- (0,1);
		\node at (-0.5,1) {\Large $k$};
		\node at (3.5,1) {\Large $k$};
		\node at (-0.5,4) {\Large $i$};
		\node at (3.5,3) {\Large $j$};
		\end{tikzpicture}
		&
        \begin{tikzpicture}[scale = 0.4,thick]
		\draw (0,0)--(0,5);
		\draw (3,0)--(3,5);
		\draw[fill] (3,4) circle [radius=0.1];
		\draw[fill] (3,2.5) circle [radius=0.1];
		\draw[fill] (0,2.5) circle [radius=0.1];
		\draw[fill] (0,1) circle [radius=0.1];
		\draw (3,2.5) -- (0,1);
		\draw (3,4) -- (0,2.5);
		\node at (-0.5,2.5) {\Large $k$};
		\node at (3.5,2.5) {\Large $k$};
		\node at (-0.5,1) {\Large $i$};
		\node at (3.5,4) {\Large $j$};
		\end{tikzpicture}
		&
        \begin{tikzpicture}[scale = 0.4,thick]
		\draw (0,0)--(0,5);
		\draw (3,0)--(3,5);
		\draw[fill] (3,1) circle [radius=0.1];
		\draw[fill] (3,2.5) circle [radius=0.1];
		\draw[fill] (0,2.5) circle [radius=0.1];
		\draw[fill] (0,4) circle [radius=0.1];
		\draw (3,2.5) -- (0,4);
		\draw (3,1) -- (0,2.5);
		\node at (-0.5,2.5) {\Large $k$};
		\node at (3.5,2.5) {\Large $k$};
		\node at (-0.5,4) {\Large $i$};
		\node at (3.5,1) {\Large $j$};
		\end{tikzpicture}
		\\
		(m) b6& (n) c6& (o) b4& (p) c4& (q) a2& (r) d2\\
	\end{tabular}
	\caption{Diagrammatic representation of $H_1^1$ (a-b) and $H_2^1$ (c-r)}
	\label{Diagrams-1}
\end{figure*}

\clearpage
\newpage

\begin{figure*}[!htp]
	\centering
	\begin{tabular}{cccccc}
		\begin{tikzpicture}[scale = 0.4,thick]
		\draw (0,0)--(0,5);
		\draw (3,0)--(3,5);
		\draw[fill] (3,4) circle [radius=0.1];
		\draw[fill] (3,2) circle [radius=0.1];
		\draw[fill] (0,3) circle [radius=0.1];
		\draw[fill] (0,1) circle [radius=0.1];
		\draw (3,2) -- (0,1);
		\draw (3,4) -- (0,3);
		\node at (-0.5,1) {\Large $i$};
		\node at (-0.5,3) {\Large $k$};
		\node at (3.5,2) {\Large $j$};
		\node at (3.5,4) {\Large $l$};
		\end{tikzpicture}
		&
		\begin{tikzpicture}[scale = 0.4,thick]
		\draw (0,0)--(0,5);
		\draw (3,0)--(3,5);
		\draw[fill] (3,4) circle [radius=0.1];
		\draw[fill] (3,3) circle [radius=0.1];
		\draw[fill] (0,2) circle [radius=0.1];
		\draw[fill] (0,1) circle [radius=0.1];
		\draw (3,3) -- (0,1);
		\draw (3,4) -- (0,2);
		\node at (-0.5,1) {\Large $i$};
		\node at (-0.5,2) {\Large $k$};
		\node at (3.5,3) {\Large $j$};
		\node at (3.5,4) {\Large $l$};
		\end{tikzpicture}
		&
		\begin{tikzpicture}[scale = 0.4,thick]
		\draw (0,0)--(0,5);
		\draw (3,0)--(3,5);
		\draw[fill] (0,1) circle [radius=0.1];
		\draw[fill] (3,4) circle [radius=0.1];
		\draw[fill] (3,2.5) circle [radius=0.1];
		\draw (0,1) -- (3,4);
		\draw (0,1) -- (3,2.5);
		\node at (3.5,2.5) {\Large $j$};
		\node at (3.5,4) {\Large $l$};
		\node at (-0.5,1) {\Large $i$};
		\end{tikzpicture}
		&
		\begin{tikzpicture}[scale = 0.4,thick]
		\draw (6,0)--(6,5);
		\draw (9,0)--(9,5);
		\draw[fill] (9,4) circle [radius=0.1];
		\draw[fill] (9,3) circle [radius=0.1];
		\draw[fill] (6,2) circle [radius=0.1];
		\draw[fill] (6,1) circle [radius=0.1];
		\draw (9,3) -- (6,2);
		\draw (9,4) -- (6,1);
		\node at (5.5,1) {\Large $i$};
		\node at (5.5,2) {\Large $k$};
		\node at (9.5,3) {\Large $j$};
		\node at (9.5,4) {\Large $l$};
		\end{tikzpicture}
		&
		\begin{tikzpicture}[scale = 0.4,thick]
		\draw (0,0)--(0,5);
		\draw (3,0)--(3,5);
		\draw[fill] (3,4) circle [radius=0.1];
		\draw[fill] (0,2.5) circle [radius=0.1];
		\draw[fill] (0,1) circle [radius=0.1];
		\draw (3,4) -- (0,2.5);
		\draw (3,4) -- (0,1);
		\node at (-0.5,1) {\Large $i$};
		\node at (3.5,4) {\Large $j$};
		\node at (-0.5,2.5) {\Large $k$};
		\end{tikzpicture}
		&
		\begin{tikzpicture}[scale = 0.4,thick]
		\draw (0,0)--(0,5);
		\draw (3,0)--(3,5);
		\draw[fill] (3,4) circle [radius=0.1];
		\draw[fill] (0,1) circle [radius=0.1];
		\draw (0,1) arc [start angle = 270,end angle = 360,radius = 3];
		\draw (3,4) arc [start angle = 90,end angle = 180,radius = 3];
		\node at (3.5,4) {\Large $j$};
		\node at (-0.5,1) {\Large $i$};
		\end{tikzpicture}
		\\
		(a) a1 & (b) a3 & (c) a4 & (d) a5 & (e) a6 & (f) a7 \\
		\begin{tikzpicture}[scale = 0.4,thick]
		\draw (0,0)--(0,5);
		\draw (3,0)--(3,5);
		\draw[fill] (3,4) circle [radius=0.1];
		\draw[fill] (3,1) circle [radius=0.1];
		\draw[fill] (0,3) circle [radius=0.1];
		\draw[fill] (0,2) circle [radius=0.1];
		\draw (3,1) -- (0,2);
		\draw (3,4) -- (0,3);
		\node at (-0.5,2) {\Large $i$};
		\node at (-0.5,3) {\Large $k$};
		\node at (3.5,1) {\Large $j$};
		\node at (3.5,4) {\Large $l$};
		\end{tikzpicture}
		&
		\begin{tikzpicture}[scale = 0.4, thick]
		\draw (0,0)--(0,5);
		\draw (3,0)--(3,5);
		\draw[fill] (0,2.5) circle [radius=0.1];
		\draw[fill] (3,1) circle [radius=0.1];
		\draw[fill] (3,4) circle [radius=0.1];
		\draw (0,2.5) -- (3,4);
		\draw (0,2.5) -- (3,1);
		\node at (3.5,1) {\Large $j$};
		\node at (3.5,4) {\Large $l$};
		\node at (-0.5,2.5) {\Large $i$};
		\end{tikzpicture}
		&
		\begin{tikzpicture}[scale = 0.4,thick]
		\draw (6,0)--(6,5);
		\draw (9,0)--(9,5);
		\draw[fill] (9,4) circle [radius=0.1];
		\draw[fill] (9,1) circle [radius=0.1];
		\draw[fill] (6,3) circle [radius=0.1];
		\draw[fill] (6,2) circle [radius=0.1];
		\draw (9,1) -- (6,3);
		\draw (9,4) -- (6,2);
		\node at (5.5,2) {\Large $i$};
		\node at (5.5,3) {\Large $k$};
		\node at (9.5,1) {\Large $j$};
		\node at (9.5,4) {\Large $l$};
		\end{tikzpicture}
		&
		\begin{tikzpicture}[scale = 0.4,thick]
		\draw (6,0)--(6,5);
		\draw (9,0)--(9,5);
		\draw[fill] (9,4) circle [radius=0.1];
		\draw[fill] (9,2) circle [radius=0.1];
		\draw[fill] (6,3) circle [radius=0.1];
		\draw[fill] (6,1) circle [radius=0.1];
		\draw (9,2) -- (6,3);
		\draw (9,4) -- (6,1);
		\node at (5.5,1) {\Large $i$};
		\node at (5.5,3) {\Large $k$};
		\node at (9.5,2) {\Large $j$};
		\node at (9.5,4) {\Large $l$};
		\end{tikzpicture}
		&

		&
		
		\\
		(g) b1& (h) b2 & (i) b3 & (j) b5 &\\
		\begin{tikzpicture}[scale = 0.4,thick]
		\draw (0,0)--(0,5);
		\draw (3,0)--(3,5);
		\draw[fill] (3,3) circle [radius=0.1];
		\draw[fill] (3,2) circle [radius=0.1];
		\draw[fill] (0,4) circle [radius=0.1];
		\draw[fill] (0,1) circle [radius=0.1];
		\draw (3,2) -- (0,1);
		\draw (3,3) -- (0,4);
		\node at (-0.5,1) {\Large $i$};
		\node at (-0.5,4) {\Large $k$};
		\node at (3.5,2) {\Large $j$};
		\node at (3.5,3) {\Large $l$};
		\end{tikzpicture}
		&
		\begin{tikzpicture}[scale = 0.4,thick]
		\draw (0,0)--(0,5);
		\draw (3,0)--(3,5);
		\draw[fill] (3,2.5) circle [radius=0.1];
		\draw[fill] (0,4) circle [radius=0.1];
		\draw[fill] (0,1) circle [radius=0.1];
		\draw (3,2.5) -- (0,4);
		\draw (3,2.5) -- (0,1);
		\node at (-0.5,1) {\Large $i$};
		\node at (-0.5,4) {\Large $k$};
		\node at (3.5,2.5) {\Large $j$};
		\end{tikzpicture}
		&
		\begin{tikzpicture}[scale = 0.4,thick]
		\draw (6,0)--(6,5);
		\draw (9,0)--(9,5);
		\draw[fill] (9,3) circle [radius=0.1];
		\draw[fill] (9,2) circle [radius=0.1];
		\draw[fill] (6,4) circle [radius=0.1];
		\draw[fill] (6,1) circle [radius=0.1];
		\draw (9,2) -- (6,4);
		\draw (9,3) -- (6,1);
		\node at (5.5,1) {\Large $i$};
		\node at (5.5,4) {\Large $k$};
		\node at (9.5,2) {\Large $j$};
		\node at (9.5,3) {\Large $l$};
		\end{tikzpicture}
		&
		\begin{tikzpicture}[scale = 0.4,thick]
		\draw (6,0)--(6,5);
		\draw (9,0)--(9,5);
		\draw[fill] (9,3) circle [radius=0.1];
		\draw[fill] (9,1) circle [radius=0.1];
		\draw[fill] (6,4) circle [radius=0.1];
		\draw[fill] (6,2) circle [radius=0.1];
		\draw (9,1) -- (6,4);
		\draw (9,3) -- (6,2);
		\node at (5.5,2) {\Large $i$};
		\node at (5.5,4) {\Large $k$};
		\node at (9.5,1) {\Large $j$};
		\node at (9.5,3) {\Large $l$};
		\end{tikzpicture}
		&
		
		&

		\\
		(k) c1& (l) c2 & (m) c3 & (n) c5 & \\
		\begin{tikzpicture}[scale = 0.4,thick]
		\draw (0,0)--(0,5);
		\draw (3,0)--(3,5);
		\draw[fill] (3,3) circle [radius=0.1];
		\draw[fill] (3,1) circle [radius=0.1];
		\draw[fill] (0,4) circle [radius=0.1];
		\draw[fill] (0,2) circle [radius=0.1];
		\draw (3,1) -- (0,2);
		\draw (3,3) -- (0,4);
		\node at (-0.5,2) {\Large $i$};
		\node at (-0.5,4) {\Large $k$};
		\node at (3.5,1) {\Large $j$};
		\node at (3.5,3) {\Large $l$};
		\end{tikzpicture}
		&
		\begin{tikzpicture}[scale = 0.4,thick]
		\draw (0,0)--(0,5);
		\draw (3,0)--(3,5);
		\draw[fill] (3,1) circle [radius=0.1];
		\draw[fill] (3,2) circle [radius=0.1];
		\draw[fill] (0,3) circle [radius=0.1];
		\draw[fill] (0,4) circle [radius=0.1];
		\draw (3,1) -- (0,3);
		\draw (3,2) -- (0,4);
		\node at (-0.5,3) {\Large $i$};
		\node at (-0.5,4) {\Large $k$};
		\node at (3.5,1) {\Large $j$};
		\node at (3.5,2) {\Large $l$};
		\end{tikzpicture}
		&
		\begin{tikzpicture}[scale = 0.4,thick]
		\draw (0,0)--(0,5);
		\draw (3,0)--(3,5);
		\draw[fill] (3,1) circle [radius=0.1];
		\draw[fill] (0,4) circle [radius=0.1];
		\draw[fill] (0,2.5) circle [radius=0.1];
		\draw (3,1) -- (0,4);
		\draw (3,1) -- (0,2.5);
		\node at (-0.5,2.5) {\Large $i$};
		\node at (-0.5,4) {\Large $k$};
		\node at (3.5,1) {\Large $j$};
		\end{tikzpicture}
		&
		\begin{tikzpicture}[scale = 0.4,thick]
		\draw (6,0)--(6,5);
		\draw (9,0)--(9,5);
		\draw[fill] (9,2) circle [radius=0.1];
		\draw[fill] (9,1) circle [radius=0.1];
		\draw[fill] (6,4) circle [radius=0.1];
		\draw[fill] (6,3) circle [radius=0.1];
		\draw (9,1) -- (6,4);
		\draw (9,2) -- (6,3);
		\node at (5.5,3) {\Large $i$};
		\node at (5.5,4) {\Large $k$};
		\node at (9.5,1) {\Large $j$};
		\node at (9.5,2) {\Large $l$};
		\end{tikzpicture}
		&
		\begin{tikzpicture}[scale = 0.4,thick]
		\draw (0,0)--(0,5);
		\draw (3,0)--(3,5);
		\draw[fill] (0,4) circle [radius=0.1];
		\draw[fill] (3,1) circle [radius=0.1];
		\draw[fill] (3,2.5) circle [radius=0.1];
		\draw (0,4) -- (3,2.5);
		\draw (0,4) -- (3,1);
		\node at (3.5,1) {\Large $j$};
		\node at (3.5,2.5) {\Large $l$};
		\node at (-0.5,4) {\Large $i$};
		\end{tikzpicture}
		&
		\begin{tikzpicture}[scale = 0.4,thick]
		\draw (0,0)--(0,5);
		\draw (3,0)--(3,5);
		\draw[fill] (3,1) circle [radius=0.1];
		\draw[fill] (0,4) circle [radius=0.1];
		\draw (3,1) arc [start angle = 0,end angle = 90,radius = 3];
		\draw (0,4) arc [start angle = 180,end angle = 270,radius = 3];
		\node at (3.5,1) {\Large $j$};
		\node at (-0.5,4) {\Large $i$};
		\end{tikzpicture}
		\\
		(o) d1 & (p) d3 & (q) d4 & (r) d5 & (s) d6 & (t) d7 \\

	\end{tabular}
	\caption{Diagrammatic representation of $H_2^2$ }
	\label{Diagrams-2}
\end{figure*}